\documentclass[pra,groupedaddress,superscriptaddress,nofootinbib,floatfix,preprintnumbers,longbibliography,notitlepage,tightenlines,twocolumn,aps]{revtex4-2}

\usepackage{amssymb,amsmath,graphicx,color,dsfont,mathrsfs,tikz-cd}
\usepackage{bm}
\usepackage[export]{adjustbox}
\usepackage{parskip}
\usepackage{braket}
\usepackage{amsfonts}
\usepackage{amssymb}
\usepackage{comment}
\usepackage{amsthm}
\usepackage{enumerate}
\usepackage{bm}
\usepackage{float}
\usepackage{array}
\usepackage{xfrac}
\usepackage{tikz}
\usepackage{quantikz}
\usepackage{url}
\usepackage{xcolor}
\usepackage{multirow}
\usepackage{makecell}
\usepackage{rotating,booktabs}
\usepackage[verbose]{placeins}
\usepackage{ragged2e}
\usepackage[titletoc]{appendix}
\usepackage{mdframed}

\usepackage[colorlinks=true,citecolor=blue,linkcolor=blue]{hyperref}
\usepackage{cleveref}

\newcommand\ha{\hat{a}}
\newcommand\phih{\hat{\phi}}
\newcommand\pih{\hat{\pi}}
\newcommand\bs[1]{\boldsymbol{#1}}

\newcommand{\norm}[1]{\left\lVert#1\right\rVert}
\newsavebox{\mstrut}

\newcommand{\ketbra}[2]{\ket {#1} \hskip -0.8ex \bra {#2}}
\newcommand{\gket}[1]{|#1\rangle}
\newcommand{\gbra}[1]{\langle#1|}
\newcommand{\gbraket}[2]{\langle#1|#2\rangle}

\dimen\footins=5\baselineskip\relax



\begin{document}

\preprint{UMD-PP-025-03}

\title{Euclidean-Monte-Carlo-informed ground-state preparation 
\\
for quantum simulation of scalar field theory}

\author{Navya Gupta}
\email{navyag@umd.edu} \thanks{corresponding author}
\affiliation{
Maryland Center for Fundamental Physics and Department of Physics, 
University of Maryland, College Park, MD 20742, USA}
\affiliation{Joint Center for Quantum Information and Computer Science,
NIST/University of Maryland, College Park, MD 20742 USA}
\affiliation{The NSF Institute for Robust Quantum Simulation, University of Maryland, College Park, Maryland 20742, USA}

\author{Christopher David White}
\email{christopher.d.white117.ctr@us.navy.mil}
\affiliation{Joint Center for Quantum Information and Computer Science,
NIST/University of Maryland, College Park, MD 20742 USA}
\affiliation{Center for Computational Materials Science, U.S. Naval Research Laboratory, 
\\
Washington, D.C. 20375, USA}

\author{Zohreh Davoudi}
\email{davoudi@umd.edu}
\affiliation{
Maryland Center for Fundamental Physics and Department of Physics, 
University of Maryland, College Park, MD 20742, USA}
\affiliation{Joint Center for Quantum Information and Computer Science,
NIST/University of Maryland, College Park, MD 20742 USA}
\affiliation{The NSF Institute for Robust Quantum Simulation, University of Maryland, College Park, Maryland 20742, USA}

\begin{abstract}
    Quantum simulators offer great potential for investigating dynamical properties of quantum field theories. However, preparing accurate non-trivial initial states for these simulations is challenging. Classical Euclidean-time Monte-Carlo methods provide a wealth of information about states of interest to quantum simulations. Thus, it is desirable to facilitate state preparation on quantum simulators using this information. To this end, we present a fully classical pipeline for generating efficient quantum circuits for preparing the ground state of an interacting scalar field theory in 1+1 dimensions. The first element of this pipeline is a variational ansatz family based on the stellar hierarchy for bosonic quantum systems. The second element of this pipeline is the classical moment-optimization procedure that augments the standard variational energy minimization by penalizing deviations in selected sets of ground-state correlation functions (i.e., moments). The values of ground-state moments are sourced from classical Euclidean methods. The resulting states yield comparable ground-state energy estimates but exhibit distinct correlations and local non-Gaussianity. The third element of this pipeline is translating the moment-optimized ansatz into an efficient quantum circuit with an asymptotic cost that is polynomial in system size. This work opens the way to systematically applying classically obtained knowledge of states to prepare accurate initial states in quantum field theories of interest in nature.
\end{abstract}

\maketitle

\section{Introduction}
\noindent
Classical-computational methods, rooted in the path-integral formulation of quantum mechanics, continue to advance non-perturbative studies of strongly-interacting quantum field theories~\cite{wilson1974confinement,creutz1983monte,montvay1994quantum,rothe2012lattice,gattringer2009quantum}. For example, several properties of hadrons and light nuclei, such as their low-lying spectra, structure, and low-energy reaction rates, and of low-density matter at finite temperatures, can now be computed from the underlying quantum chromodynamics (QCD)~\cite{aoki2024flag,usqcd2019hot,davoudi2022report,kronfeld2022lattice}. These studies are performed in Euclidean spacetime to be amenable to Monte-Carlo sampling methods---methods which only scale polynomially with system size. Two-point correlation functions decay as a function of Euclidean time, and at late times are dominated by the ground-state contribution. Three- and higher-point functions are also accessible via Euclidean Monte-Carlo methods. Using theoretical frameworks such as finite-volume formalisms~\cite{luscher1991two,lellouch2001weak}, such Euclidean correlation functions can be related to scattering amplitudes in the low-energy, low-inelasticity regimes~\cite{briceno2015nuclear,briceno2018scattering,hansen2019lattice,davoudi2021nuclear}. The Euclidean signature used in these computations, however, hinders access to general dynamical, Minkowski-time observables, and direct Minkowski-time Monte-Carlo computations suffer from a sign problem~\cite{gattringer2016approaches,cohen2015taming}. Sign problems also hinder simulations of systems at finite baryon density, even in Euclidean spacetime~\cite{troyer2005computational,goy2017sign,nagata2022finite}.

Quantum simulators are expected to overcome these hurdles: they enable a sign-problem-free simulation of real-time dynamics, and offer an efficient encoding of the Hilbert space of the simulated quantum system~\cite{klco2022standard,bauer2023quantum,bauer2023quantum2,di2024quantum,beck2023quantum}. Several algorithms for simulating quantum field theories, including gauge theories of relevance to the Standard Model of particle physics~\cite{byrnes2006simulating,shaw2020quantum, ciavarella2021trailhead, kan2021lattice,lamm2019general,haase2021resource,davoudi2022general,murairi2022many,rhodes2024exponential,lamm2024block,balaji2025quantum}, have been proposed in recent years. 
These algorithms have mainly focused on concrete resource analyses for simulating time evolution of a generic quantum state. One major bottleneck, however, is the preparation of non-trivial initial states, which are often ground and low-energy excited states or their superpositions~\cite{jordan2011quantum, jordan2012quantum,jordan2018bqp}. 
In fact, ground-state preparation has been shown to be QMA-complete\footnote{QMA-complete complexity class is the quantum counterpart of classical Merlin-Arthur (MA)-complete (i.e., probabilistic NP-complete) class~\cite{bookatz2012qma}.} for certain quantum systems~\cite{kempe2006complexity,kitaev2002classical}. In the context of quantum-field-theory simulations, the adiabatic preparation of the initial state in the scattering algorithms of Refs.~\cite{jordan2011quantum, jordan2012quantum,jordan2018bqp,marshall2015quantum,marshall2015quantum} constitutes the most computationally demanding step. One often needs to resort to preparing approximate ground states, and to using the knowledge of states and symmetries to find efficient routes to their preparation. 

More precisely, preparing a state involves determining and implementing the quantum circuit which realizes that state on a quantum simulator (assuming some exact or approximate mapping between the theory's and simulator's degrees of freedom). In many cases, the wavefunction of the desired state is already known, and the main state-preparation task is to approximate this known wavefunction using a quantum circuit~\cite{bagherimehrab2022nearly,klco2020minimally, klco2020systematically, klco2020fixed, kitaev2008wavefunction}. In this work, we are primarily interested in cases where the wavefunction is not known and in fact, is not expected to possess an analytical form (as is the case for general interacting quantum field theories). Thus, the state-preparation task also includes direct or indirect computation of this state. Different state-preparation algorithms can be classified by the source(s) of their dominant computational cost (quantum, classical, or both) together with their guarantees for accuracy (provably accurate or heuristic).

On the one hand, there are quantum-cost-dominated algorithms, which often offer provable guarantees on the accuracy of the prepared ground state. In such cases, modest classical resources can be used to determine quantum circuits which implement certain quantum routines. For example, real-time evolution is used in adiabatic quantum computing to turn the easily prepared ground state of an initial Hamiltonian into that of a final, desired Hamiltonian by slowly varying Hamiltonian parameters
~\cite{jordan2011quantum, jordan2012quantum,jordan2018bqp,marshall2015quantum,marshall2015quantum}; quantum imaginary-time evolution is used to suppress the excited-state components of an initial, trial state at long imaginary times~\cite{nishi2021implementation}; eigenvalue-estimation methods such as quantum phase estimation are used to post-select the low-lying eigenstates of the system by measuring the phase of the state~\cite{ge2019faster}; iterative filtering methods are used to suppress the excited-state components of a trial initial state upon repetitive action of a suitable operator~\cite{
dong2022ground,lin2020near}; 
quantum measurements are used in projective ground-state preparations~\cite{choi2021rodeo,cohen2024efficient, luis2001quantum} and reservoir-based methods~\cite{zhan2025rapid, ding2024single} to approach the ground state. 
While these methods often enjoy provable success and accuracy guarantees, they require deep coherent quantum circuits. For this reason, such quantum algorithms are often untenable for implementation on near-term quantum hardware.
Most algorithms, further, require a suitable, easily prepared trial state, and the depth or the success probability of the algorithm may depend on the overlap between the trial state and the desired state to be prepared~\cite{dong2022ground, choi2021rodeo, lin2020near, ge2019faster}. 

In contrast, variational quantum algorithms trade these deeper quantum circuits for shallower, parametrized quantum circuits, which are optimized by a quantum-classical feedback loop: the quantum simulator estimates the value of some loss function while the classical computer optimizes it. In the context of quantum chemistry, nuclear physics, and quantum field theories, the parametrized quantum circuit is typically based upon physics-inspired ansatzes (such as Hartree Fock, Unitary Coupled Cluster, approximate full configuration interaction~\cite{feniou2024sparse, tilly2022variational,grimsley2019adaptive,google2020hartree,romero2018strategies,anand2022quantum,romero2022solving,dumitrescu2018cloud,kiss2022quantum,rethinasamy2024neutron}, etc.). One popular variational quantum algorithm is the Variational Quantum Eigensolver (VQE), which minimizes the energy of the output state. It has been used to prepare ground states~\cite{klco2018quantum,atas20212,farrell2024scalable,crippa2024analysis,fromm2024simulating}, scattering wave packets~\cite{farrell2024quantum,davoudi2024scattering,davoudi2025quantum}, and thermal states~\cite{than2024phase,xie2022variational} of gauge theories in recent years. The variational algorithms, nonetheless, face several challenges: the classical optimization may run into barren plateaus~\cite{larocca2024review, tilly2022variational, cerezo2021variational}, and encounter a multitude of local minima~\cite{anschuetz2022beyond} (unless the circuit ansatz and its initialization are chosen very carefully); statistical fluctuations can hinder quantum-assisted classical estimates of the loss function~\cite{larocca2024review, tilly2022variational, cerezo2021variational}; it is not guaranteed that the optimized circuit represents the target ground state accurately; and VQE typically only optimizes the energy, potentially at the cost of reducing accuracy of other physically relevant quantities. Thus, while such variational algorithms offer promising state-preparation routes on near-term quantum computers, their success often hinges upon a good choice for the ansatz.

When additional information about the target wavefunction is available, one may want to perform a more fine-tuned optimization depending upon subsequent tasks. We desire an ansatz that is \textit{accurate}, i.e., it can get sufficiently close to the ground state, is \textit{circuit-translatable}, i.e., it  can be efficiently mapped to a quantum circuit, and is \textit{circuit-efficient}, i.e., the size of the resulting quantum circuit scales with a low-degree polynomial in system size. We will demonstrate an ansatz possessing these qualities for an example bosonic quantum field theory. In fact, we require that both the computation and optimization of the loss function for this ansatz be performed \textit{classically}. Thus, we will fully determine the quantum circuit for preparing ground states in this theory using classical computing, and eliminate any quantum-computing costs beyond implementing this fixed predetermined quantum circuit. We will refer to this property of the ansatz as \textit{classical tractability}. This property is desirable since it avoids the statistical noise associated with quantum computations of the loss function.

\begin{figure*}
    \centering
    \includegraphics[width=\textwidth]{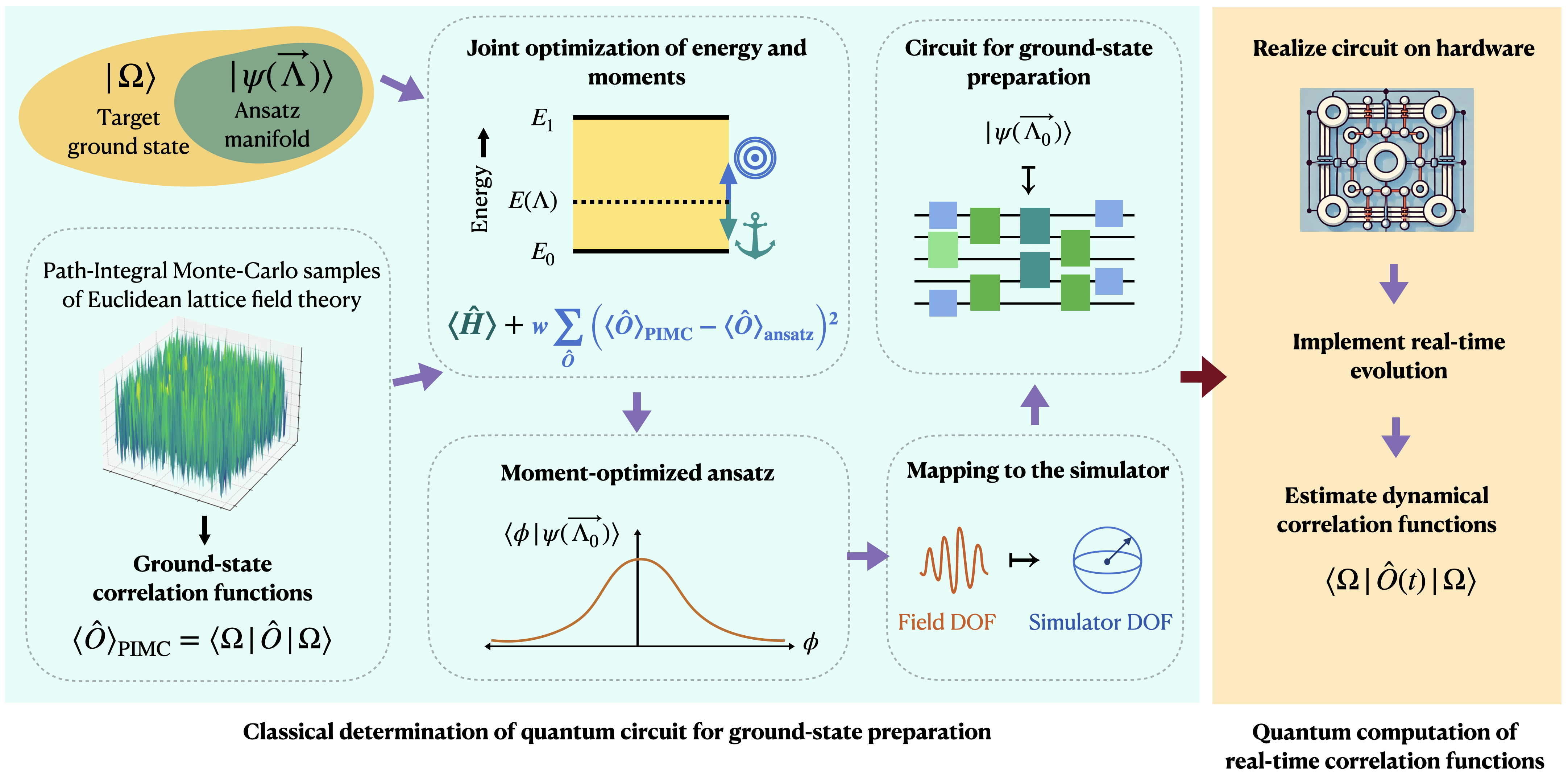}
    \caption{\textit{Schematic overview of classically informed ground-state preparation.} Using ground-state correlation functions in an interacting scalar field theory, sourced from Euclidean path-integral Monte Carlo, a joint optimization of the energy and moments of the ansatz wavefunction is performed. Based on a mapping between the field and simulator degrees of freedom, the optimized ansatz is translated into a quantum circuit using a classical algorithm. This classically determined circuit can thereafter be implemented on quantum hardware, and be used as the starting point for performing other tasks, such as simulating real-time dynamics and estimating dynamical correlation functions. The goal of this paper is to demonstrate the classical determination of the quantum circuit for preparing the ground state.
    }
    \label{fig:pipeline}
\end{figure*}

We would further like to use classically computed properties of ground states to inform the optimization of our variational ansatz. Typically, variational algorithms minimize the ansatz energy and, therefore, assume minimal knowledge about the target ground state. However, one often has access to some special information about ground states. Notably, while Euclidean path-integral Monte-Carlo (PIMC) computations do not provide direct access to the state's wavefunction, they yield static ground-state correlation functions, or ``moments.'' As depicted in Fig.~\ref{fig:pipeline}, we will incorporate this knowledge in our loss function to jointly optimize, on a classical computer, the energy and a set of pre-specified ground-state moments of the ansatz. Once such a ``moment-optimized'' ansatz is obtained, we map the degrees of freedom of the theory to those of the quantum simulator, and determine the quantum circuit for preparing this optimized wavefunction. This classically determined ground-state quantum circuit could then be implemented on quantum hardware.  For example, such a circuit can be used as the starting point for the quantum computation of dynamical properties of the state, which are often classically intractable but are quantumly accessible. In this paper, we will focus on only the classical pipeline for determining the ground-state circuit in a quantum field theory, and will leave studies of real-time dynamics to future work.

We demonstrate this procedure using the simplest non-trivial quantum field theory (QFT), a $\phi^4$ theory in $1+1$ dimensions (D). Despite their simplicity, scalar fields in (3+1)D describe, among other things, pions, the Higgs boson, and scalar inflation fields in nature. Furthermore, scalar field theories are important testing grounds for the development of quantum-simulation algorithms~\cite{jordan2011quantum,jordan2012quantum,jordan2018bqp,marshall2015quantum,klco2019digitization,klco2020fixed,klco2020systematically,klco2020minimally,liu2022towards,barata2021single,hardy2024optimized,yeter2019scalar,thompson2023quantum,li2023simulating,turco2024quantum,zemlevskiy2024scalable,ingoldby2025real,watson2023quantum}. These theories have also been widely studied using sign-problem-free lattice techniques rooted in Euclidean PIMC; see, e.g.,~\cite{loinaz1998monte}. Hence, static ground-state correlation functions can, in principle, be computed using classical computing. As befits a bosonic theory, we will use bosonic ansatzes---states with finite stellar rank~\cite{chabaud2020stellar,chabaud2022holomorphic}---to first estimate and then circuitize the ground-state wavefunction. We show that the structure of this ansatz enables an efficient translation to a quantum circuit on a discrete-variable, i.e., qubit-based platform, and the resulting quantum circuit is efficient, i.e., scales polynomially in the system size. We also comment on the feasibility of preparing this state on a continuous-variable, i.e., bosonic quantum simulator.\\

To summarize, the two main pillars of our work are: a classically tractable, circuit-translatable, and circuit-efficient ansatz for interacting scalar field theory, and the PIMC-informed moment-optimization procedure for optimizing this ansatz. The efficiency of our ansatz stems from the fact that it is directly expressed in terms of the degrees of freedom of the target field theory, making it straightforward to encode features such as correlations and non-Gaussianity in the ansatz. This is in contrast to the more hardware-oriented variational-circuit ansatzes. Our ansatz is further amenable to optimization on a classical computer, i.e., is classically tractable. Given that any ansatz will only express limited features of a nontrivial interacting ground state, we use PIMC-informed moment optimization to tune the ansatz to reproduce the most relevant features of the ground state. Such features depend on the subsequent quantum-simulation goals, e.g., studies of certain excitations and their dynamics. \\

This paper is organized as follows. In Sec.~\ref{sec:euclidean}, we introduce the optimization procedure---Euclidean-Monte-Carlo-informed moment optimization---which is independent of the choice of theory and ansatz. In Secs.~\ref{sec:singlemode} and~\ref{sec:multimode}, we describe the ansatz and optimization procedure for the (0+1)D and (1+1)D $\phi^4$ ground states, respectively. The discussion in the above sections is independent of quantum simulations, and is applicable to Hamiltonian simulations at large. In Sec.~\ref{sec:circ}, we specialize the discussion to quantum simulation by describing the translation of these optimized ansatzes to discrete variable, i.e., qubit-based quantum simulators. We conclude in Sec.~\ref{sec:sum} with a summary and outlook. Several Appendices are presented to provide further details on our methodology, and to supplement our discussions, including a discussion of the viability of continuous variable, i.e., bosonic quantum simulators, for our state-preparation scheme.

\section{Euclidean-Monte-Carlo-informed moment optimization}\label{sec:euclidean}
\noindent

Consider an ansatz family $\ket{\psi(\vec{\Lambda})}$ parametrized by some finite-dimensional vector $\vec{\Lambda}$. 
In general, a finite-dimensional ansatz 
family cannot exactly represent the ground state $\ket{\Omega}$ (assumed to be non-degenerate) of a (lattice-regularized Hilbert-space truncated) quantum-field-theory Hamiltonian $\hat H$. Therefore, the aim is to determine $\vec{\Lambda}$ such that $\ket{\psi(\vec{\Lambda})}$ is a good proxy for  $\ket{\Omega}$. Broadly, we seek a value of $\vec{\Lambda}$ for which the ansatz can be expressed as 
\begin{align}
    \ket{\psi(\vec{\Lambda})}=e^{i\phi(\vec{\Lambda})}\sqrt{1-\big|\epsilon(\vec{\Lambda})\big|^2} \ \ket{\Omega}+\epsilon(\vec{\Lambda}) \ \ket{\Omega^{\perp}(\vec{\Lambda})},
\end{align}
where $\epsilon(\vec{\Lambda})$ has a sufficiently small magnitude, and $\ket{\Omega^{\perp}(\vec{\Lambda})}$ is orthogonal to the ground state. Assuming a non-zero mass gap, $\Delta E$, the ratio of the energy discrepancy to the spectral gap is bounded as:
\begin{align}
    \delta_E(\vec{\Lambda}) &\coloneq\frac{\bra{\psi(\vec{\Lambda})}\hat H\ket{\psi(\vec{\Lambda})}-\bra{\Omega}\hat H\ket{\Omega}}{\Delta E} \nonumber \\
    &= \big|\epsilon(\vec{\Lambda})\big|^2\frac{\bra{\Omega^{\perp}(\vec{\Lambda})}\hat H\ket{\Omega^{\perp}(\vec{\Lambda})}-\bra{\Omega}\hat H\ket{\Omega}}{\Delta E}\nonumber \\
    &\geq \big|\epsilon(\vec{\Lambda})\big|^2.
\end{align}
This relation leads to a lower bound on the fidelity $F(\vec{\Lambda}) \coloneq \big|\langle\Omega|\psi(\vec{\Lambda})\rangle\big|^2\leq 1$:
\begin{align}
    F(\vec{\Lambda})=1-\big|\epsilon(\vec{\Lambda})\big|^2\geq 1-\delta_E(\vec{\Lambda}).
    \label{eq:fid_bound}
\end{align}
Thus, a natural strategy is to minimize $\delta_E(\vec{\Lambda})$ [and hence the ansatz energy $\gbra{\psi(\vec{\Lambda})}\hat H\gket{\psi(\vec{\Lambda})}$], yielding the familiar variational principle. While foundational, this is not a unique prescription. For instance, the relation~\eqref{eq:fid_bound} only provides a lower bound on the fidelity, and the maximum-fidelity ansatz can differ from the minimum-energy ansatz, as demonstrated through an analytically solvable example in Appendix~\ref{app:en_v_fid}. Furthermore, for many applications, achieving an energy discrepancy $\delta_E(\vec{\Lambda})$ below a certain tolerance $\delta_E^{\rm tol}$ is sufficient. This is especially true in the context of numerical simulations, where systematic errors (due to e.g., finite volume, discretization, truncation cutoffs) naturally define such a non-zero threshold. This threshold then identifies a \emph{region} of acceptable parameter space rather than a single point. Within this region, ansatzes can exhibit varied characteristics despite having similar values of energy. What constitutes a good choice of $\vec{\Lambda}$ within this region, and how one arrives at it, depend upon various factors: the information available about $\ket{\Omega}$, the intended use for  $\ket{\psi(\vec{\Lambda})}$, and the computational resources available for optimizing $\ket{\psi(\vec{\Lambda})}$. 

When no information about the target ground state is available (except for the Hamiltonian $\hat H$), the variational principle provides an upper bound on the ground-state energy upon minimizing $\bra{\psi({\vec{\Lambda}})}\hat H \ket{\psi{(\vec{\Lambda}})}$. The state $\ket{\psi(\vec{\Lambda}_0)}$ which achieves this minimum energy can then serve as a proxy for the target ground state $\ket{\Omega}$. In some cases, $\ket{\Omega}$ may be known exactly, and one may instead maximize the fidelity $F$. Oftentimes, one operates at neither extreme of minimal or maximal information about $\ket{\Omega}$. 
In particular, PIMC techniques efficiently determine (static) ground-state correlation functions $\bra{\Omega}\hat O \ket{\Omega}$ and the spectral gap $\Delta E$ (when the computation is not subject to a sign problem). Thus, one could optimize the values of specific correlation functions of the ansatz depending upon the physics one wants to study. In the context of Hamiltonian simulations, the ground-state representative $\ket{\psi(\vec{\Lambda})}$ serves as an initial state for real-time evolution. Based on the pre-existing knowledge about the correlation functions of $\ket{\Omega}$ (sourced from methods like PIMC) and the dynamical properties one wishes to investigate with $\ket{\psi(\vec{\Lambda})}$, one can discriminate between the values of $\vec{\Lambda}$ in regions of low energy discrepancy. This is unlike energy minimization and fidelity maximization, which offer limited control over the features of $\ket{\Omega}$ that can be reproduced by the ansatz.\\

There are several ways in which one could incorporate pre-existing knowledge about ground-state correlation functions, i.e., moments, in the search for $\vec{\Lambda}$. Here, we will work with a loss function which is close to the energy loss function. Suppose that it is important for $\ket{\psi(\vec{\Lambda})}$ to accurately reproduce the ground-state expectation values for a given set of operators $\mathcal{T} = \{\hat O_i\}$, which we refer to as the ``target set.'' We propose to find the appropriate value of $\vec{\Lambda}$ by minimizing the loss function
\begin{align}\label{eq:mom_obj}
    \vec{\Lambda}_0 = &\mathrm{argmin}_{\vec{\Lambda}} \ \bigg[\bra{\psi(\vec{\Lambda})}\hat H \ket{\psi(\vec{\Lambda})} \nonumber\\
    &+ \sum_{\hat O \in \mathcal{T}} \ w_{\hat O} \bigg|\bra{\psi(\vec{\Lambda})}\hat O \ket{\psi(\vec{\Lambda})} - \bra{\Omega} \hat O \ket{\Omega}\bigg|^2\bigg],
\end{align}
where the real weights $w_{\hat O}\geq 0$ are determined by physical intuition and experimentation.\footnote{We either work with Hermitian target operators, or non-Hermitian target operators whose expectation value is guaranteed to be real by symmetry constraints. Thus, our loss function is real-valued even in the absence of the absolute value surrounding the terms proportional to $w_{\hat O}$.} We will refer to this procedure of minimizing the above loss function as ``moment optimization.'' When $w_{\hat O}=0$, one recovers energy minimization. As the values of the $w_{\hat O}$ are increased, the ansatz better captures the expectation values $\langle \hat O\rangle$---but the energy of the optimized ansatz grows. In an effective instance of moment optimization, the behavior of the ansatz target moments will improve at sufficiently small values of the weights, such that the energy penalty paid is small with respect to the spectral gap. In a less favorable attempt,
target-moment behavior would only improve at high weight values, implying that a bigger energy penalty is paid, which in turn may lower the fidelity with the ground state. As will become clear through examples of this work, the efficacy of this procedure will depend upon the characteristics of the ansatz, the Hamiltonian parameters, and the chosen set of target moments. We will refer to the resulting ansatz as the ``moment-optimized ansatz.'' \\

In summary, we will quantify the success of the moment-optimization procedure by comparing the behavior of target moments, and by ensuring that $\delta_E\ll 1$. In what follows, we focus on the case of an interacting scalar field theory. Since fidelity is only efficiently computable for systems with small Hilbert spaces, we will compute both energy discrepancy and fidelity for the single-mode, i.e., the (0+1)D case, but restrict the analysis to just energy discrepancy for the multimode, i.e., the (1+1)D case. 

\section{The single-mode case: (0+1)D $\phi^4$ ground states}\label{sec:singlemode}
\noindent
Consider a single bosonic mode governed by the dimensionless Hamiltonian
\begin{align}\label{eq:N=1_ham}
    \hat H^{(0+1)}_\sigma = \frac{\pih^2}{2} + \sigma\frac{\phih^2}{2} + \lambda \phih^4,
\end{align}
where $\lambda \geq 0$ and $\sigma = \pm 1$. This Hamiltonian is invariant under the parity, $(\phih,\pih) \mapsto (-\phih,-\pih)$, and 
anti-unitary time-reversal, $(\phih,\pih) \mapsto (\phih,-\pih)$, transformations. When $\sigma = 1$, Eq.~\eqref{eq:N=1_ham} describes a quantum anharmonic oscillator which possesses a non-degenerate ground state. When $\sigma=-1$, the Hamiltonian can be written in terms of a double-well potential, $V(\hat{\phi})$:
\begin{align}\label{eq:dwp}
\begin{split}
    \hat H^{(0+1)}_{-1}&=\frac{\pih^2}{2} + \lambda\left(\phih^2-\frac{1}{4\lambda}\right)^2 - \frac{1}{16\lambda}\\
    & \equiv \frac{\pih^2}{2} + V(\phih)-\frac{1}{16\lambda}.
\end{split}
\end{align}
The minima for the two wells are located at $\phi= \pm \frac{1}{\sqrt{4\lambda}}$ while the height of the well is given by $V(\phi=0)=\frac{1}{16\lambda}$. Thus, as $\lambda$ becomes smaller, the barrier separating the two wells becomes taller. However, as long as $\lambda$ is nonzero, there is a finite gap between the lowest-energy parity-symmetric and parity-antisymmetric eigenstates, and hence the Hamiltonian is non-degenerate. 

When $\lambda\ne 0$, the theory is interacting, and its ground state does not have a known analytical solution in either regime. In the following, we introduce the \emph{finite-stellar-rank} ansatz to approximate the ground state of this theory.

\subsection{Single-mode stellar hierarchy and the finite-rank ansatz}\label{subsec:singlemode_stellar}

In terms of the dimensionless quadrature operators introduced in Eq.~\eqref{eq:N=1_ham}, one can define the ladder operators
\begin{equation}
    \hat a \coloneq \frac{\phih + i\pih}{\sqrt{2}}, \ \hat{a}^{\dagger} \coloneq \frac{\phih - i\pih}{\sqrt{2}},
\end{equation}
which satisfy the commutation $[\hat a, \hat a^{\dagger}]=1$. The Fock states, i.e., the eigenstates of the number operator $\hat a^{\dagger}\hat a$, will be denoted by $\ket{n}, $ where $n \in \mathbb{N} \cup \{0\}$. Unitary transformations acting on the bosonic Hilbert space can either be Gaussian or non-Gaussian. Gaussian unitary transformations are generated by Hamiltonians which are at most quadratic in the bosonic ladder operators. All other unitary transformations are non-Gaussian, and are generated by Hamiltonians with a greater than quadratic degree in the ladder operators.\\

Our choice of ground-state ansatz is based on the \textit{stellar hierarchy} of Ref.~\cite{chabaud2020stellar}. A single-mode bosonic state $\ket{\psi}$ is said to have a finite stellar rank $R_{\psi} \in \mathbb{N} \cup \{0\}$ if it admits the decomposition
\begin{align}\label{eq:N=1_stellar}
    &\ket{\psi} = \hat U_{G_{\psi}} \hat{C}_{\psi}     \ket{0} \equiv \hat U_{G_{\psi}} \ket{C_{\psi}}, 
\end{align}
where
\begin{align}\nonumber \\ &\hat{C}_{\psi} \coloneq C_{\psi}(\hat a^{\dagger})=\sum_{n=0}^{R_{\psi}} \ c'_{\psi,n} \ (\hat a^{\dagger})^n.
\end{align}
Here, $\hat U_{G_{\psi}}$ is a Gaussian unitary transformation, $C_{\psi}(\cdot)$ is a polynomial of degree $R_{\psi}$ with complex coefficients $c'_{\psi,n}$, and $\hat{C}_{\psi}$ is the operator obtained by evaluating this polynomial with the argument $\hat a^{\dagger}$. The coefficients of the polynomial are chosen so that the state is normalized. The state $\ket{C_{\psi}}\equiv \hat{C}_{\psi} \ket{0}$ is called the \textit{core state} corresponding to $\ket{\psi}$. This decomposition of a finite-rank state into a Gaussian unitary acting on a core state is unique. If a single-mode bosonic state does not admit the decomposition in Eq.~\eqref{eq:N=1_stellar}, it is said to have an infinite stellar rank~\cite{chabaud2020stellar}. \\

\begin{figure}
    \centering   \includegraphics[width=0.8\columnwidth]{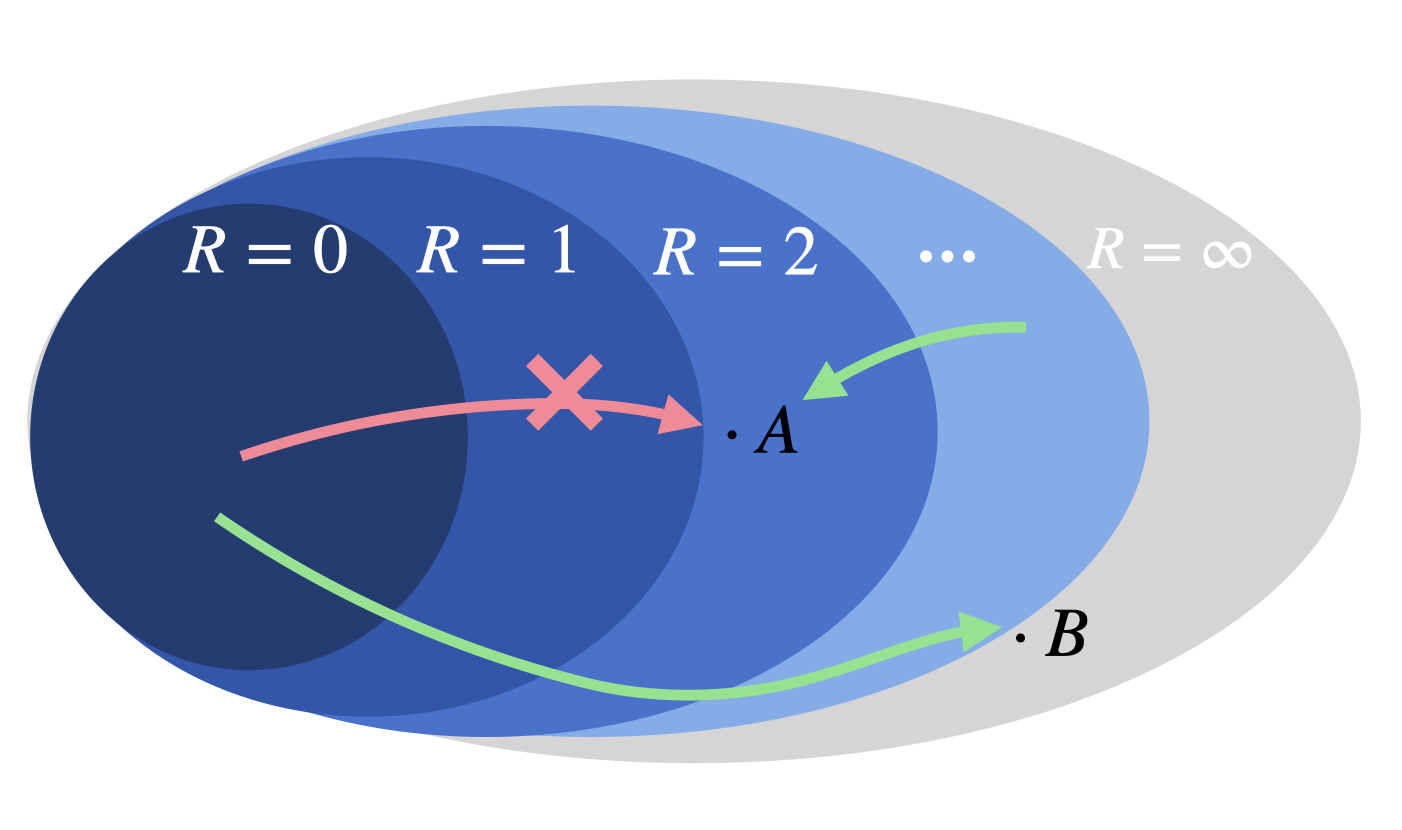}
    \caption{\textit{Schematic representation of the stellar hierarchy inspired by Fig. 1 of Ref.~\cite{chabaud2022holomorphic}.} Stellar rank partitions the single-mode bosonic Hilbert space into disjoint subspaces with different ranks. Finite-rank states are dense in this Hilbert space. Thus, an infinite-rank state, such as the one marked by the point `B' in this figure, can be approached by a sequence of finite-rank states (depicted by the green arrow), i.e., it can be approximated with arbitrary precision in trace distance using finite-rank states. Similarly, a state with finite-rank $R$, such as the one marked by point `A', can be approached by a sequence of states which have a rank greater than or equal to $R$ (indicated by the green arrow). However, the state `A' cannot be approached by a sequence of states with rank strictly less than $R$ (indicated by the orange arrow). In other words, there is a finite trace distance between states with finite rank $R$, and all lower-rank states.}
    \label{fig:stellar_hierarchy}
\end{figure}

As shown in Fig.~\ref{fig:stellar_hierarchy}, the stellar rank partitions the single-mode Hilbert space into disjoint subspaces with different ranks $R_{\psi} \in \mathbb{\bar{N}} \cup \{0\}$ ($\bar{\mathbb{N}}\equiv \mathbb{N}\cup \{\infty\}$), and this is referred to as stellar hierarchy. Finite-rank states, i.e., states with $R_{\psi} \in \mathbb{N} \cup \{0\}$ form a dense subset of the Hilbert space under the trace norm. In other words, infinite-rank states (such as the one marked by $B$ in Fig.~\ref{fig:stellar_hierarchy}) are not isolated from finite-rank states in trace distance. Thus, one can find a sequence of finite-rank states which gets arbitrarily close to any given infinite-rank state. Finite-rank states, on the other hand, are isolated from states with lower stellar ranks, and this property is referred to as \textit{stellar robustness}. Thus, there is a finite trace distance between any specific rank-$R_{\psi}$ state (such as the one marked by $A$ in Fig.~\ref{fig:stellar_hierarchy}) and all the states with rank less than $R_{\psi}$. A rank-$R_{\psi}$ state, nonetheless, can be approached by a sequence of states with rank greater than or equal to $R_{\psi}$. Thus, for a finite-rank state, the rank indicates the minimal cost of boson additions needed to engineer that state with arbitrary precision in trace distance.\\

The ground state of the Hamiltonian in Eq.~\eqref{eq:N=1_ham} is expected to have an infinite rank~\cite{blahnik2024natural}. Thus, we choose to approximate this state by a finite-rank $R_{\psi}$ states which obey the symmetries of the Hamiltonian. The $\psi$ subscript on quantities will be dropped from now on for simplicity. Instead of demanding the full rank-$R$ state, i.e., $\hat U_{G}\ket{C}$, to possess the desired symmetries, we will impose the potentially stronger condition that both $\ket{C}$ and $\hat U_{G}$ possesses these symmetries. An arbitrary single-mode Gaussian unitary $\hat U_G$ can be expressed as~\cite{serafini2023quantum} $\hat U_G = \hat S(\xi)\hat D(\alpha)$,
up to a phase. Here, $\hat S(\xi)\coloneq e^{\frac{1}{2}\left(\xi (\hat a^{\dagger})^2-\xi^*\hat a^2\right)}$ with $ \xi \in \mathbb{C}$ is the squeezing operator, and $\hat D(\alpha)\coloneq e^{\alpha \hat a^{\dagger}-\alpha^*\hat a} $ with $\alpha \in \mathbb{C}$ is the displacement operator. $\hat S(\xi)$ is invariant under a parity transformation while under the same transformation, $\hat D(\alpha)\mapsto \hat D(-\alpha)$. This means that a parity-invariant Gaussian is just the squeezing operator $\hat S(\xi)$. Under time reversal, $\hat S(\xi)\mapsto \hat S(\xi^*)$, so time-reversal invariance further restricts the squeezing operator to have a real-valued parameter $\xi=r\in \mathbb{R}$. This results in the symmetric Gaussian unitary transformation
\begin{align}
\hat{U}_{G}=\hat S(r) = e^{\frac{1}{2}r(\hat a^{\dagger 2}-\hat{a}^2)},
\end{align}
with $ r\in \mathbb{R}$. Similarly, the core state is constrained to have the decomposition 
\begin{align}
\ket{C}_R\coloneq\sum_{n=0,2,\ldots, R} \ c'_{n} (\hat a^{\dagger})^n\ket{0},
\end{align}
with $c'_{n} \in \mathbb{R}$ and even rank $R$. Defining $c_{n}\coloneq c'_{n}\sqrt{n!}$, the symmetric rank-$R$ single-mode ansatz can finally be expressed as
\begin{align}\label{eq:single_mode_decomp}
    \ket{\psi}_R
    = \hat S(r) \left(c_0\ket{0} + c_2\ket{2} + \ldots + c_R\ket{R}\right).
\end{align}
This ansatz is parametrized by the $2+\frac{R}{2}$ real parameters
$\vec{\Lambda}=(r,c_0,c_2,\ldots,c_{R})$. These $2+R/2$ real parameters are not independent due to the normalization constraint. During optimization, we will allow all these parameters to vary freely within the normalized manifold of states.

\subsection{Moment optimization}\label{subsec:moment_opt}
Consider the Weyl-ordered basis $\{\hat{\mathcal{O}}_{p,q} \coloneq \phih^p\pih^q\}$ with $p,q \in \mathbb{N} \cup \{0\}$ for the operator Hilbert space. Then:
\begin{itemize}
    \item[i)] The lowest-order expectation value of the form $\langle \phih^{2p+1}\pih^{2q+1}\rangle$ is $\langle \phih \pih\rangle$. Time-reversal invariance fixes its value as follows. First, one notes that:
    \begin{align}
        \langle \phih \pih\rangle &= -\langle \pih \phih\rangle = -\langle \phih \pih\rangle+i.
    \end{align}
    In the first equality, the time-reversal transformation is applied. In the second equality, the canonical bosonic commutation relation is used. Therefore, $\langle \phih \pih\rangle = \frac{i}{2}$.
    \item[ii)] All other expectation values of the form $\langle \phih^{2p+1}\pih^{2q+1}\rangle$ with $p,q\neq 0$ can be expressed as a sum of expectation values of lower-order basis elements. To see this, one notes that
    \begin{align}
        \langle \phih^{2p+1}\pih^{2q+1}\rangle &= -\langle \pih^{2q+1}\phih^{2p+1}\rangle \nonumber \\
        &= -\langle \phih^{2p+1}\pih^{2q+1}\rangle+\cdots\,.
    \end{align}
    Time-reversal invariance is used in the first line to equate the two expectation values. In the second line, the canonical bosonic commutation relation is employed. The ellipses represent a linear combination of the expectation values of lower-order basis elements, i.e., expectation values of the form $\langle\phih^{p'}\pih^{q'}\rangle$ with $p'<2p+1$ and $q'<2q+1$, considering the commutation relation $[\hat \phi , \hat \pi]=i$. Thus, $\langle \phih^{2p+1}\pih^{2q+1}\rangle$ can be expressed as a sum of expectation values of lower-order basis elements, which are eventually related to the trivial value $\langle \phih \pih\rangle = \frac{i}{2}$ and to expectation values with $p'$ and $q'$ being both even.
    \item[iii)] When $p$ and $q$ are not both even or odd, $\langle \hat{\mathcal{O}}_{p,q} \rangle $ vanishes as a result of parity invariance.
\end{itemize}
Thus, an arbitrary (parity and time-reversal) symmetric state is characterized by independent correlation functions of the form $\langle \hat{\mathcal{O}}_{p,q}\rangle=\langle \phih^{p}\pih^{q} \rangle$, where both $p$ and $q$ are even.\footnote{Note that $\langle \phih^{2p}\pih^{2q} \rangle$ is real-valued. This is because the complex conjugate $\langle \phih^{2p}\pih^{2q} \rangle^*=\langle \pih^{2p}\phih^{2q} \rangle$ is equal to $\langle \phih^{2p}\pih^{2q} \rangle$ due to time-reversal invariance.} As expected, this set of independent correlation functions is infinite-dimensional (unless further constraints such as Gaussianity or stationarity are imposed~\cite{han2020bootstrapping,berenstein2021bootstrapping,ozzello2023bootstrap}). Thus, we will restrict our attention to bosonic operators $\hat{\mathcal{O}}_{p,q}$ with some fixed maximum degree.

\subsubsection{Energy minimization}

\begin{figure}
    \includegraphics[scale=0.51]{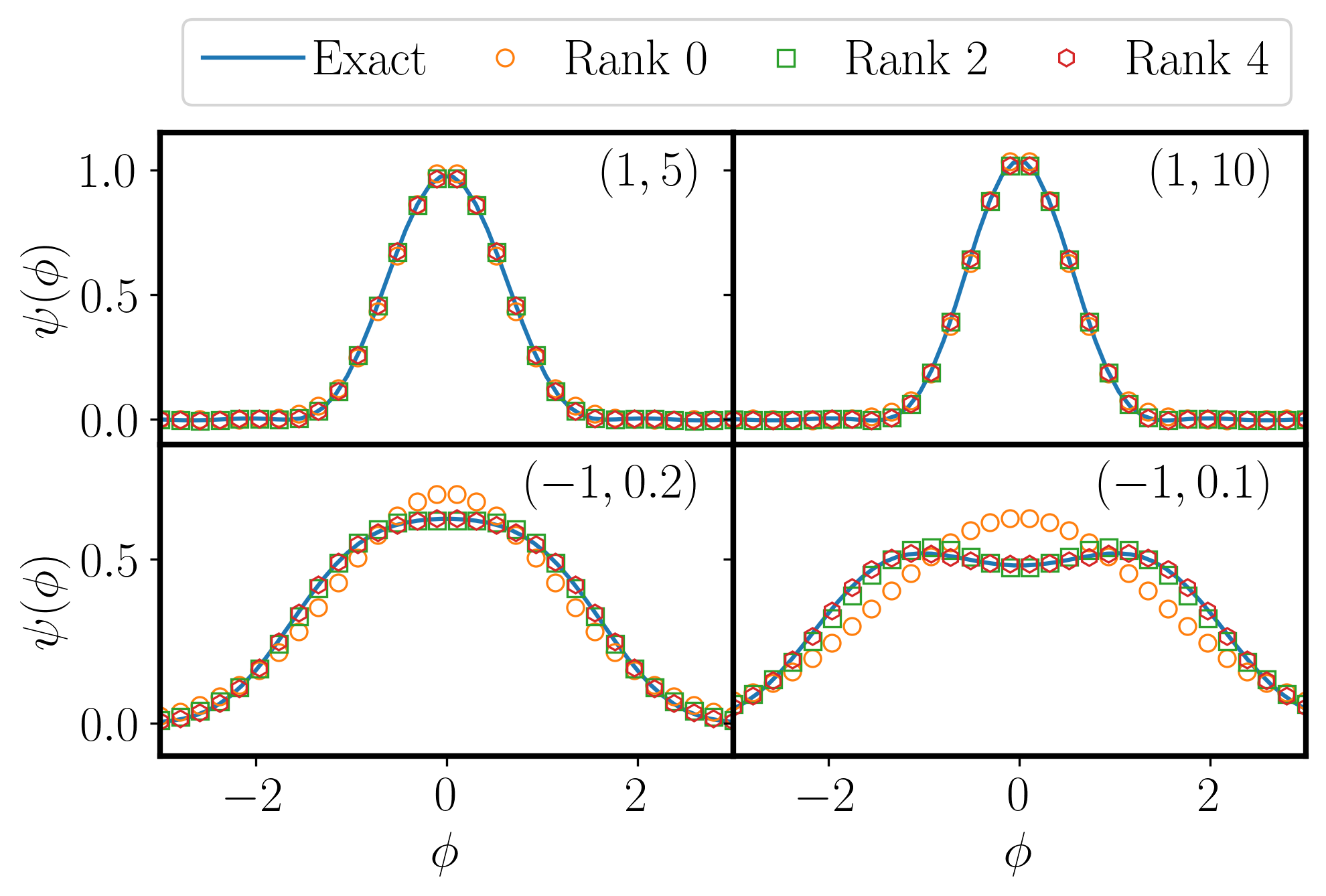}
    \caption{Minimum-energy wavefunctions for the     (0+1)D $\phi^4$ Hamiltonian starting from the ansatz in Eq.~\eqref{eq:single_mode_decomp} for various values of rank and Hamiltonian parameters, $(\sigma,\lambda)$ (indicated on the top right of each plot). All wavefunctions show relatively good agreement with the exact wavefunction except for the $R=0$ ansatz deep in the double-well regime, i.e., for $(\sigma,\lambda)=(-1,0.1)$.}
    \label{fig:min_en_wavefunctions}
\end{figure}

\begin{figure}[t!]
    \includegraphics[scale=0.51]{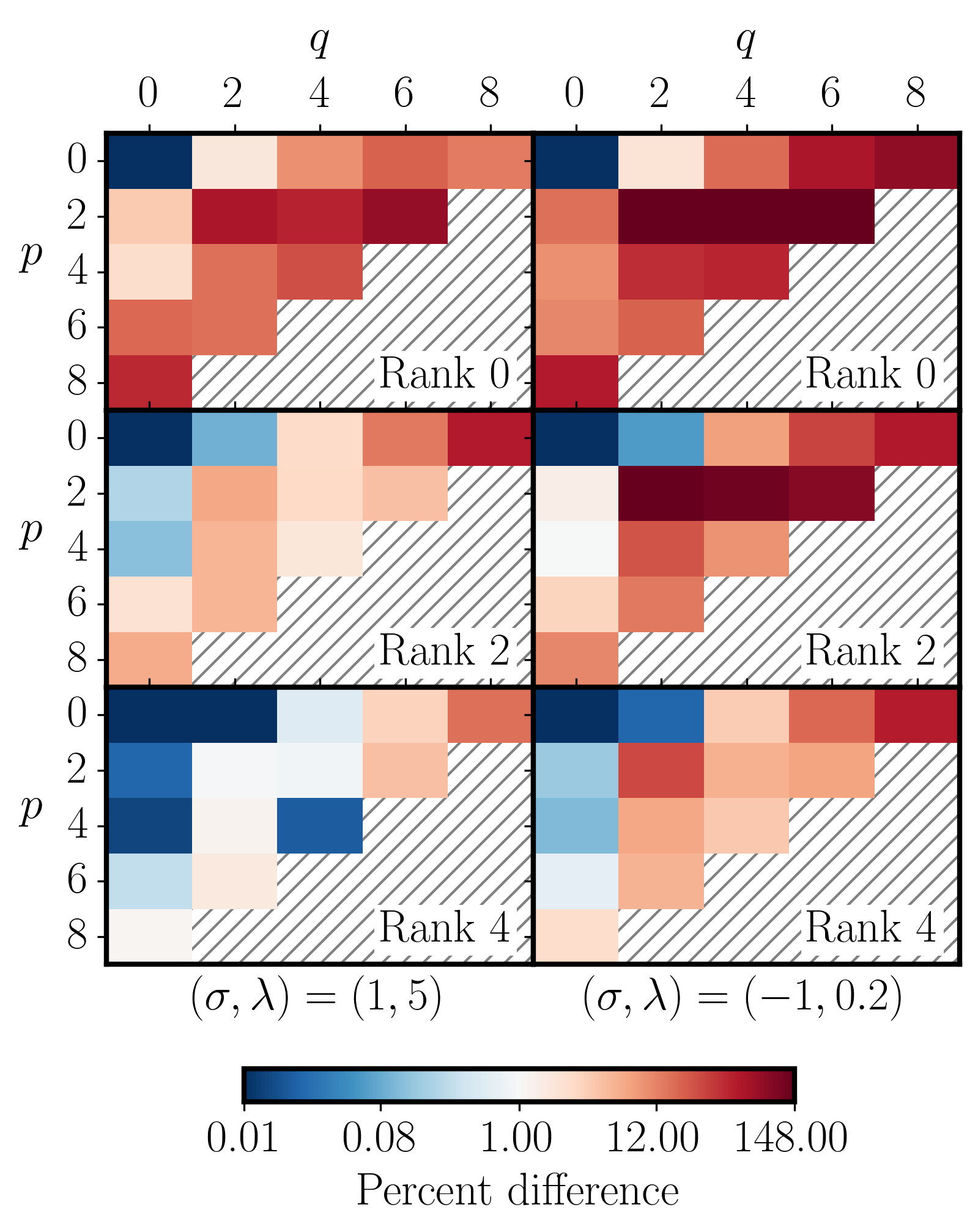}
    \caption{Percent errors in the moments $\langle \hat{\mathcal{O}}_{p,q} \rangle = \langle \phih^p\pih^q \rangle$ with $\ p+q\leq 8$ for minimum-energy states with various ansatz ranks in Eq.~\eqref{eq:single_mode_decomp}, and for $(\sigma,\lambda)=(1,5)$ and $(-1,0.2)$ in the left and right columns, respectively. 
    In all cases, moment errors are the least for the operators that feature in the Hamiltonian, i.e., $\hat \phi^2, \hat \pi^2, \hat \phi^4$ [corresponding to $(p,q)=(2,0),(0,2),(4,0)$, respectively], and they increase as the value of $p+q$ increases. Errors decrease as the ansatz rank increases, and are worse for the double-well case compared with the anharmonic-oscillator case.}
    \label{fig:min_en_mom_errs}
\end{figure}

We first minimize energy alone, i.e., we set $w_{\hat O}=0$ in Eq.~\eqref{eq:single_mode_decomp}. The resulting wavefunctions are shown in Fig.~\ref{fig:min_en_wavefunctions} while the errors in the moments $\langle \hat{\mathcal{O}}_{p,q} \rangle$ with $\ p+q\leq 8$ are presented in Fig.~\ref{fig:min_en_mom_errs}. The values of the fidelity $F$ and the energy-discrepancy to spectral-gap ratio $\delta_E$ are provided in Table~\ref{tab:singlemode_minen_fid_en}. For a fixed value of $(\sigma,\lambda)$, the behavior of all three metrics---$F$, $\delta_E$, and moment errors---improves with increasing the ansatz rank. In the double-well regime ($\sigma=-1$), the behavior of ansatzes of all three lowest ranks is worse compared with that in the $\sigma=1$ case. This is reflected in the values of $\delta_E$, $F$, the moment errors, as well as the shape of the wavefunction. The rank-0 ansatz obviously fails to reproduce the bimodal form of the ground-state wavefunction. Even though the rank-2 and rank-4 ansatzes are able to achieve this bimodality, the fidelities obtained in this case are not as high as those obtained for the anharmonic-oscillator case. Increasing the ansatz rank would, therefore, be necessary to achieve a better approximation of the ground-state wavefunction. \\

\begin{table}[b!]
\centering
\begin{tabular}{cccrr}
\hline\hline
$\sigma$ & $\lambda$ & Rank & Fidelity & $\delta_E$ (\%)\\
\hline\hline
1   & 5   & 0 & 0.9986896 & 0.6415 \\
    &     & 2 & 0.9999647 & 0.0315 \\
    &     & 4 & 0.9999981 & 0.0024 \\
    \hline
1   & 10  & 0 & 0.9985955 & 0.6885 \\
    &     & 2 & 0.9999620 & 0.0343 \\
    &     & 4 & 0.9999979 & 0.0027 \\
    \hline
    \hline
-1  & 0.2 & 0 & 0.9885943 & 5.8026 \\
    &     & 2 & 0.9996527 & 0.4166 \\
    &     & 4 & 0.9999771 & 0.0418 \\
    \hline
-1  & 0.1 & 0 & 0.9557221 & 26.5263 \\
    &     & 2 & 0.9983677 & 2.5773 \\
    &     & 4 & 0.9998871 & 0.3021 \\
\hline\hline
\end{tabular}
\caption{Fidelity and the energy-discrepancy to spectral-gap ratio for the minimum-energy ansatz for various $(\sigma,\lambda)$ and ranks $R$.}
\label{tab:singlemode_minen_fid_en}
\end{table}
\begin{figure}
    \centering
    \includegraphics[scale=0.51]{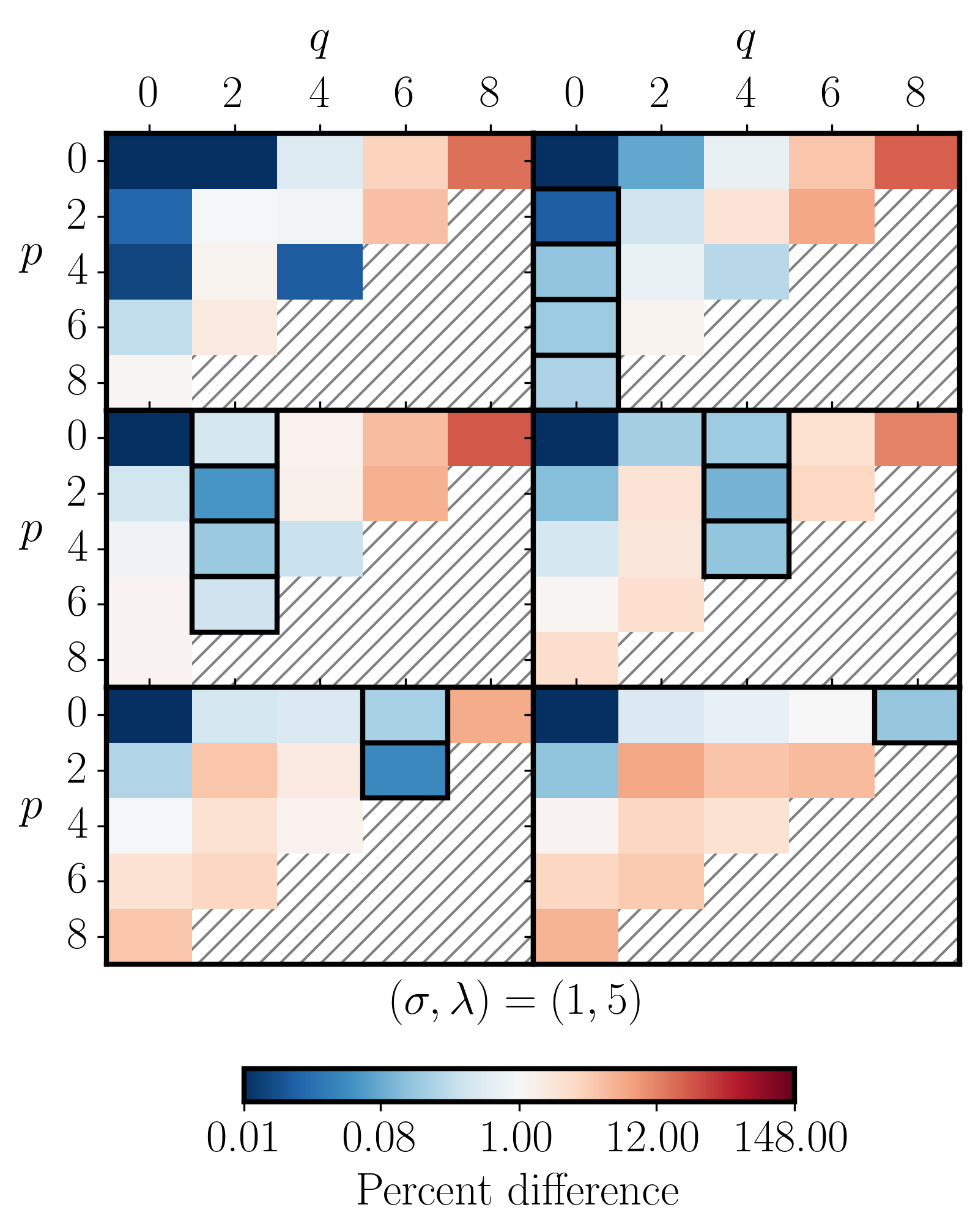}
    \caption{Percent errors in the moments $\langle \hat{\mathcal{O}}_{p,q} \rangle = \langle \phih^p\pih^q \rangle$ with $\ p+q\leq 8$ resulting from column-moment optimization for  $(\sigma,\lambda)=(1,5)$ with the rank-4 ansatz in Eq.~\eqref{eq:single_mode_decomp}. The top left plot shows the moment errors for the minimum-energy ansatz (also shown as a subplot in Fig.~\ref{fig:min_en_mom_errs}), and the rest of the plots show the errors with the boxed column moments chosen as target moments. For the columns $q=6,8$, all errors in the boxed moments are significantly lowered when the corresponding columns moments are optimized (as compared to mere energy minimization). For the columns $q=0,2,4$, the error in some of the boxed moments is raised (relative to the minimum-energy case) while the errors in other boxed moments is lowered. However, the errors in all boxed moments are less than 1\% (unlike the minimum-energy case for the same columns).   
    Thus, there is a redistribution of moment errors so that the overall behavior of target moments is improved.
    }
    \label{fig:col_opt}
\end{figure}
\begin{figure}[t]
    \centering    \includegraphics[width=0.98\columnwidth]{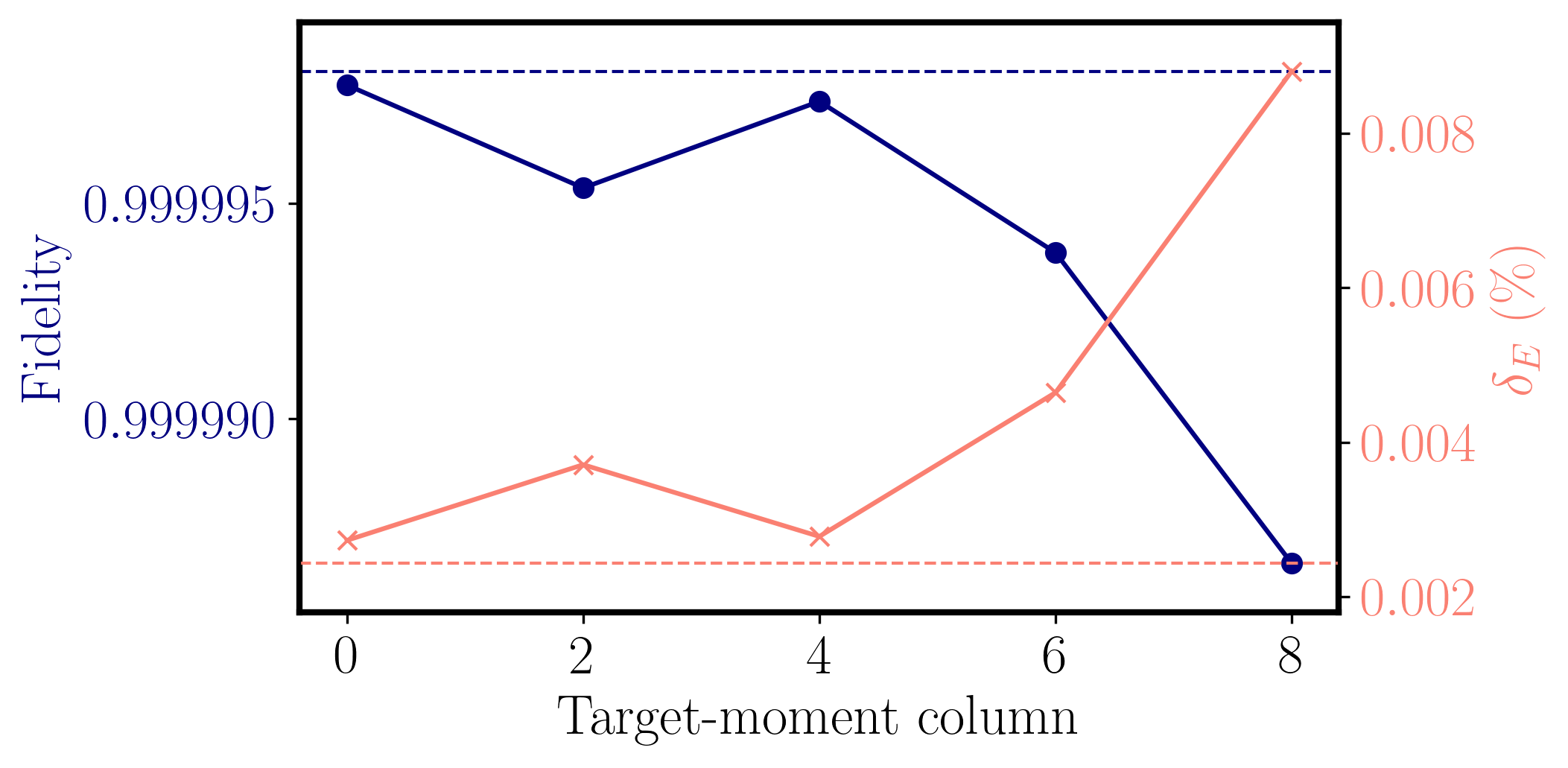}
    \caption{Fidelity $F$ and the energy-discrepancy to spectral-gap ratio $\delta_E$ for the column-moment-optimized ansatz plotted as a function of $q$, where $q$ is the column labeling the target set. The optimization is performed for a rank-4 ansatz in Eq.~\eqref{eq:single_mode_decomp} with $(\sigma,\lambda)=(1,5)$. Dashed horizontal lines mark the value of $F$ and $\delta_E$ corresponding to the minimum-energy ansatz.}
    \label{fig:col_fiden}
\end{figure}

The moment-error plots in Fig.~\ref{fig:min_en_mom_errs} show that minimizing the energy results in some natural distribution of moment errors across different values of $p$ and $q$. In particular, the errors are small for $(p,q)=(2,0),(4,0),(0,2)$, i.e., for the moments $\langle \phih^2 \rangle$, $\langle \phih^4 \rangle$, and $\langle \pih^2 \rangle$ (corresponding to operators that are present in the Hamiltonian). However, the moment errors increase as $p$ and $q$ increase. It may be desirable to improve the behavior of higher-order moments. In the following, we show how this can be done at a minimal cost in $F$ and $\delta_E$.

\subsubsection{Column-moment optimization}

Suppose one wishes to accurately reproduce the following ``column" moments
\begin{align}
    \mathcal{T}_q=\{\hat{\mathcal{O}}_{p,q} 
 \text{ for even $p$ satisfying $p+q\leq 8$}\}.
\end{align}
We take $\mathcal{T}_q$ to be the target set for the moment-optimization loss function in Eq.~\eqref{eq:mom_obj}, and the weight accompanying the target operator $\hat O$ to be
\begin{align}
    w_{\hat O} = \frac{1}{(\bra{\Omega}\hat O \ket{\Omega})^2}.
\end{align}
Thus, the loss function in Eq.~\eqref{eq:mom_obj} becomes the sum of the ansatz energy and the squared fractional discrepancies in the target moments.

The errors in the moments $\langle \hat{\mathcal{O}}_{p,q} \rangle$ with $\ p+q\leq 8$ resulting from this optimization are shown in Fig.~\ref{fig:col_opt} for $(\sigma,\lambda)=(1,5)$ with the rank-4 ansatz. We observe that the moment optimization redistributes moment errors such that the overall behavior of target moments is improved. More quantitatively, for the minimum-energy ansatz, the average percent errors in each column from left to right (i.e., from $q=0$ to $q=8$) are given by 0.36\%, 0.95\%, 0.46\%, 3.9\%, and 14.9\%, respectively. When the corresponding column moments are optimized, these errors go down to  0.14\%, 0.24\%, 0.13\%, 0.11\%, and 0.15\%, respectively. While the errors in the first three columns, i.e., $q=0,2,4$, were already small for the minimum-energy ansatz, dramatic improvements are seen in the behavior of the last two columns $q=6,8$. These improvements only occur at the cost of increasing non-target moment errors. For example, when moments in the columns corresponding to $q=2,4,6,8$ are optimized, the average percent errors in the moments in the $q=0$ column are 0.88\%, 1.02\%, 1.85\%, and 2.43\%, respectively. The redistribution of errors is achieved at an insignificant cost in $F$ and $\delta_E$, as plotted in Fig.~\ref{fig:col_fiden}. The improvement in higher-order target moments $q\geq 4$ is much more pronounced than that in lower-order moments. This could be due to higher-order moments being more \emph{independent} of the lower-order moments featuring in the expression for energy, and also the simpler fact that the size of the target set grows smaller as $q$ increases, imposing less demand on the optimization. 

\section{The multimode case: (1+1)D $\phi^4$ ground states}\label{sec:multimode}
\noindent
Our aim in this section is to construct and constrain approximate wavefunctions for the ground state of the (1+1)D lattice $\phi^4$ theory. The continuum Hamiltonian for this theory is:
\begin{align}
    \hat{H}_{\rm cont} = \int dx \ & \left[ \frac{\pih(x)^2}{2} + \frac{\partial_x \phih(x)^2}{2}+ \right. \nonumber \\
    &\left. ~~ \frac{1}{2}m_0^2\phih(x)^2 + \frac{\lambda_0}{4}\phih(x)^4 \right]. 
\end{align}
Here, $m_0^2$ and $\lambda_0$ are the bare mass and coupling constant, respectively. $\phih(x)$ and $ \pih(x)$ are the field and its conjugate, respectively, obeying the canonical bosonic commutation relation $[\phih(x),\pih(x')]=i\delta(x-x')$. We discretize this theory on a finite periodic lattice with lattice spacing $a$ and size $L=Na$, to obtain the dimensionless lattice Hamiltonian
\begin{align}\label{eq:ham_multi}
   \hat H &\equiv a\hat H_{a,N} \nonumber\\
   &= \sum_{j=0}^{N-1} \left[\frac{\pih_j^2}{2} + \frac{(\phih_{j+1}-\phih_j)^2}{2} + \frac{1}{2}m^2\phih_j^2 + \frac{\lambda}{4}\phih_j^4 \right]. \nonumber \\
\end{align}
Here, we have introduced the dimensionless lattice mass $m^2\equiv m_0^2a^2$, the dimensionless coupling $\lambda\equiv\lambda_0 a^2$, and dimensionless lattice-field operators $\phih_n\equiv \phih(na)$ and $\pih_n\equiv a\pih(na)$. These field operators obey the commutation relations $[\phih_j,\pih_k]=i\delta_{j,k}$, where $\delta_{j,k}$ is the Kronecker delta function. Due to periodicity, $\phi_{j+N}=\phi_j$ and similarly for the conjugate field. \\

This Hamiltonian in Eq.~\eqref{eq:ham_multi} possesses ($\mathbb{Z}_2$) parity [i.e., $(\hat \phi, \hat \pi) \mapsto (-\hat \phi,-\hat \pi)$], time reversal, lattice inversion (i.e., spatial parity), and ($\mathbb{Z}_N$) lattice translation symmetries.
In the thermodynamic limit and for $m^2 < 0$, the system undergoes a phase transition, in which  the parity symmetry is spontaneously broken beyond a critical value $\lambda_c$. The bare mass $m_0$ diverges logarithmically $m_0^2 \sim m_r^2\mathrm{log}(\frac{1}{a})$~\cite{loinaz1998monte}, where $m_r$ is some finite renormalized mass, while the coupling $\lambda_0$ stays finite as the lattice spacing is reduced. Thus, the continuum theory can be characterized by the dimensionless coupling constant $\lambda_0/m_r^2$ with a suitable choice for mass renormalization (as discussed in Ref.~\cite{loinaz1998monte}). The continuum limit is achieved upon taking $N \rightarrow \infty$ followed by $a\rightarrow 0$, i.e., $(m^2,\lambda)=(m_0^2(a^2)a^2,\lambda_0(a)a^2)\rightarrow (0,0)$ while holding the coupling $\lambda_0/m_r^2$ fixed.

To keep the subsequent PIMC computations manageable, we work with a fixed small value of $N=10$. This value suffices to demonstrate the conclusions of this section. The values of $(m^2,\lambda)$ are chosen such that $N\Delta E > 1$ and $\Delta E < \pi$, where $\Delta E$ is the (dimensionless) spectral gap. The first condition ensures that the Compton wavelength of the scalar excitation is contained within the system volume while the second condition ensures that the scalar particle is light compared to the UV cutoff imposed by the lattice spacing. To qualitatively study the effects toward the continuum limit, we choose the values of $(m^2,\lambda)$ to be on a straight line approaching the origin with $m^2>0$. This study concerns only the symmetric regime but the discussions that follow can be straightforwardly extended to the full phase diagram.

Using Euclidean PIMC techniques, we estimate the values of various ground-state correlation functions. In particular, we will study the two-point function $\langle \phih_{j'} \phih_{j}\rangle$ with $j' \neq j$,
which characterizes the strength of non-local correlations in the theory. 
We will also study the non-Gaussianity of the $\phi_j^{2n}$-moments resulting from the quartic coupling. While the raw moments $\langle\phi_j^{2n}\rangle$ are useful for examining the scale of fluctuations, information on shape (and hence non-Gaussianity) can be more cleanly extracted through relations among different moments. In particular, we will compute the moment ratio:\footnote{This measure is inspired by Ref.~\cite{blahnik2024natural}, which performs a detailed characterization of the non-Gaussianity of single-mode bosonic systems with a quasi-solvable sextic potential. Nonetheless, despite adopting the same terminology and notation, our definition of moment ratio in Eq.~\eqref{eq:moment-ratio} is different from that introduced in Ref.~\cite{blahnik2024natural}.}
\begin{align}
\label{eq:moment-ratio}
    R_{2n}(j) \coloneq \frac{1}{(2n-1)!!}\frac{\langle \phih_j^{2n}\rangle}
    {
    \langle \phih_j^2\rangle^n}.
\end{align}
For Gaussian states, $R_{2n}(j)=1$. Thus, any deviations from Gaussianity can be readily seen in the moment ratio. 
Note that due to translational invariance, $\langle \phih_{j'} \phih_{j}\rangle=\langle \phih_{0} \phih_{j-j'}\rangle$. Thus, we will simply refer to the two-point functions by $\langle \phih_{0} \phih_{j}\rangle$. Also, one single quantity $R_{2n} \equiv R_{2n}(j)$ characterizes the moment ratio across the translationally invariant lattice.
More details about the PIMC calculations of these quantities can be found in Appendix~\ref{app:pimc}.

\subsection{Multimode stellar hierarchy and finite-rank ansatzes
\label{sec:multi-ansatze}}

Consider the ladder operators
\begin{equation}
    \hat a_j \coloneq \frac{\phih_j + i\pih_j}{\sqrt{2}}, \ \hat{a}_j^{\dagger} \coloneq \frac{\phih_j - i\pih_j}{\sqrt{2}},
    \label{eq:mode-op-def}
\end{equation}
for each mode $j \in \{0,\ldots,N-1\}$, satisfying the commutation relations $[\hat a_j, \hat a_k^{\dagger}]=\delta_{j,k}$ and $[\hat a_j,\hat a_k]=[\hat a_j^{\dagger}, \hat a_k^{\dagger}]=0$. We will collectively denote all the mode operators with the notation $\bs{\hat a^{\dagger}} \coloneq(\ha_0^{\dagger},\ldots,\ha_{N-1}^{\dagger})$, and a multimode Fock state with $\ket{\bs{n}} \coloneq (n_0!\ldots n_{N-1}!)^{-1/2}(\bs{\hat a^{\dagger}})^{\bs{n}}\ket{\bs{0}} \coloneq \ket{n_0}_0\otimes \ldots \otimes \ket{n_{N-1}}_{N-1}$, where $\ket{\bs{0}}$ is the Fock vacuum and $(\bs{\hat a^{\dagger}})^{\bs{n}}\equiv (\hat a_0^{\dagger})^{n_0}\ldots(\hat a_{{N-1}}^{\dagger})^{n_{N-1}}$. Just as in the single-mode case, multimode Gaussian unitary transformations are defined to be those which can be generated by Hamiltonians quadratic in the ladder operators. \\

A multimode state $\ket{\psi}$ with rank $R\in \mathbb{N} \cup \{0\}$ can be expressed as
\begin{align}\label{eq:multimode_stellar}
    \ket{\psi} &= \hat U_{G}\hat C
    \ket{\bs{0}} = \hat U_{G}\ket{C}.
\end{align}
$\hat U_{G}$ is a multimode Gaussian operation, while $\ket{C}\equiv C(\hat{\bs{a}}^{\dagger})\ket{\bs{0}}$ is the multimode core state. $C(\cdot)$ is the degree $R$ $N$-variate polynomial

\begin{align}\label{eq:multimode_polynomial}
    C(\bs{\hat{a}^{\dagger}}) &= \sum_{R'=0}^{R} \ \sum_{\substack{\boldsymbol{n} \\ \mathrm{sum}(\bs{n})= R'}} \ c'_{\bs{n}} (\bs{\hat a^{\dagger}})^{\bs{n}},
\end{align}
where $\mathrm{sum}(\bs{n})\equiv n_0+n_1+\ldots+n_{N-1}$, and the complex coefficients $c'_{\bs{n}}$ are chosen to ensure $\ket{\psi}$ is normalized to unity. The core state is a unique state with bounded support over multimode Fock states which can be rotated into $\ket{\psi}$ by means of a Gaussian unitary transformation. \\

States which do not admit the decomposition presented in Eq.~\eqref{eq:multimode_stellar} are said to have an infinite rank. The properties of the single-mode stellar hierarchy are applicable in the multimode case too: finite-rank states can get arbitrarily close to infinite-rank states in trace distance~\cite{chabaud2022holomorphic}. The (1+1)D $\phi^4$ ground state is expected to have an infinite rank, and thus, we will approximate it using finite-rank states. \\

Similar to the single-mode case, we demand that both the multimode core state $\ket{C}$ and Gaussian unitary transformation $\hat U_{G}$ obey the symmetries of the lattice $\phi^4$ Hamiltonian, i.e., parity, time-reversal, lattice translations, and lattice inversion. To simplify calculations, we will introduce two additional restrictions to our ansatz. First, we will demand the Gaussian operation to not entangle the modes---thus, all cross-mode entanglement in the ansatz will arise from the core state. Second, we will introduce an additional truncation in the core-state polynomial. These restrictions are discussed in more detail below.\\

An arbitrary multimode Gaussian can be expressed as~\cite{serafini2023quantum}
\begin{align}
    \hat U_G = \hat R(\Phi_1) \otimes_{j=0}^{N-1}\left[\hat S_j(\xi_j)\hat D_j(\alpha_j)\right]\hat R(\Phi_2),
\end{align}
where $\hat S(\xi_j)$ and $\hat D(\alpha_j)$ are the single-mode squeezing and displacement operators introduced earlier, while $\hat R(\Phi) \coloneq e^{i\sum_{j,j'=0}^{N-1}\hat a_j^{\dagger}\Phi_{j,j'}\hat a_{j'}}$ is a passive rotation parametrized by the Hermitian rotation matrix $\Phi$. The passive rotations are responsible for entangling different modes, and we will drop them from our ansatz. This will result in a tensor product of single-mode Gaussian operations. Due to translation invariance, and following the arguments in the single-mode section, this would leave us with the simplified symmetric Gaussian operation:
\begin{align}
\hat U_G=\otimes_{j=0}^{N-1}\hat S_j(r),
\end{align}
that is, a tensor product of single-mode squeezing operators.\\

Next, we impose the Hamiltonian symmetries on the coefficients $c'_{\bs{n}}$ of the  polynomial $C(\bs{\hat{a}^{\dagger}})$, which generates the core state $\ket{C}$. As a result of the translation and inversion symmetries, $c'_{\bs{n}}$ is invariant under cyclic and anticylic permutations of $\bs{n}$. Parity constrains $c'_{\bs{n}}=0$ for $\bs{n}$ such that $\mathrm{sum}(\bs{n})$ is odd (and thus the rank is constrained to be even). Finally, $c'_{\bs{n}}\in \mathbb{R}$ due to time-reversal invariance. Despite the symmetry constraints imposed above, the space of symmetric multimode rank-$R$ core states grows rapidly with $R$. Thus, we will introduce an additional truncation in the ansatz. Define the \emph{span} of a Fock state $\ket{\bs{n}}$ (on a translation- and inversion-symmetric lattice) to be the smallest integer $Q'$ such that all bosonic excitations are contained within the sites $\{j,j+1,\cdots,j+Q'\}$ for some site $j$. Consequently, define the span of a superposition of such Fock states to be the largest value of span associated with the Fock states in the superposition. We will restrict our ansatz to rank-$R$, span-$Q\leq \lfloor\frac{N}{2}\rfloor$ core states generated by: 
\begin{align}
    \hat C_{R,Q} \equiv C_{R,Q}(\bs{\hat a^{\dagger}}) = \sum_{j=0}^{N-1} \ \ \hat c_{R,Q}^{(j)},
\end{align}
where 
\begin{widetext}
    \begin{align}
    \hat c_{R,Q}^{(j)} \equiv c_{R,Q}^{(j)}(\bs{a^{\dagger}}) & = \sum_{R'=0,2,\ldots,R} \, \sum_{Q'=0}^Q \, \sum_{\substack{n'_0,\ldots,n'_{Q'} \\ n'_0 +\ldots + n'_{Q'}=R' \\ n'_0,n'_{Q'}\geq 1}}d_{n'_0,\ldots,n'_{Q'}} (a_j^{\dagger})^{n'_0}(a_{j+1}^{\dagger})^{n'_1} \ldots (a_{j+Q'}^{\dagger})^{n'_{Q'}},
    \label{eq:c-RQ-def}
\end{align}
\end{widetext}
where $d_{n'_0,\ldots,n'_{Q'}} \in \mathbb{R}$ and $ d_{n'_0,\ldots,n'_{Q'}} = d_{n'_{Q'},\ldots,n'_{0}}$. Using straightforward combinatorics, the number of terms in the polynomial $C_{R,Q}(\cdot)$ is $|C_{R,Q}|=N|c_{R,Q}|$, where
\begin{align}\label{eq:coresize}
    |c_{R,Q}| = 1+\sum_{R'=2,\ldots,R} \ \sum_{Q'=0}^Q \ \binom{R'+Q'-2}{Q'}. 
\end{align}

\begin{figure*}[t!]
    \includegraphics[scale = 0.675]{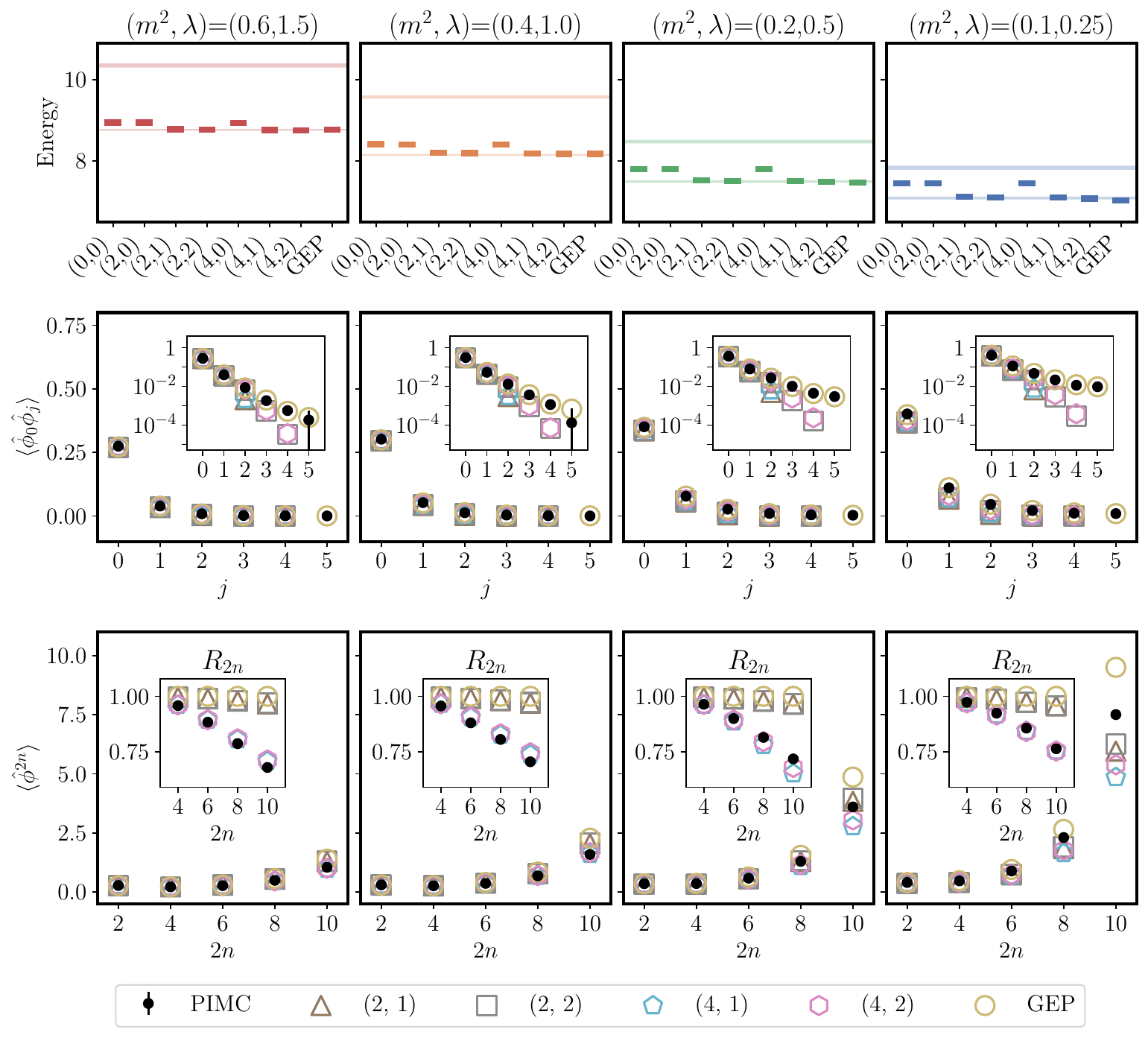}
    \caption{Minimum Hamiltonian expectation values (top panels), two-point functions (middle panels), and local $\hat \phi$ moments and moment ratios (bottom panels) of various ansatzes compared with the Monte-Carlo estimates for several values of $(m^2,\lambda)$. The two bands in the top panels indicate the corresponding Monte-Carlo energy estimate for the ground state and first excited state. Black points with error bars in the middle and bottom panels  show the values from PIMC. A log scale is adopted for the inset plots in the middle panels. Error bars and error bands are obtained from bootstrap resampling of the Monte-Carlo estimates, as explained in Appendix~\ref{app:pimc}.
    }
    \label{fig:multimode_min_en}
\end{figure*}

Incorporating both the simplification in the Gaussian unitary transformation and the $Q$-truncation of the core state, the complete form for our simplified ansatz reads
\begin{align}
        |\psi\rangle_{R,Q} \coloneq \left[\otimes_{j=0}^{N-1} \hat S_j(r)\right] |C\rangle_{R,Q},
        \label{eq:psi-QR}
\end{align}
\noindent where $\ket{C}_{R,Q}\equiv C_{R,Q}(\bs{\hat a^{\dagger}})\ket{\bs{0}}$. We shall refer to this as the $(R,Q)$ ansatz. \\

In order to compute local expectation values involving the operators $(\hat \phi_j,\hat \pi_j)$, it will be useful to express the core state using a polynomial generator with the argument $\hat{\bm{\phi}}$, i.e., $\ket{C}_{R,Q}=C_{R,Q}(\bs{\hat a^{\dagger}})\ket{\bs{0}}\equiv C^{\phi}_{R,Q}(\hat{\bm{\phi}})\ket{\bs{0}}$, with $\hat{\bm{\phi}} \coloneq (\hat \phi_0, \cdots, \hat \phi_{N-1})$. It is always possible to deduce a unique polynomial $C^{\phi}_{R,Q}(\cdot)$ given some core state $\ket{C}_{R,Q}$, such that there is a one-to-one mapping between the terms and the number of independent coefficients of the polynomials $C_{R,Q}(\cdot)$ and $C^{\phi}_{R,Q}(\cdot)$; see Appendix~\ref{app:core_quad}. %
For simplicity, we will drop the $\phi$ superscript in $C^{\phi}_{R,Q}(\cdot)$; the argument of the polynomial clarifies whether $C_{R,Q}(\bs{\hat{\phi}})$ or  $C_{R,Q}(\bs{\hat{a}^{\dagger}})$ is assumed.

Our subsequent numerical study assumes maximum values of $R=4$ and $Q=2$ (and the number of modes $N=10 \geq 2Q$). Thus, it is instructive to explicitly enumerate all the terms in core-state polynomial associated with these maximum $R$ and $Q$ values:
\begin{widetext}
    \begin{align}
     C_{4,2}(\hat{\bm{\phi}}) &= \sum_{j=0}^{N-1} \left\{A + B_0 \phih_j^2 + B_1 \phih_j\phih_{j+1} + B_2 \phih_j\phih_{j+2} + C_0 \phih_j^4 + C_{11}\left(\phih_j^3\phih_{j+1}+\phih_j\phih_{j+1}^3\right) + C_{12}\phih_j^2\phih_{j+1}^2 \right. \nonumber \\
     &\hspace{1.25 cm} \left.+ C_{21}\left(\phih_j^3\phih_{j+2}+\phih_j\phih_{j+2}^3\right) + C_{22}\left(\phih_j^2\phih_{j+1}\phih_{j+2}+\phih_j\phih_{j+1}\phih_{j+2}^2\right) + C_{23}\phih_j\phih_{j+1}^2\phih_{j+2}+C_{24}\phih_j^2\phih_{j+2}^2\right\} \nonumber \\
     &= \sum_{j=0}^{N-1} \ c^{(j)}_{4,2}(\phih_j,\phih_{j+1},\phih_{j+2}).
     \label{eq:C42-ansatz}
\end{align}
\end{widetext}
Here, all coefficients are real. Recall the cyclic translation invariance of the theory which imposes $\phi_j=\phi_{j+N}$. There are $|c_{4,2}|=14$ independent terms in the $c_{4,2}^{(j)}(\cdot)$ polynomial, consistent with Eq.~\eqref{eq:coresize}. There are, however, only 10 independent polynomial coefficients, due to the inversion symmetry and normalization constraint. 

To determine the complexity of computing expectation values with respect to the $(R,Q)$ ansatz, consider the expectation value of some operator $
\hat{\mathcal{O}} =\mathcal{O}(\hat{\bm{\phi}},\hat{\bm{\pi}})$:
\begin{align}
    {}_{R,Q}&\langle \psi| 
    \hat{\mathcal{O}}|\psi\rangle_{R,Q}\nonumber \\
    &={}_{R,Q}\bra{C}\left[\otimes_{j=0}^{N-1} \hat S^{\dagger}_j(r)\right] 
    \hat{\mathcal{O}} \left[\otimes_{j=0}^{N-1} \hat S_j(r)\right] \ket{C}_{R,Q}.
    \label{eq:exp-value-RQ}
\end{align}
Since $\hat S(r)\hat{\bm{\phi}} \hat S^{\dagger}(r)=e^{-r}\hat{\bm{\phi}}$ and $\hat S(r)\hat{\bm{\pi}} \hat S^{\dagger}(r)=e^{r}\hat{\bm{\pi}}$, one can apply the squeezing operators to $
\hat{\mathcal{O}}$ to express the expectation value in Eq.~\eqref{eq:exp-value-RQ} as:
\begin{align}
    {}_{R,Q}&\langle \psi| 
    \hat{\mathcal{O}}|\psi\rangle_{R,Q} \nonumber \\
    &= \bra{\bs{0}}C_{R,Q}(\hat{\bm{\phi}})\ 
    \mathcal{O}(e^r\hat{\bm{\phi}},e^{-r}\hat{\bm{\pi}}) \ C_{R,Q}(\hat{\bm{\phi}})\ket{\bs{0}}.
\end{align}
Upon writing the terms in the $C_{R,Q}(\hat{\bm{\phi}})$ polynomial explicitly and defining $
\hat{\mathcal{O}}_r\coloneq 
\mathcal{O}(e^r\hat{\bm{\phi}},e^{-r}\hat{\bm{\pi}}) $, the full expectation value will break into a sum of $|c_{R,Q}|^2$ expectation values of the form
\begin{align}\sum_{j,j'=0}^{N-1}\bra{\bs{0}}\phih_{j}^{n_0}\cdots\phih_{j+Q}^{n_Q} \ 
\hat{\mathcal{O}}_r\ \phih_{j'}^{n_0'}\cdots\phih_{j'+Q}^{n'_Q}\ket{\bs{0}}.
\end{align}
We will be interested in computing expectation values of operators that feature in the Hamiltonian, i.e., the local operators $\phih_j^2, \phih_j^4, \pih_j^2$, and the non-local operators $\phih_0\phih_j$. Additionally, we will choose local operators such as $\phih_j^{2n}$ with $n>2$, and non-local operators such as $\phih_0\phih_j$ with $j\geq 2$ to feature in the target-moment set. In all these cases, 
only $O(Q^2)$ expectations value in the above double-sum need to be evaluated independently, and each such expectation value breaks into a trivial product of local expectation values. Thus, expectation values with respect to the $(R,Q)$ ansatz can be computed with classical time complexity $O(|c_{R,Q}|^2Q^2)$. Notably, this complexity is independent of the system size $N$. \\

The ground state of an interacting $\phi^4$ scalar field theory with bare mass $m$ can also be approximated by that of a free scalar field theory, but with a \emph{different} mass $\mu$. This constitutes another finite-rank ansatz, namely, the Gaussian effective potential (GEP)~\cite{stevenson1984gaussian, stevenson1985gaussian, thompson2023quantum}. Here, $\mu$ is a
single parameter specifying the ansatz.\footnote{The full definition of the GEP introduced in Ref.~\cite{stevenson1984gaussian} includes a displacement. The displacement is dropped here because we are only interested in states that respect parity.}
For the sake of comparison, we will also consider this ansatz. Recall that the free scalar-field-theory Hamiltonian can be diagonalized as 
\begin{align}
    \hat H_{\rm free}
    &= \sum_{k=0}^{N-1}\omega_k \left(\hat A_k^{\dagger}\hat A_k+\frac{1}{2}\right),
\end{align}
where 
$\omega_k^2\coloneq \mu^2+4 \,\mathrm{sin}^2\left(\frac{\pi k}{N}\right)$ and 
\begin{align}
    \hat A_k &\coloneq \sqrt{\frac{\omega_k}{2}}\hat{\tilde{\phi}}_k + \frac{i}{\sqrt{2\omega_k}}\hat{\tilde{\pi}}_k \nonumber \\
    &= \frac{1}{\sqrt{N}}\sum_{j=0}^{N-1} \left(\mathrm{cosh}(\chi_k)\hat a_j + \mathrm{sinh}(\chi_k)\hat a_j^{\dagger}\right)e^{-i\frac{2\pi jk}{N}}\nonumber\\
    & \equiv \hat U_{\rm GEP} \ \hat a_k \ \hat U_{\rm GEP}^{\dagger},
\end{align}
with $\hat{\tilde{O}}_k\coloneq \frac{1}{\sqrt{N}}\sum_{j=0}^{N-1}\hat O_j e^{-i\frac{2\pi jk}{N}}$ and $\chi_k\coloneq\frac{1}{2}\mathrm{log}(\omega_k)$. The equality in the second line arises from replacing the $\hat{\phi}_j$ and $\hat{\pi}_j$ operators with the $\hat a_j$ and $\hat a_j^\dagger$ operators using Eq.~\eqref{eq:mode-op-def}. We have then used the standard form of a Gaussian unitary transformation  in Eq.~\eqref{eq:multimode_gaussian} to define $\hat U_\text{GEP}$ with $\alpha_k=0, \ u^*_{kj}=\frac{\mathrm{cosh}(\chi_k)e^{-i\frac{2\pi jk}{N}}}{\sqrt{N}}$, and $v^*_{kj}=\frac{\mathrm{sinh}(\chi_k)e^{-i\frac{2\pi jk}{N}}}{\sqrt{N}}$, which yield the equivalence in the third line. The ground state $\ket{\psi}_{\rm GEP}$ is given by the vacuum of the $\hat A_k$ modes. Using $\hat A_k\ket{\psi}_{\rm GEP}=\hat U_{\rm GEP}\hat a_k\hat{U}_{\rm GEP}^{\dagger}\ket{\psi}_{\rm GEP}=0$, we deduce that $\hat{U}_{\rm GEP}^{\dagger}\ket{\psi}_{\rm GEP}$ is the vacuum of the original modes $\hat a_k$. Thus, $\ket{\psi}_{\rm GEP}=\hat U_{\rm GEP}\ket{0}$, and this is a rank-0 ansatz. Due to translation invariance, two-point correlation functions in momentum space have the form $_{\rm GEP}\langle\psi|\hat{\tilde{\phi}}_{k_1}\hat{\tilde{\phi}}_{k_2}|\psi\rangle_{\rm GEP}= f(\omega_{k_1})\delta_{k_1+k_2,0}$. In this instance, $f(\omega_k)=\frac{1}{2\omega_k}$. Thus, the cost of computing the (low-order) position-space correlation functions is equivalent to the cost of taking a discrete Fourier transform of $f(\omega_k)$. For example, for the two-point function,
\begin{align}
    \langle \hat \phi_{j} \hat \phi_{j'}\rangle = \frac{1}{N}\sum_{k} f(\omega_k)e^{i\frac{2\pi(j-j')k}{N}}.
\end{align}
This computation has, therefore, an $O(N\mathrm{log}N)$ classical complexity.

\subsection{Euclidean-Monte-Carlo-informed moment optimization}

We aim to optimize the multimode ansatzes introduced in Sec.~\ref{sec:multi-ansatze} by minimizing the loss function in Eq.~\eqref{eq:mom_obj}. Ground-state moments $\bra{\Omega} \hat O \ket{\Omega}$ are sourced from Euclidean PIMC computations. Monte-Carlo data exhibit statistical uncertainties; these uncertainties need to be propagated to the optimized parameters $\vec{\Lambda}_0$. Appendix~\ref{app:pimc} presents details of this uncertainty propagation.

We first minimize the energy of various ansatzes, and observe the behavior of the two-point functions and local non-Gaussianity of the resulting optimized wavefunctions. Thereafter, we demonstrate moment optimization for target sets consisting of local and non-local $\phih$ moments.
\begin{figure}
    \includegraphics[width=0.8\columnwidth]{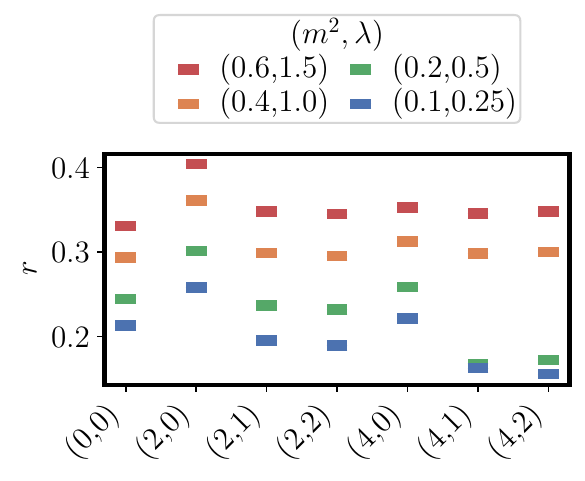}
    \caption{Optimized value of the squeezing parameter $r$ for the minimum-energy $(R,Q)$ ansatzes.
    }
    \label{fig:multisqueeze}
\end{figure}

\subsubsection{Energy minimization}
\begin{figure}
    \centering
    \includegraphics[width=0.95\columnwidth]{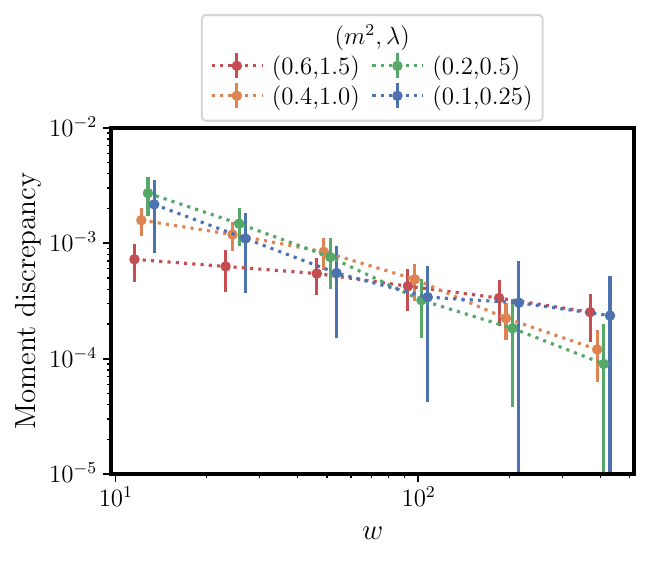}
    \caption{Sum of squared discrepancies in the target moments plotted as a function of the weight $w$ for moment-ratio optimization. For comparison, the corresponding values for the minimum-energy ansatz for $(m^2,\lambda)=(0.6,1.5)$, $(0.4,1.0)$, $(0.2,0.5)$, and $(0.1,0.25)$ are 0.07, 0.25, 0.19, and 2.10, respectively. The error bars on the moment discrepancy are obtained by propagating the Monte-Carlo uncertainties using the procedure described in Appendix~\ref{app:pimc}. The values associated with different $(m^2,\lambda)$ are slightly offset in the horizontal direction to improve visibility of the error bars.}
    \label{fig:sqd_diff_ratio_opt}
\end{figure}
\begin{figure*}
    \includegraphics[width=\textwidth]{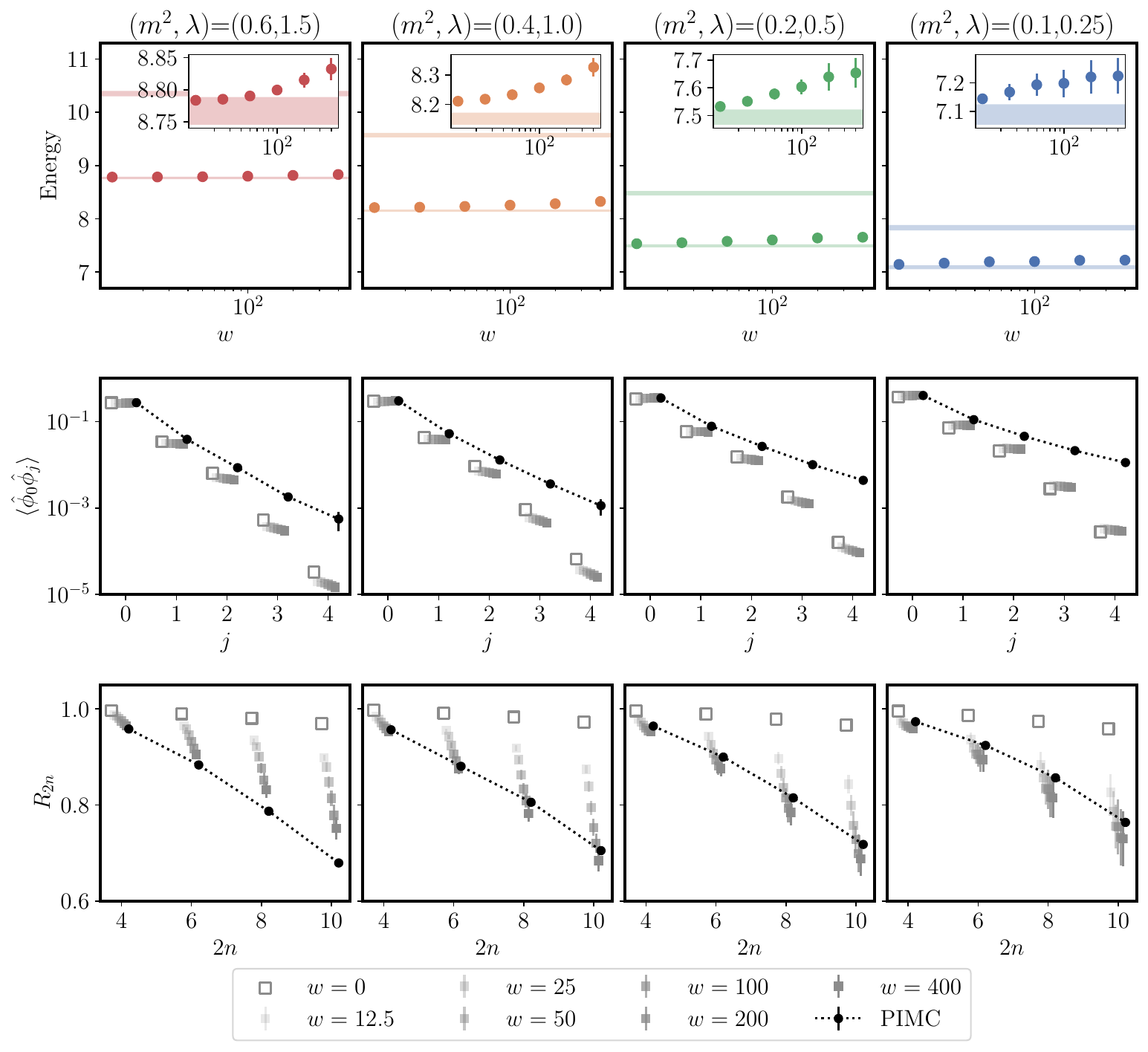}
    \caption{
    Optimization of the moment ratio for the $(R,Q)=(2,2)$ ansatz for various values of $(m^2,\lambda)$. Moment optimization is performed for the target set $\mathcal{T}=\{\phih_j^6,\phih_j^8,\phih_j^{10}\}$ for various values of $w$. The bands in the plots on the top panels represent the PIMC values of the ground- and first excited-state energies and their associated uncertainties. In the central and bottom panels, the values associated with different weights and PIMC are slightly offset in the horizontal direction to improve visibility of the error bars. Error bars and error bands on PIMC values are obtained from bootstrap resampling of the Monte-Carlo estimates, as explained in Appendix~\ref{app:pimc}. The error bars on the expectation values from moment optimization are obtained by propagating the Monte-Carlo uncertainties using the procedure described in Appendix~\ref{app:pimc}.
    }
    \label{fig:ratio_mom_opt}
\end{figure*}
The top row in Fig.~\ref{fig:multimode_min_en} plots the energy obtained upon minimizing the Hamiltonian expectation value using the $(R,Q)$ and GEP ansatzes for various values of $(m^2,\lambda)$. The bands show the PIMC estimates for the ground and first excited-state energies. The $Q \neq 0$ $(R,Q)$ ansatzes and the GEP ansatz all lead to a comparable energy close to the PIMC estimate. The fact that the GEP leads to a good energy estimate illustrates that the quartic coupling behaves like an effective mass. This can also be seen in the behavior of the two-point function shown in the center row of Fig.~\ref{fig:multimode_min_en}. The GEP does a good job at reproducing the two-point function for all values of $j$, while the $(R,Q)$ ansatzes do a worse job with increasing $j$. The $Q$ truncation in the $(R,Q)$ ansatz results in $\langle \phih_0\phih_j\rangle=0$ for $j> 2$ when $Q=1$ and for $j > 4$ when $Q=2$.\\

The quartic coupling, however, induces non-Gaussianity in the ground state. As mentioned previously, we inspect non-Gaussianity by studying the $\phi^{2n}$-moments and their ratios. The bottom row of Fig.~\ref{fig:multimode_min_en} compares the PIMC values of the $\phi^{2n}$-moments with the corresponding ansatz values. The $\langle \phih^{2n}\rangle$ values for all ansatzes gradually deviate from the Monte-Carlo values as $n$ increases, and this deviation gets more pronounced as $(m^2,\lambda) \to (0,0)$. The values of these moments by themselves do not clearly convey any information on the states' non-Gaussianity.  
However, when one computes moment ratios, even small discrepancies among different ansatzes' $\phi^2$-moments (not visible in Fig.~\ref{fig:multimode_min_en} given the scale of the moment plots) contribute significantly to the ratio in Eq.~\eqref{eq:moment-ratio}, and a clear hierarchy emerges: The moment ratios for the GEP ansatz are pinned to one. For the $R=2$ ansatzes, the value only deviates slightly from one. In contrast, the $R=4$ ansatzes get much closer to reproducing the Monte-Carlo values of the moment ratios. 
Thus, while the purely Gaussian effect of readjusting the mass can lead the GEP close to the ground state energetically, the GEP fails to reproduce the non-Gaussian moment ratios. The $(R,Q)$ ansatz, on the other hand, can reproduce the expected values of the non-Gaussian moment ratios for sufficiently large rank $R$.

\begin{figure*}
    \includegraphics[scale=0.65]{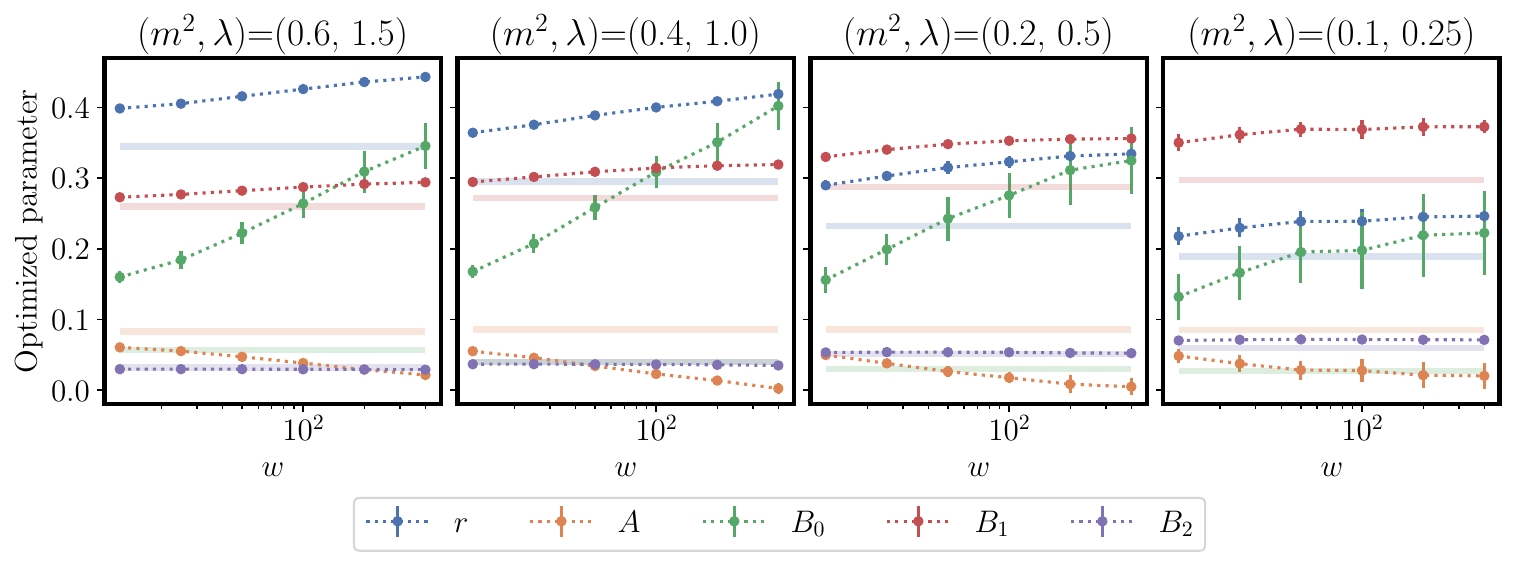}
    \caption{
    Optimized values of the $(R,Q)=(2,2)$ ansatz parameters [with its core state polynomial given by Eq.~\eqref{eq:C42-ansatz} when all parameters except $A, B_0, B_1$, and $B_2$ are set to zero] plotted as a function of $w$ for moment-ratio optimization. The dotted lines passing through the data points are to guide the eye and do not represent a fit. The error bars on the moment-optimized parameters are obtained by propagating the Monte-Carlo uncertainties using the procedure described in Appendix~\ref{app:pimc}. The thick light-colored lines show the parameter values for the minimum-energy ansatz. These lines are thickened to improve visibility and do not imply error bars.
    }
    \label{fig:ratio_mom_opt_params}
\end{figure*}

Importantly, even though energy minimization leads to comparable values of energy for different ansatz families, the resulting minimum-energy ansatzes behave differently in terms of their non-local correlations and non-Gaussianity. Given this distinctive behavior of moments for different ansatz families, in the next section, we will explore how moment errors can be varied within a \textit{fixed} ansatz family using moment optimization.\\

Before moving on, we plot the values of the squeezing parameter $r$ for the minimum-energy $(R,Q)$ ansatzes in Fig.~\ref{fig:multisqueeze}. The value of $r$ drops as the continuum limit $(m^2,\lambda)\rightarrow (0,0)$ is approached. When the squeezing operator acts on the core state, it extends its support from a finite domain in the local Hilbert space to the entire local Hilbert space. The value of the squeezing parameter $r$ determines how the state's Fock-basis amplitudes are distributed across the the Hilbert space. The 
$r$ value, therefore, becomes important when considering Hilbert-space truncations for implementing the state on qubit-based platforms. We will discuss this point in more detail in Sec.~\ref{sec:qsqueeze}. \\ 

\subsubsection{Moment-ratio optimization}

We observed that the minimum-energy $(R,Q)=(2,2)$ ansatz captures the ground-state energy rather accurately. However, its moment ratios are quite Gaussian. Thus, we will employ moment optimization to look for alternative low-energy solutions which reproduce the theory's non-Gaussianity more accurately. We take the target set to be $\mathcal{T}=\{\phih_j^6,\phih_j^8,\phih_j^{10}\}$ (\emph{any} value of $j$ works because of translation invariance) with uniform weights $w_{\hat O}=w \in \{12.5,25,50,100,200,400\}$, see Eq.~\eqref{eq:mom_obj}. These weight values are chosen by experimentation. We display the results from this optimization in Figs~\ref{fig:sqd_diff_ratio_opt},~\ref{fig:ratio_mom_opt}, and~\ref{fig:ratio_mom_opt_params}. 

Figure~\ref{fig:sqd_diff_ratio_opt} shows the value of the optimized target-moment discrepancy, i.e., the term proportional to $w$ in the loss function, $\sum_{\hat O \in \mathcal{T}} \left(\langle \psi(\vec{\Lambda})|\hat O|\psi(\vec{\Lambda})\rangle-\langle\Omega|\hat O|\Omega\rangle\right)^2$, as a function of $w$. One observes that the value of this moment discrepancy decreases as a function of $w$, indicating a steady improvement in the behavior of the target moments $\mathcal{T}=\{\phih_j^6,\phih_j^8,\phih_j^{10}\}$. 

Crucially, this improvement happens at a negligible cost in energy relative to the spectral gap, as shown in the top panel of Fig.~\ref{fig:ratio_mom_opt}. Additionally, the middle panel in Fig.~\ref{fig:ratio_mom_opt} shows that the behavior of the two-point functions is not altered significantly. As a consequence of the improvement in the expectation value of $\phih_j^6,\phih_j^8$, and $\phih_j^{10}$, there is an initial improvement in the behavior of their moment-ratios as shown in the bottom panel of Fig.~\ref{fig:ratio_mom_opt}. However, as $w$ becomes larger, the accuracy of the values of individual moments $ \langle\hat\phi_j^n\rangle$ is prioritized over their ratios. This results in the moment ratios for $(m^2,\lambda)=(0.4,1.0), \ (0.2,0.5)$, and $(0.1,0.25)$ overshooting their PIMC value at some value of $w$. In other words, the value of $w$ at which the optimal behavior of moment ratios is recovered decreases as $(m^2,\lambda)\rightarrow (0,0)$. 

Finally, Fig.~\ref{fig:ratio_mom_opt_params} shows how the optimized parameters evolve as a function of the weight $w$. Interestingly, for all values of $(m^2,\lambda)$, the moment optimization guides the ansatz in a roughly similar direction in parameter space---$B_2$ stays mostly constant, $r$ and $B_1$ increase slowly, $A$ decreases slowly, and there is a more dramatic increase in $B_0$ with $w$.

Since the ground-state moments in the moment-optimization loss function are sourced from PIMC, they carry statistical uncertainty. A greater uncertainty is observed in the moment-optimized quantities corresponding to smaller values of $(m^2,\lambda)$, as seen in all plots in Figs~\ref{fig:sqd_diff_ratio_opt},~\ref{fig:ratio_mom_opt}, and~\ref{fig:ratio_mom_opt_params}. Furthermore, statistical errors on moment-optimized quantities increase with $w$, as the moments carrying statistical error are weighed more.

\subsubsection{Two-point function optimization}
\begin{figure}
    \centering
    \includegraphics[width=0.95\columnwidth]{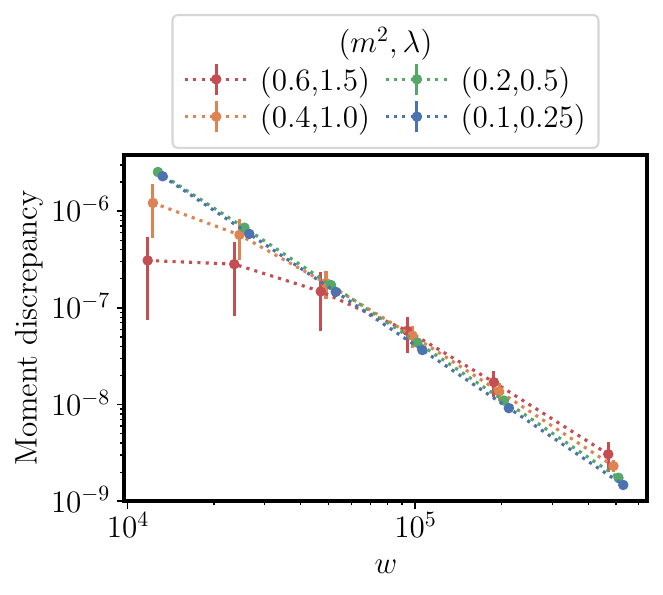}
    \caption{Sum of squared discrepancies in the target moments plotted as a function of the weight $w$ for two-point function optimization. For comparison, the corresponding values for the minimum-energy ansatz for $(m^2,\lambda)=(0.6,1.5)$, $(0.4,1.0)$, $(0.2,0.5)$, and $(0.1,0.25)$ are $7.7\times 10^{-7}$, $7.8 \times 10^{-7}$,$1.7 \times 10^{-5}$, and $1.4\times 10^{-4}$, respectively. The error bars on the moment discrepancy are obtained by propagating the Monte-Carlo uncertainties using the procedure described in Appendix~\ref{app:pimc}. The values associated with different $(m^2,\lambda)$ are slightly offset in the horizontal direction to improve visibility of the error bars.}
    \label{fig:sqd_diff_twopt_opt}
\end{figure}
\begin{figure*}
    \includegraphics[width=\textwidth]{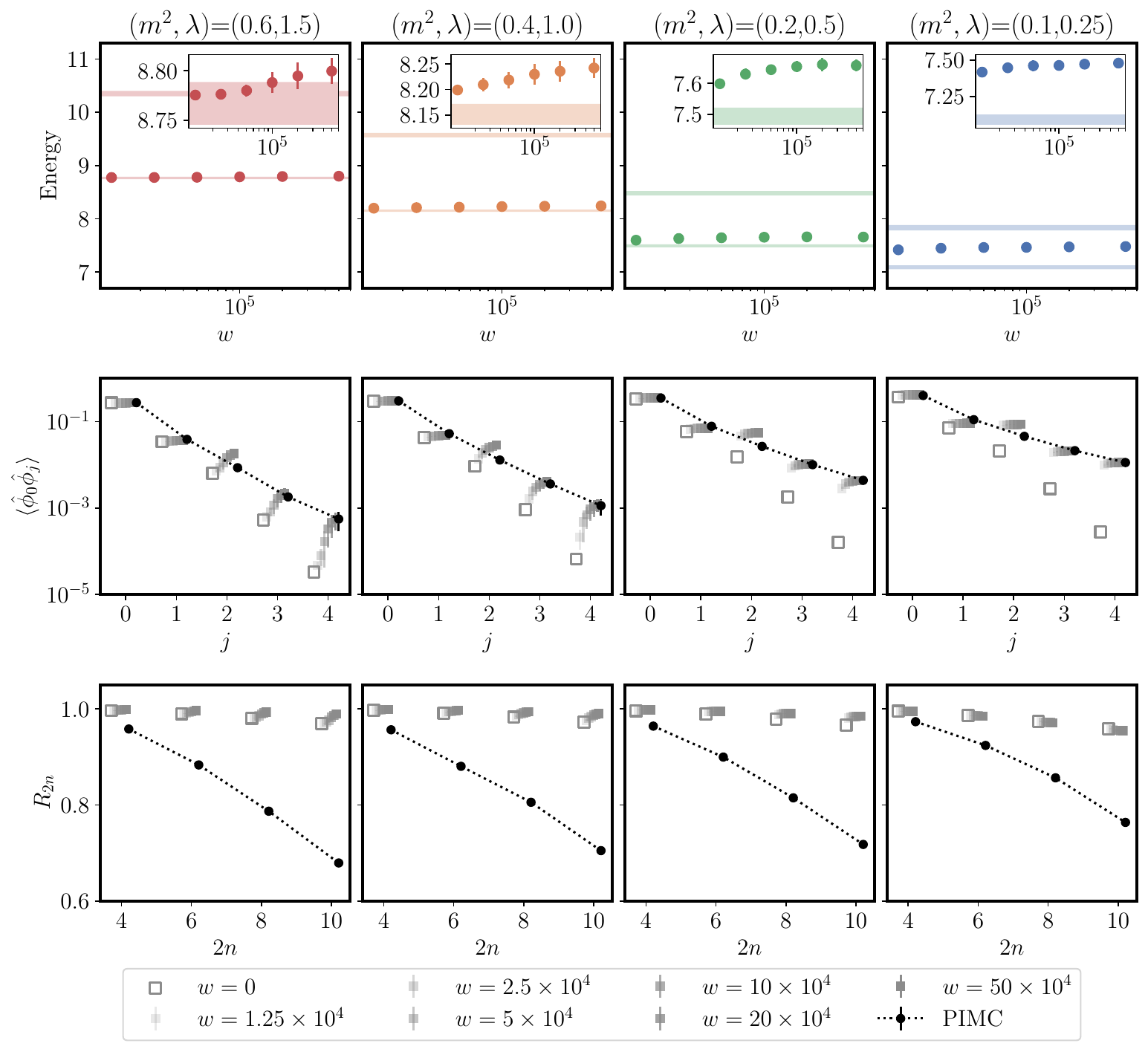}
    \caption{
    Optimization of the two-point function for the $(R,Q)=(2,2)$ ansatz for various values of $(m^2,\lambda)$. Moment optimization is performed for the target set $\mathcal{T}=\{\phih_0\phih_4\}$ for various values of $w$. The bands in the plots on the top panels represent the PIMC values of the ground- and first excited-state energies and their associated uncertainty. In the central and bottom panels, the values associated with different weights and PIMC are slightly offset in the horizontal direction to improve visibility of the error bars. Error bars and error bands on PIMC values are obtained from bootstrap resampling of the Monte-Carlo estimates, as explained in Appendix~\ref{app:pimc}. The error bars on the expectation values from moment optimization are obtained by propagating the Monte-Carlo uncertainties using the procedure described in Appendix~\ref{app:pimc}.
    }
    \label{fig:twopt_mom_opt}
\end{figure*}
\begin{figure*}
    \includegraphics[scale=0.65]{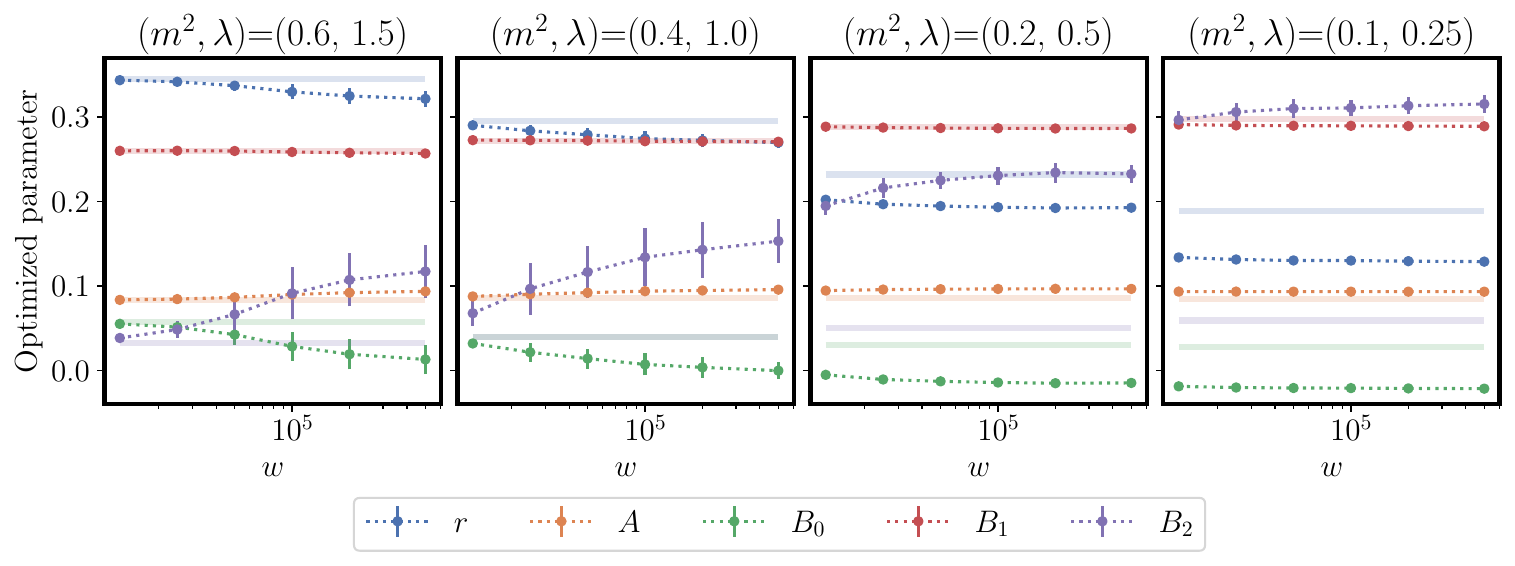}
    \caption{
    Optimized values of the $(R,Q)=(2,2)$ ansatz parameters
    [with its core state polynomial given by Eq.~\eqref{eq:C42-ansatz} when all parameters except $A, B_0, B_1$, and $B_2$ are set to zero] plotted as a function of $w$ for two-point function optimization. The dotted lines passing through the data points are to guide the eye and do not represent a fit. The error bars on the moment-optimized parameters are obtained by propagating the Monte-Carlo uncertainties using the procedure described in Appendix~\ref{app:pimc}. The thick light-colored lines show the parameter values for the minimum-energy ansatz. These lines are thickened to improve visibility and do not imply error bars.
    }
    \label{fig:twopt_mom_opt_params}
\end{figure*}

Next, we explore alternative low-energy solutions with improved behavior for the two-point function of the $(R,Q)=(2,2)$ ansatz. Towards this goal, we take the target set to be $\mathcal{T}=\{\phih_0\phih_4\}$ with the weight $w\in \{1.25,2.5,5,10,20,50\} \times 10^4$, see Eq.~\eqref{eq:mom_obj}. These weight values have been chosen by experimentation. It is observed that much bigger values of weight are required to shift the two-point functions as compared to the moment ratios. The results of the consequent optimization are shown in Figs.~\ref{fig:sqd_diff_twopt_opt},~\ref{fig:twopt_mom_opt}, and~\ref{fig:twopt_mom_opt_params}.

Figure~\ref{fig:sqd_diff_twopt_opt} shows that the value of the optimized target-moment discrepancy, i.e., the term proportional to $w$ in the loss function, $\sum_{\hat O \in \mathcal{T}} \left(\langle \psi(\vec{\Lambda})|\hat O|\psi(\vec{\Lambda})\rangle-\langle\Omega|\hat O|\Omega\rangle\right)^2$, decreases as a function of $w$. 

Even though only the $\phih_0\phih_4$ moment belongs to the target set, the behavior of both $\langle\phih_0\phih_3\rangle$ and $\langle\phih_0\phih_4\rangle$ improves while that of $\langle\phih_0^2\rangle$ and $\langle\phih_0\phih_1\rangle$ stays intact as shown in the bottom panel of Fig.~\ref{fig:twopt_mom_opt}. The $\langle\phih_0\phih_2\rangle$ value hits its PIMC value at a smaller value of $w$, and overshoots the PIMC value at bigger values of $w$ (i.e., at values where $\phih_0\phih_j$ with $ j=3,4$ hit their PIMC values). This peculiar behavior of the ansatz could be potentially fixed by incorporating $\phih_0\phih_2$ into the target set. 
The moment ratios stay largely intact in all cases.  However, the energy penalty paid rises steadily as $(m^2,\lambda)\rightarrow (0,0)$, and it is significant for $(m^2,\lambda)=(0.1,0.25)$. Thus, this example illustrates the limits of performing moment optimization with this ansatz. As one approaches the continuum limit, larger values of $R$ and $Q$ are needed to accommodate the correlations. 

Figure~\ref{fig:twopt_mom_opt_params} displays the values of the optimized parameters as a function of the weight $w$. The largest shifts and errors are associated with the parameter $B_2$ (which accompanies the $\phih_j\phih_{j+2}$ term in the core-state polynomial). This is perhaps not surprising because this parameter controls the strength of non-local correlations. However, the parameter $B_1$ (which accompanies $\phih_j\phih_{j+1}$) hardly shifts. \\

In contrast with the previous section, the propagated errors in the moment-optimized quantities now decrease as $(m^2,\lambda)\rightarrow(0,0)$. Also in contrast with the previous section, while the sizes of the errors increase with weight in the moment-optimized energy, such a dependence is not observed for the moment-optimized two-point functions and moment discrepancy.

To summarize, one observes several distinct behaviors of ansatz moments in the above examples. First, the examples reveal completely rigid moments, e.g., for the GEP ansatz, the moment ratio is pinned to one while for the $(R,Q)$ ansatz, $\langle \phih_0\phih_j\rangle$ is pinned to be zero for $j> 2Q$ (when $Q<N/2$). Minimizing the ansatz energy imposes a natural distribution of moment errors in the remaining non-rigid moments. When one tries to control these moment errors by using moment optimization, one observes more or less rigid-moment behaviors. For example, the local $\phih$ moments for the $(R,Q)$ ansatz appear to be less rigid as they can be continuously improved with $w$ with only a steadily but slowly rising energy penalty. The $\phih_0\phih_j$ moments, on the other hand, behave more rigidly as $(m^2,\lambda)\rightarrow (0,0)$ because they are improved at a greater energy penalty. \\

This section, therefore, equipped us with an ansatz for the (1+1)D interacting scalar field theory that
possesses at most $|c|_{R,Q}$ independent parameters, and is systematically improvable. Importantly, it can be optimized with the aid of Euclidean-Monte-Carlo-informed moment optimization, to exhibit accurate energies, non-local correlation functions, and local non-Gaussianity. The efficiency of this ansatz for quantum simulation will be discussed in the next section.

\section{Quantum-circuit representation}\label{sec:circ}
\noindent
In this section, we encode the bosonic wavefunctions obtained in the previous sections into efficient quantum circuits. We consider a circuit design for discrete-variable, i.e., qubit-based, platforms, and discuss algorithms for core-state preparation and Gaussian-unitary implementation, as well as their associated complexities by analyzing the theoretical and algorithmic errors. In Appendix~\ref{app:CV}, we further discuss opportunities for continuous-variable, i.e., bosonic, platforms.

\subsection{Encoding bosonic states on a discrete-variable platform
\label{sec:DV}}

To represent bosonic degrees of freedom using qubits, one must cut off the infinite-dimensional bosonic Hilbert space. Thus, in addition to thermodynamic and continuum-limit extrapolations, a discrete-variable quantum simulation must first remove the dependence on this cutoff in observables. Several digitization and truncation schemes for bosons have been considered in the literature~\cite{jordan2012quantum,somma2015quantum,macridin2018bosonic,macridin2018digital,klco2019digitization,barata2021single,brennen2015multiscale,bagherimehrab2022nearly}. Here, we present one straightforward choice, and describe its quantum and classical computing costs. For the finite-rank ansatzes considered in this work, a natural choice is to encode bosonic Fock states in the qubits' computational basis states, and to introduce a cutoff in the total number of bosons or the number of bosons per mode. Such Fock-space Hamiltonian truncation has been employed for the two-dimensional Euclidean $\phi^4$ theory in Ref.~\cite{rychkov2015hamiltonian} to make predictions about the theory's spectral and critical properties, and for bosonic systems in Refs~\cite{somma2005quantum,encinar2021digital} in the context of quantum simulations.\\

We assume a cutoff $\Lambda$ on the number of bosons per mode. Thus, the truncated single-mode Fock space is defined as the span of the Fock states $\ket{n}_j$ with $n=0,\cdots,\Lambda$, for each mode $j \in \{0,\cdots,N-1\}$. The full truncated multimode Hilbert space is the tensor product of these Hilbert spaces. Following Ref.~\cite{somma2005quantum}, we introduce the ladder operators
\begin{align}
    &\hat a^{\Lambda}_j \coloneq \sum_{n=1}^{\Lambda} \ \sqrt{n} \ket{n-1}_j\bra{n}_j ,    \\
    &\hat{a}^{\Lambda \dagger}_j
    \coloneq \sum_{n=0}^{\Lambda-1} \ \sqrt{n+1} \ket{n+1}_j\bra{n}_j ,
\end{align}
on the truncated Fock space, with the commutation relation
\begin{align}
    &[\hat a^{\Lambda}_j,\hat a^{\Lambda \dagger}_k] = \delta_{j,k}\left(1-\frac{\Lambda +1}{\Lambda !}\left(\hat a_{j}^{\Lambda\dagger}\right)^{\Lambda}\left(\hat a^{\Lambda}_{j}\right)^{\Lambda}\right). 
\end{align}
$(\cdot)^\Lambda$ denotes the $\Lambda$th powers of the operator enclosed. The canonical bosonic commutation relation is recovered in the limit $\Lambda \to \infty$. In the following, when working with a single mode, the mode index $j$ will be dropped for brevity.
\\

Denote the binary ($q=\text{b}$) or unary ($q=\text{u}$) representation of $n$ by $q^{\Lambda}(n)$. Both of these representations can be encoded in the computational basis of a register of $n_q(\Lambda)$ qubits given by $\ket{q^{\Lambda}(n)}$, where 
\begin{align}
    n_q(\Lambda) =
    \begin{cases}
        \Lambda + 1 & \text{if}~~~ q=\text{u}, \\
        \lceil \mathrm{log}(\Lambda+1)\rceil & \text{if}~~~ q = \text{b}.
    \end{cases}
\end{align}
Label the qubits in each register by $i \in \{0,1,\ldots,n_q(\Lambda)-1\}$. Further denote the ladder operators on the $i$th qubit as
\begin{align}
    \hat L_i^{\pm} \coloneq \frac{1}{2}(\hat X_i \mp i\hat Y_i),
\end{align}
with Pauli operators $\hat X_i$ and $\hat Y_i$ acting on the $i$th qubit. Using these ladder operators, the mapping from Fock states to the qubits' computational basis states can be defined as
\begin{align}
    \ket n \mapsto \ket{\text{u}^\Lambda(n)} &= 
    \hat L^+_n \ket{0}^{\otimes\, n_\text{u}(\Lambda)}
    \label{eq:unary-state}
\end{align}
for the unary representation, and
\begin{align}    
    \ket n \mapsto \ket{\text{b}^\Lambda(n)} &= \sum_{i=0}^{n_\text{b}(\Lambda)-1}\left(\hat L^+_i\right)^{b_i} \ket{0}^{\otimes\, n_\text{b}(\Lambda)}
    \label{eq:binary-state}
\end{align}
for the binary representation. For the latter, we we have assumed the 0th qubit to represent the least significant digit and introduced the binary notation $n=\sum_{n=0}^{n_\text{b}(\Lambda)-1} \ b_i2^i$. \\

An $N$-mode Fock state $\ket{\bs{n}}$ can be encoded into a collection of $N$ registers of the form in Eqs.~\eqref{eq:unary-state} or \eqref{eq:binary-state} with $n_q(\Lambda)$ qubits each. We define
\begin{align}
    \ket{\bm{n}} \mapsto \ket{q^{\Lambda}(\bs{n})}
    &\coloneq \ket{q^{\Lambda}(n_0)}\otimes\ldots\otimes\ket{q^{\Lambda}(n_{N-1})}.
\end{align}

In the following, we describe the preparation of the $(R,Q)$ ansatz $\ket{\psi}_{R,Q}=\hat U_G\ket{C}_{R,Q}$, where $\hat U_G = \otimes_{j=0}^{N-1} \hat S_j(r)$. We assume that the qubit registers are initialized in the state $\ket{0}^{\otimes\, n_q(\Lambda)}$.

\subsection{Core-state preparation
\label{sec:core-state-prep}}
 
Since the core state has a finite support over the Fock space, it can be represented exactly in a qubit register provided that the cutoff is set to $\Lambda\geq R$. The core state is mapped to the qubit state
\begin{align}
    \ket{C}_{R,Q} =\bar{\sum}_{\bs{n}} c_{\bs{n}}\ket{\bs{n}} \mapsto \ket{C}_{R,Q}^{q,\Lambda}=\bar{\sum}_{\bs{n}} c_{\bs{n}}\ket{q^{\Lambda}(\bs{n})}, 
\end{align}
where the notation $\bar{\sum}$ indicates that the summation is restricted such that it obtains the symmetric rank-$R$ and $Q$-truncated core state described in the previous section.\footnote{This preparation method, nonetheless, works for arbitrary core states.} Since the state is only non-trivial over the first $n_q(R)$ qubits in each register, we focus our attention to this part of the qubit register. Recall from Eq.~\eqref{eq:coresize} that the core state is mapped to a superposition of $N|c_{R,Q}|$
multi-qubit computational basis states, the number of which is much smaller than the total Hilbert-space dimension of $2^{Nn_q(R)}$ (provided that $R,Q \ll N$).  Thus, one can invoke sparse state-preparation methods such as the ones proposed in Refs.~\cite{tubman2018postponing,gleinig2021efficient, de2022double,zhang2022quantum,fomichev2024initial,feniou2024sparse}.\\

In particular, the method presented in Ref.~\cite{gleinig2021efficient} can prepare the core state using $O\left(N^2n_q(R)|c_{R,Q}|\right)$ two-qubit entangling (CNOT) gates and $O\left(Nn_q(R) + N|c_{R,Q}|\mathrm{log}(N|c_{R,Q}|)\right)$ single-qubit rotation gates. The key idea is to determine the circuit decomposition of a unitary transformation $\hat{\mathcal{U}}_{R,Q}^{q,\Lambda}$ which maps $\ket{C}^{q,\Lambda}_{R,Q}$ to the multi-qubit computational basis state $\ket{q^{\Lambda}(\bs{0})}$, i.e., $\ket{C}^{q,\Lambda}_{R,Q}=\hat{\mathcal{U}}_{R,Q}^{q,\Lambda\,\dagger}\ket{q^{\Lambda}(\bs{0})}$. The transformation works by successively rotating two elements in the superposition $\ket{C}^{q,\Lambda}_{R,Q}$ into one. This operation is performed with suitable controls so that the merging of two computational basis states is not accompanied by introducing additional basis states in the superposition. Thus, by performing these merges $N|c_{R,Q}|-1$ times, $\ket{C}^{q,\Lambda}_{R,Q}$ is mapped to a single computational basis state, which can be converted to $\ket{q^{\Lambda}(\bs{0})}$ using NOT gates. The decomposition of $\hat{\mathcal{U}}_{R,Q}^{q,\Lambda}$ into elementary quantum gates is obtained using a classical algorithm with an $O\left(N^3n_q(R)|c_{R,Q}|^2\mathrm{log}(N|c_{R,Q}|)\right)$ runtime. More details about this algorithm, together with an example, can be found in Appendix~\ref{app:sparse_prep}.\\

Once the ansatz, i.e., the values of $R$ and $Q$ as well as the system size $N$, are fixed, the circuit template 
that prepares the core state is known. The values of the couplings $(m^2,\lambda)$ only affect the magnitudes of the rotation angles in the single-qubit rotation gates within this fixed template. 

\subsection{Gaussian unitary transformation}\label{sec:qsqueeze}

Once the core state $\ket{C}_{R,Q}$ is prepared, the next step is to the apply Fock-space-truncated squeezing operators [recall Eq.~\eqref{eq:psi-QR}]. This step introduces two kinds of error:
truncation error from restriction to a finite Fock space, and error in the implementation of the resulting truncated operator.\footnote{Experimental errors arising from environmental noise, imperfections in gate implementations, etc. will not be considered here.}

The core state $\ket{C}_{R,Q}$ has a bounded support over Fock space, i.e., the maximum number of bosons in any mode is the rank $R$. Thus, as already mentioned, it can be represented exactly in the truncated Fock space with a cutoff of $R$ bosons per mode, which is in turn isomorphic to the multi-qubit Hilbert space (or to a subspace of it in case of the unary map). In contrast, the Gaussian unitary transformation, $\hat U_G = \otimes_{j=0}^{N-1} \hat S_j(r)$, extends the support of the $(R,Q)$ ansatz $\ket{\psi}_{R,Q}=\hat U_G \ket{C}_{R,Q}$ to the entire Hilbert space; thus, 
it can only be represented approximately. Following Refs.~\cite{encinar2021digital,somma2002simulating}, we map the single-mode bosonic squeezing operator $\hat S(r)$ to an approximate squeezing operator on a truncated Hilbert space consisting of at most $\Lambda$ bosons per mode:
\begin{align}\label{eq:qsqueeze_map}
    \hat S(r) = e^{\frac{r}{2}\left(\hat a^{\dagger 2} - \hat a^2\right)}
    \mapsto & \hat S^{\Lambda}(r) \coloneq e^{\frac{r}{2}\left[(\hat a^{\Lambda \dagger})^2 - (\hat a^{\Lambda})^2\right]} 
    \nonumber \\
    \mapsto & \hat S^{\Lambda}(r) \coloneq e^{-ir\hat s^{\Lambda}},
\end{align}
where 
\begin{align}\label{eq:s_def}
    \hat s^{\Lambda} & \coloneq \sum_{n=0}^{\Lambda-2} \hat T_n,
\end{align}
with
\begin{align}
\label{eq:T-def}
\hat T_n  \coloneq \frac{i \ell_n}{2}&\left(\ket{n+2}\bra{n}- \ket{n}\bra{n+2}\right) 
\end{align}
and $\ell_n \coloneq \sqrt{(n+1)(n+2)}$. \\

The application of this approximate squeezing operator to the core state results in the approximate $(R,Q)$ state $\gket{\psi^{\Lambda}}_{R,Q}\coloneq \otimes_{j=1}^{N-1}\hat S_j^{\Lambda}(r)\gket{C}_{R,Q}$, which has support only on the truncated Hilbert space. As shown in Appendix~\ref{app:truncation}, a suitable value of the cutoff, $\Lambda$, can be determined by upper-bounding the distance between $\gket{\psi}_{R,Q}$ and $\gket{\psi^{\Lambda}}_{R,Q}$: 
\begin{align}
    &\norm{\left(\otimes_{j=0}^{N-1}\hat S_j(r)-\hat \otimes_{j=0}^{N-1}S_j^{\Lambda}(r)\right)\ket{C}_{R,Q}} 
    \nonumber\\
    & \hspace{4 cm}\leq \epsilon^{(N)}_{\rm trunc}(r,R,\Lambda,Q),
\end{align}
where
\begin{align}
    &\epsilon^{(N)}_{\rm trunc}(r,R,\Lambda,Q)\coloneq N\sqrt{N|c_{R,Q}|}\Big[G(r,R,\Lambda)+
    \nonumber\\
    & \hspace{4.5 cm} g^{\rm leak}(r,R,\Lambda)\Big].
    \label{eq:trunc-error-expresssion}
\end{align}
Here, the term proportional to $g^{\rm leak}(r,R,\Lambda)$ bounds the error resulting from discarding the weight of $\gket{\psi}_{R,Q}$ outside the truncated Hilbert space, with
\begin{align}
    g^{\rm leak}(r,R,\Lambda)\coloneq\mathrm{argmax}_{n_1\leq R}  \ \left[ \sum_{n_2=\Lambda+1}^{\infty}\left|\langle n_2 |\hat S(r)|n_1\rangle\right|^2\right]^{1/2}.
\end{align}
For the small values of the squeezing parameter $r$ in this work, the squeezing operator $\hat S(r)$ distributes most of the Fock-basis amplitude among the lower Fock states, and this limits the value of the above leakage parameter; see Fig.~\ref{fig:squeeze_spread} in Appendix~\ref{app:truncation}. The term proportional to $G(r,R,\Lambda)$ bounds the discrepancy between the matrix elements of the exact and approximate squeezing operators within the truncated Hilbert space, and the expression for $G(r,R,\Lambda)$ can be found in Appendix~\ref{app:truncation}. Once a truncation-error tolerance is set, Eq.~\eqref{eq:trunc-error-expresssion} determines the smallest value of $\Lambda$ that achieves such an error tolerance.

It is important to note that the above analysis focuses only on the accuracy of the initial-state representation. However, as this state evolves in time, its support on higher Fock states is expected to grow, and thus, eventually larger cutoff values are needed. As a result, additional qubits need to be appended to the state register to account for this expansion of the state's domain within the Hilbert space. One could use the strategy of Refs.~\cite{tong2022provably,peng2025quantum} to achieve an analytical bound on the error accumulated throughout the evolution due to constraining the evolved state to a truncated Hilbert space. \\

The implementation error depends on the capabilities of the device and the computation mode. The most common mode of quantum computation, which we used for core-state preparation as well, is the digital mode. This mode relies on applying a discrete set of universal single and two-qubit gates, hence requiring Trotterization of, or other digitization schemes for, the squeezing operator.
The squeezing operator in Eq.~\eqref{eq:qsqueeze_map} can be expressed as
\begin{align}
\label{eq:S-multi-full}
    &\hat S^{\Lambda}(r) = e^{-ir(\hat s_0^{\Lambda}+\hat s_2^{\Lambda})}e^{-ir(\hat s_1^{\Lambda}+\hat s_3^{\Lambda})},
\end{align}
where
\begin{align}
\label{eq:s-q-lambda-m-def}
    &s_m^{\Lambda} \coloneq \sum_{n=0}^{\lfloor \frac{\Lambda-2-m}{4}\rfloor} \hat T_{4n+m},
\end{align}
with $m \in \{0,1,2,3\}$.\footnote{It is assumed that $s_m^{\Lambda}=0$ when $\lfloor \frac{\Lambda-2-m}{4}\rfloor<0$.} With this splitting of the squeezing operator, we define the first-order Trotter approximation to $\hat S^{\Lambda}$ as
\begin{align}
\label{eq:T-trotter-decom}
    \hat S^{\Lambda,K}(r) & \coloneq \left(e^{-i\frac{r}{K}\hat s_0^{\Lambda}} e^{-i\frac{r}{K}\hat s_2^{\Lambda}}\right)^K \left(e^{-i\frac{r}{K}\hat s_1^{\Lambda}} e^{-i\frac{r}{K}\hat s_3^{\Lambda}}\right)^K.
\end{align}
As shown in Appendix~\ref{app:trotter}, this results in the following bound on the distance between the state $\gket{\psi^{\Lambda}}_{R,Q}$ defined previously, and the state, $\gket{\psi^{\Lambda,K}}_{R,Q}\coloneq \otimes_{j=0}^{N-1}\hat S_j^{\Lambda,K}(r)\gket{C}_{R,Q}$, 
\begin{align}
    &\norm{\left(\otimes_{j=0}^{N-1}\hat S^{\Lambda}_j(r)-\hat \otimes_{j=0}^{N-1}S_j^{\Lambda,K}(r)\right)\ket{C}_{R,Q}}\nonumber \\
    &\hspace{4.5 cm} \leq \epsilon^{(N)}_{\rm trott}(r,\Lambda,K),
\end{align}
where
\begin{align}
    \epsilon^{(N)}_{\rm trott}(r,\Lambda,K)\coloneq \frac{Nr^2}{2K}\left(\norm{[\hat s_0^{\Lambda},\hat s_{2}^{\Lambda}]}+\norm{[\hat s_1^{\Lambda},\hat s_{3}^{\Lambda}]}\right).
\end{align}
Using triangle inequality, a looser but analytical bound $\tilde\epsilon^{(N)}_{\rm trott}(r,\Lambda,K)>\epsilon^{(N)}_{\rm trott}(r,\Lambda,K)$ can also be obtained:
\begin{align}
    \tilde\epsilon^{(N)}_{\rm trott}(r,\Lambda,K)\coloneq \frac{Nr^2\beta(\Lambda)}{2K},
\end{align}
where
\begin{align}
    \beta(\Lambda)\coloneq \frac{1}{2}\sum_{n=0}^{\lfloor \frac{\Lambda-2-m}{4}\rfloor} \ell_{4n+m}(\ell_{4n+m-2}+\ell_{4n+m+2}),
\end{align}
resulting in an $O\left(\frac{Nr^2\Lambda^3}{K}\right)$ error. Unlike the truncation error bound, this bound is obtained using a \emph{state-independent} first-order product-formula error bound~\cite{childs2021theory}, which is quite loose. 

It is useful to enumerate the minimum value of the cutoff, $\Lambda$, and the number of Trotter layers, $K$, which guarantee a minimum fidelity $F_0$ between $\gket{\psi}_{R,Q}$ and $\gket{\psi^{\Lambda,K}}_{R,Q}$ based on these bounds (we use the tighter commutator-norm bound for the Trotter error). Assuming an equal value for the maximum permissible error resulting from truncation and Trotterization, we list these minimum $\Lambda$ and $K$ values in Table~\ref{tab:cutoff_layers} for the minimum-energy $(R,Q)=(4,2)$ ansatz studied previously. 

\begin{table}[t!]
\centering
\begin{tabular}{c|c|c|c|c|c}
\hline\hline
\multirow{2}{*}{$(m^2, \lambda)$} & \multirow{2}{*}{$r$} 
& \multicolumn{2}{c|}{$F_0 = 0.9$} 
& \multicolumn{2}{c}{$F_0 = 0.95$} \\
\cline{3-6}
& & $\Lambda$ & $K$ & $\Lambda$ & $K$ \\
\hline 
(0.6,1.5) & 0.348 & 33 & 2375 & 35 & 3876 \\
(0.4,1) & 0.300  & 27 & 1102 & 27 & 1569 \\
(0.2,0.5) & 0.172 & 15 & 87 & 15 & 123\\
(0.1,0.25) & 0.155 & 14 & 59 & 15 & 100 \\
\hline\hline
\end{tabular}
\caption{Minimum cutoff $\Lambda$ and number of Trotter layers $K$ which guarantee a minimum fidelity, $F_0$, for the Trotterized and truncated multimode product squeezing operator with $N=10$ and various mass and coupling values $(m^2,\lambda)$. $r$ is the squeezing parameter obtained upon minimizing the energy of the $(R,Q)=(4,2)$ ansatz.}
\label{tab:cutoff_layers}
\end{table}

In the above example, the number of Trotter layers needed to guarantee high fidelity in the implementation of the squeezing operator is larger for larger values of $(m^2,\lambda)$. This is because of the larger value of the associated squeezing parameter, which in turn leads to a higher value for the Fock-space cutoff. The value of the optimized squeezing parameter decreases as $(m^2,\lambda) \rightarrow (0,0)$, resulting in the much smaller number of required Trotter layers in this regime. One should, however, note that the accuracy with which a fixed $(R,Q)$ ansatz represents the ground state declines as $(m^2,\lambda)\rightarrow (0,0)$ (where the gap becomes smaller while the correlations grow stronger). This feature can be captured, for instance, by a growing value of $\delta_E$. To maintain a constant value of $\delta_E$ as one takes $(m^2,\lambda)\rightarrow (0,0)$, the value of both $R$ and $Q$ must be increased. Consequently, the complexity of implementing the core state would increase. The complexity of implementing the squeezing operator, on the other hand, would depend upon the values of $R$, $Q$, and the squeezing parameter $r$, and could be determined empirically. \\

One can now faithfully map each of the single-mode Trotterized squeezing operators on the truncated Hilbert space to an operator acting on a register of $n_q(\Lambda)$ qubits:
\begin{align}
    \hat S^{q,\Lambda,K}(r)\coloneq \left(e^{-i\frac{r}{K}\hat s_0^{q,\Lambda}} e^{-i\frac{r}{K}\hat s_2^{q,\Lambda}}\right)^K \left(e^{-i\frac{r}{K}\hat s_1^{q,\Lambda}} e^{-i\frac{r}{K}\hat s_3^{q,\Lambda}}\right)^K,
\end{align}
Here,
\begin{align}
    &s_m^{q,\Lambda} \coloneq \sum_{n=0}^{\lfloor \frac{\Lambda-2-m}{4}\rfloor} \hat T^q_{4n+m}, 
\end{align}
with
\begin{align}
    &\hat T^q_n  \coloneq \frac{i \ell_n}{2}\left(\ket{q(n+2)}\bra{q(n)}- \ket{q(n)}\bra{q(n+2)}\right). 
\end{align}
In the unary representation, to implement each Trotter layer $e^{-i\frac{r}{K}\hat s_m^{u,\Lambda}}$ with $ m=0,1,2,3$, one first notes that
\begin{align}
    e^{-i\frac{r}{K}\hat s_m^{u,\Lambda}}&=\prod_{n=0}^{\lfloor \frac{\Lambda-2-m}{4} \rfloor} e^{-i\frac{r}{K}\hat T_{4n+m}} \nonumber \\
    &=\prod_{n=0}^{\lfloor \frac{\Lambda-2-m}{4} \rfloor} e^{\frac{
    r\ell_{4n+m}}{2K}\left(\ket{u^{\Lambda}_\text{u}(4n+m+2)}\bra{u^{\Lambda}_\text{u}(4n+m)}-\mathrm{h.c.}\right)} \nonumber \\
    &= \prod_{n=0}^{\lfloor \frac{\Lambda-2-m}{4} \rfloor} \left(e^{i\frac{ir\ell_{4n+m}}{4K}\hat X_{4n+m+2}\hat Y_{4n+m}}\times\right. \nonumber \\
    & \hspace{1.8 cm} \left. e^{-i\frac{r\ell_{4n+m}}{4K}\hat Y_{4n+m+2}\hat X_{4n+m}}\right).
\end{align}
Here, we have used Eq.~\eqref{eq:unary-state} for the unary representation of the qubitized bosonic states. Thus, for the single-mode operator, each Trotter layer can be implemented with $O(\Lambda)$ CNOT and single-qubit rotation gates. For the multimode tensor-product operator, this gate count becomes $O(N\Lambda)$. Switching to the binary map, and recalling the representation of the qubitized bosonic states in Eq.~\eqref{eq:binary-state}, one readily observes a gate complexity $O(\Lambda \,  \log\Lambda)$ for each Trotter layer of the single-mode squeezing operator, following steps similar to those outlined for the unary map. This implementation, however, introduces additional Trotter error and, thus 
is less preferred.
\\

Alternatively, we rely upon the singular-value-decomposition (SVD) approach of Ref.~\cite{davoudi2023general} which works both for the binary and unary representations, as detailed in Appendix~\ref{app:svd}. Importantly, this approach avoids any additional Trotter error beyond that from Eq.~\eqref{eq:T-trotter-decom}. SVD enables the implementation of the single-mode squeezing operator in the binary representation using $O(\Lambda \, \mathrm{log}\Lambda)$ CNOT and single-qubit rotation gates. However, the SVD representation for the single-mode operator in the unary representation relies upon $O(\Lambda \, 2^{\Lambda})$ CNOT and single-qubit rotation gates. Thus, the direct implementation discussed in the previous paragraph is favored for the unary representation.\\

With a hybrid analog-digital platform, such as that in Ref.~\cite{andersen2025thermalization}, the squeezing operator $\hat S^{q,\Lambda}(r)$ can be implemented without incurring a Trotter error. The operator $\hat s^{q,\Lambda}$ in the exponent of Eq.~\eqref{eq:qsqueeze_map} can be put into the form of a nearest-neighbor XY Hamiltonian by
rearranging the qubits in the unary representation, and upon performing a local basis change. Such a Hamiltonian can be natively generated on this platform, upon careful calibration and basis choice~\cite{andersen2025thermalization}.
Details of such a scheme are provided in Appendix~\ref{app:fsim}. The accuracy of this implementation is controlled by spurious next-to-nearest-neighbor and farther terms in the effective spin Hamiltonian implemented on the device, characterization of which is beyond the scope of our work. One may still need a fully digital scheme to implement the core state on such a platform.\\

Even more powerful is the use of continuous-variable quantum platforms, e.g., those taking advantage of bosonic degrees of freedom. While the discrete-variable approach is more standard and is commonly practiced, the continuous-variable approach is more natural for bosonic theories and may ultimately present practical benefits. As shown in Appendix~\ref{app:CV}, our single-mode 
finite-rank ansatz for the ground state of an interacting scalar field theory can be implemented using standard operations in bosonic platforms, namely single-mode Gaussian (squeezing $\hat S(\xi)=e^{\frac{1}{2}\left[\xi(\hat a^{\dagger})^2-\xi^*\hat a^2\right]}$ and displacement $\hat D(\alpha)=e^{\alpha\hat a^{\dagger}-\alpha^*\hat a}$) operations, and single-boson addition ($\hat a_j^\dagger$). However, more work needs to be done to arrive at a bosonic preparation recipe for the multimode $(R,Q)$ ansatz; see Appendix~\ref{app:CV} for more details.

\section{Discussion and Outlook}\label{sec:sum}
\noindent
In this work, we demonstrated a classical Euclidean-Monte-Carlo-informed method for determining quantum circuits that prepare states close to the ground state of the (1+1)D $\phi^4$ theory. The crucial components for translating Euclidean ground-state information to Minkowski wavefunctions are the finite stellar-rank ansatz, which we specialize to the $(R,Q)$ ansatz, and a Euclidean-Monte-Carlo-informed moment-optimization procedure. The $(R,Q)$ ansatz is an $N$-mode bosonic ansatz consisting of a tensor-product of single-mode squeezing operations acting on a superposition of Fock states, which is referred to as the $(R,Q)$ core state. The rank, $R$, specifies the maximum total number of bosons while $Q$ specifies the largest span of occupied modes featuring in this superposition. Roughly, the rank $R$ is associated with the level of non-Gaussianity in the ansatz, while the maximum span $Q$ is associated with its level of locality. Given a fixed $(R,Q)$ ansatz, we showed how the moment-optimization procedure can systematically improve the ansatz's estimates of the ground-state's two-point correlation functions as well as its non-Gaussianity. This procedure works by augmenting the familiar variational energy minimization by penalizing deviations in selected target sets of ground-state moments (whose values are sourced from Euclidean path-integral Monte-Carlo computations). 

The extent to which various target-moment sets can be optimized depends upon the structure and expressiveness of the ansatz, together with the properties of the theory itself (such as its spectral gap). Based on this, we conclude that the $(R,Q)$ ansatz possesses the following properties for the tested range of couplings and masses:
\begin{itemize}
    \item[--] \textit{Classical tractability:} 
    We showed that expectation values with respect to the $(R,Q)$ ansatz can be computed in $O(|c_{R,Q}|^2Q^2)$ classical-computing time,    with $|c_{R,Q}|$ defined in Eq.~\eqref{eq:coresize}. Away from the continuum limit, this cost is independent of the system size $N$. However, close to the continuum limit, it is expected that greater values of $R$ and $Q$ would be needed to reproduce the diverging correlation length, and in fact, $R$ and $Q$ may no longer be small compared to $N$, increasing the classical-computing cost. 
    \item[--] \textit{Circuit translatability:} 
    For a discrete-variable quantum simulator, the core state can be translated into a quantum circuit with classical time complexity $O\left(N^3 R|c_{R,Q}|^2\mathrm{log}(N|c_{R,Q}|)\right)$ using a sparse quantum-state preparation algorithm (and assuming a unary mapping of the bosons). The circuit for the squeezing operator, on the other hand, can be determined using $O(1)$ classical computing time (for the direct Trotterization recipe with the unary map). 
    \item[--] \textit{Circuit efficiency:} 
    With a near-term implementation in mind, we characterized circuits based on their CNOT and single-qubit gate counts. The core-state circuit consists of $O(N^2R|c_{R,Q}|)$ CNOT gates while the squeezing circuit consists of $O(N\Lambda)$ CNOT gates per Trotter layer (assuming a unary encoding of bosons). Here, $\Lambda$ is the maximum number of bosons in each mode in the truncated Hilbert space. While the core state is implemented exactly, the qubitized squeezing operator is subject to two sources of error, namely, the truncation of the bosonic Hilbert space and Trotterization. We show how suitable values for the cutoff and the number of Trotter layers can be chosen to meet specified fidelity goals. 
\end{itemize}
We ascribe the above features (i.e., classical tractability, circuit translatability, and circuit efficiency) to an ansatz suitable for the purpose of classical determination of quantum circuits that prepare ground states. \\

A comprehensive field-theory study necessitates a detailed examination of the thermodynamic and continuum limits. As these limits are approached, the spectral gap closes and correlation lengths diverge, imposing significant demands on the ansatz, and implying system-dependent scalings for both $R$ and $Q$. Critical slowing down also poses challenges for sourcing ground-state correlation functions from Monte-Carlo simulations. However, since practical computations often operate at points slightly removed from the continuum limit, extrapolations using finite-lattice-spacing results remain a feasible strategy. Furthermore, as the ansatz becomes less classically tractable toward the continuum limit, one could potentially switch to quantum-computing ansatz moments, as in more conventional variational-quantum-eigensolver settings. In particular, it would be interesting to use ADAPT-VQE methods to systematically grow the $(R,Q)$ ansatz~\cite{farrell2024scalable, grimsley2019adaptive}. \\

Optimizing the choice of weights in the moment-optimization loss function is another area needing further study. In our current work, these weights were selected through experimentation. However, higher-order moments in the target set have greater statistical error. Thus, developing a principled approach for selecting moment-dependent weights could enhance the effectiveness of the optimization process. Moreover, since ground-state correlation functions are sourced from Monte-Carlo simulations, the moment-optimized parameters inherently carry statistical errors. Understanding the implications of these errors for quantum simulations is crucial. Specifically, one may need to perform multiple runs of the quantum simulation to propagate these errors to subsequent real-time quantities.

Our introduction of the moment-optimization procedure was aimed at obtaining ansatzes that are more dynamics-aware. A deeper and more comprehensive understanding of the relationship between dynamics and target-moment sets, utilizing both analytical and numerical tools, remains an open question. Studying dynamics using these moment-optimized ansatzes as initial states would be particularly exciting. Interestingly, the construction of the $(R,Q)$ ansatz is reminiscent of matrix product states. Making this relationship more concrete could facilitate the development of more efficient ansatz-optimization methods, or even a density matrix renormalization group (DMRG) analog. Additionally, finite-rank ansatzes hold relevance for the far-term fault-tolerant era. 
The reason is that 
the efficiency of near-optimal ground-state preparation strategies~\cite{lin2020near,dong2022ground} is often related to the overlap of an initial trial state with the ground state. Our method, therefore, could be used to prepare such a suitable trial state, and subsequent quantum computation could evolve this state even closer to the ground state.

Another approach to leveraging Euclidean PIMC information has been proposed in Refs.~\cite{harmalkar2020quantum,gustafson2021toward}: One stochastically samples the density matrix classically, and then passes the basis states with proper weights to the quantum computer for subsequent computation of real-time expectation values. This method, therefore, avoids preparing the state on the quantum computer but is inherently stochastic. Our approach, on the other hand, constrains an ansatz wavefunction with the aid of static correlation functions obtained from Euclidean PIMC. It would be worthwhile to explore other alternative routes to leveraging Euclidean PIMC results in the quantum computation of quantum field theories.

There exist quantum-state preparation strategies that rely upon the knowledge of the exact ground-state wavefunction. For example, Refs.~\cite{klco2019digitization,klco2020minimally,klco2020systematically,klco2020fixed} introduce a quantum-circuit ansatz for preparing the scalar-field-theory ground state. The locality of the digitized free lattice-scalar-field theory enables a systematic truncation of the circuit elements. The circuit elements for preparing large-scale ground states can then be determined using classically computed ground states in small spatial volumes. It would be interesting to apply the results of Refs.~\cite{klco2019digitization,klco2020minimally,klco2020systematically,klco2020fixed} to the interacting wavefunctions computed in our work. 
Notably, the concept of fixed-point circuits introduced in Ref.~\cite{klco2020fixed} could be used to trade off system-size dependence in our translated quantum circuits with correlation-length dependence. More generally, it would be interesting to see how the structure imposed by the $(R,Q)$ ansatz on digital quantum circuits compares with the structures obtained in these works. 

Since the $(R,Q)$ ansatz is primarily designed to enable efficient computation of expectation values using classical computers, its translatability to bosonic quantum simulators is compromised. Investigating alternative modifications to the finite-rank ansatz that retain classical tractability, and can also be efficiently prepared on bosonic quantum simulators (hence bypassing truncation errors), is an intriguing direction for future work. Alternatively, one could develop variational-quantum-eigensolver methods based upon the $(R,Q)$ ansatz.\footnote{Recently, a VQE computation using a multimode finite-stellar-rank ansatz for the Hubbard model was demonstrated in Ref.~\cite{stornati2023variational}.}\\

In this work, we adopted a single measure, namely the $\phi^{2n}$-moment ratio, to assess deviations from Gaussianity. This quantity can be easily computed using only the
$\langle\hat\phi^{2n}\rangle$ and $\langle\hat \phi^{2}\rangle$ values, and cleanly isolates shape effects from the overall scale. One can also study alternative probes, such as connected cumulants, their variance-normalized versions (e.g., Binder cumulants), or variants thereof as introduced in Ref.~\cite{blahnik2024natural}. 
Furthermore, it may be interesting to study non-Gaussianity in other operator bases, and for non-local correlation functions as well. The choice of the probe may ultimately be informed by the type of phenomenon to be studied. For example, non-Gaussianity in scalar fields is physically relevant in the study of Cosmic Microwave Background (CMB) anisotropies arising from slow-roll inflation driven by scalar fields~\cite{linde1982new, albrecht1982cosmology}. Future work could explore how the $(R,Q)$ ansatz fares on measures of non-Gaussianity in these cosmological settings.

Extending our methods to higher spatial dimensions, gauge theories, and fermionic degrees of freedom represents another set of clear directions for future exploration. For example, it will be interesting to investigate if the finite-rank bosonic ansatz of this work has any relevance to bosonic states in gauge theories, and what extensions of this ansatz best describe fermionic and coupled fermionic-bosonic ground states. \\

\section*{Acknowledgements}
\noindent
The authors would like to thank Jacob Bringewatt, Mohammad Hafezi, Chung-Chun Hsieh, Matthew Johnson, Saurabh Kadam, En-Jui Kuo, Yannick Meurice, Christine Muschik, David Roberts, Sarah Shandera, Brayden Ware, and Jing-Chen Zhang for discussions about various aspects of this work. 
N.G., C.D.W., and Z.D. were supported by
the National Science Foundation (NSF) Quantum Leap
Challenge Institutes (QLCI) (award no. OMA-2120757).
Z.D. and N.G. further acknowledge funding by the Department of Energy (DOE), Office of Science, Early Career Award (award no. DESC0020271), as well as by the Department of Physics; Maryland Center for Fundamental Physics; and College of Computer, Mathematical, and Natural Sciences at the University of Maryland, College Park. Z.D. further acknowledges support from the U.S. DOE, Office of Science, Office of Advanced Scientific Computing Research (ASCR), program in Accelerated Research in Quantum Computing, Fundamental Algorithmic Research toward Quantum Utility (FAR-Qu). Z.D. and N.G. are also grateful for the hospitality of Perimeter Institute where part of this work was carried out. Research at Perimeter Institute is supported in part by the Government of Canada through the Department of Innovation, Science, and Economic Development and by the Province of Ontario through the Ministry of Colleges and Universities. Z.D. was also supported in part by the Simons Foundation through the Simons Foundation Emmy Noether Fellows Program at Perimeter Institute.
Z.D. is further grateful for the hospitality of Nora Brambilla, and of the Excellence Cluster ORIGINS at the Technical University of Munich, where part of this work was carried out.
The research at ORIGINS is supported by the Deutsche Forschungsgemeinschaft (DFG, German Research Foundation) under Germany's Excellence Strategy (EXC-2094–-390783311).
C.D.W. further acknowledges support from the U.S. DOE, Office of Science, Office of Advanced Scientific Computing Research (ASCR) Quantum Computing Application Teams program, under fieldwork proposal number ERKJ347, as well as DOE Quantum Systems Accelerator program (award no. DE-AC02-05CH11231), AFOSR MURI FA9550-22-1-0339, ARO grant W911NF-23-1-0242, and ARO grant W911NF-23-1-0258; the work was completed while C.D.W. held an NRC Research Associateship award at the United States Naval Research Laboratory.

\appendix
\onecolumngrid

\section{Energy minimization versus fidelity maximization}\label{app:en_v_fid}
\noindent
In Sec.~\ref{sec:singlemode}, we argued that the minimum-energy and maximum fidelity ansatzes may not coincide. To make this point more transparent, we provide in this appendix a simple example for which both the energy and fidelity are analytically computable.

Consider the following squeezed harmonic-oscillator  Hamiltonian. (Section~\ref{sec:singlemode} introduces some of the notation used in this appendix and Ref.~\cite{gong1990expansion}
discusses the squeezed harmonic-oscillator states.) 
\begin{align}
\label{eq:squeezed-H}
    \hat{H}_\text{S}&\coloneq \hat S(r) \, \hat a^{\dagger}\hat a \, \hat S(r)^{\dagger} \nonumber \\
    &= \mathrm{cosh}(2r) \, \hat a^{\dagger}\hat a- \frac{1}{2}\mathrm{sinh}(2r)((\hat a^{\dagger})^2+\hat a^2)+\mathrm{sinh}^2(r),
\end{align}
where in the second line, we have used the relation $\hat S(r)\hat a \hat S(r)^{\dagger}=\mathrm{cosh}(r)\hat a-\mathrm{sinh}(r)\hat a^{\dagger}$. This relation can be derived using the Baker-Campbell-Hausdorff formula, and the definition of the squeezing operator $\hat S(r)\coloneq e^{\frac{r}{2}\left(\hat a^{\dagger 2}-\hat a^2\right)}$, with $ r \in \mathbb{R}$. The ground state of this Hamiltonian is given by
\begin{align}
    \ket{\Omega}=\hat S(r)\ket{0}=\frac{1}{\sqrt{\mathrm{cosh}(r)}}\sum_{n=0}^{\infty}\mathrm{tanh}(r)^n \frac{\sqrt{(2n)!}}{2^n n!}\ket{2n}.
\end{align}
The above expansion can be obtained by noting that $\hat S(r)\hat a \hat S(r)^{\dagger} \hat S(r)\ket{0}=\left(\mathrm{cosh}(r)\hat a-\mathrm{sinh}(r)\hat a^{\dagger}\right)\hat S(r)\ket{0}=0$. By inserting the expansion $\hat S(r)\ket{0}=\sum_{n=0}^{\infty} c_n\ket{n}$, the resulting recursive relation between the coefficients $c_n$ gives the desired result. The ground-state energy is $\bra{\Omega}\hat H_\text{S}\ket{\Omega}=0$, and the spectral gap is $\Delta E=1$. 

\begin{figure}
    \centering
    \includegraphics[width=\linewidth]{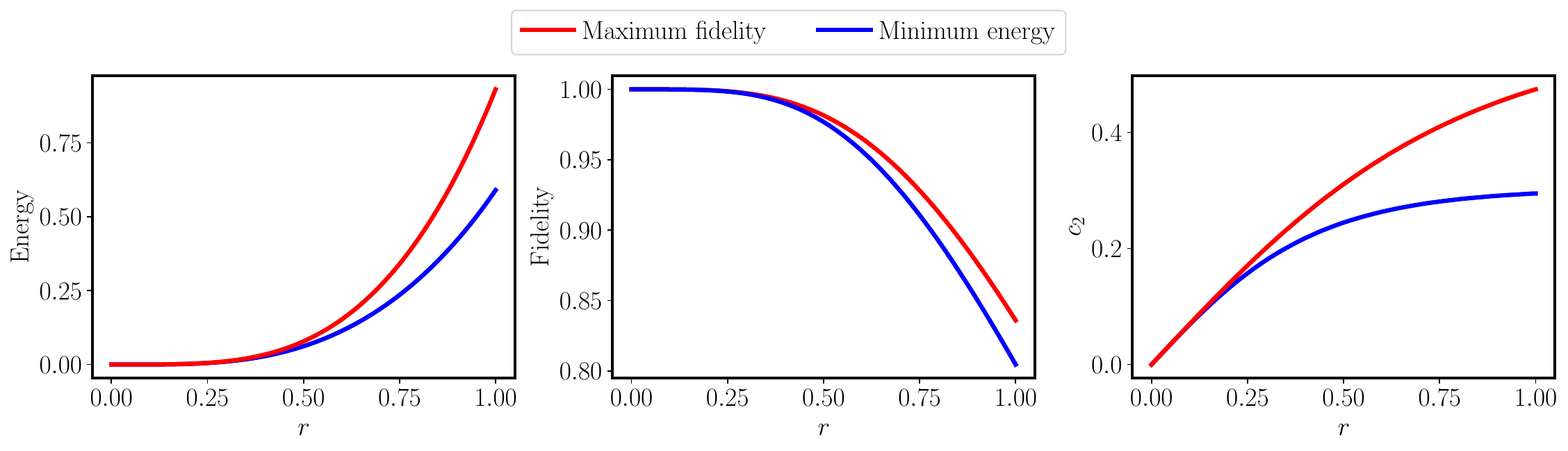}
    \caption{Behavior of the minimum-energy and maximum-fidelity rank-2 ansatz for the ground state of the squeezed Hamiltonian in Eq.~\eqref{eq:squeezed-H}. The exact value of ground-state energy is zero while the spectral gap is 1.}
    \label{fig:fid_v_en}
\end{figure}

Let us now take the rank-2 symmetric core state 
\begin{align}
    \ket{C}\coloneq c_0\ket{0}+c_2\ket{2}; \ c_0, c_2 \in \mathbb{R}; \ c_0^2+c_2^2=1,
\end{align}
\noindent to be the ansatz for the ground state. This leads to the energy and fidelity loss functions
\begin{align}
    &\bra{C}\hat H \ket{C}=2c_2^2 \ \mathrm{cosh}(2r)-\sqrt{2}c_0c_2 \ \mathrm{sinh}(2r) + \mathrm{sinh}^2(r), \nonumber \\
    &\big|\langle C|\Omega\rangle\big|^2=\frac{1}{\mathrm{cosh}(r)}\left[c_0+\frac{c_2}{\sqrt{2}}\mathrm{tanh}(r)\right]^2.
\end{align}
\noindent Figure~\ref{fig:fid_v_en} depicts the result of minimizing (maximizing) the energy (fidelity) loss functions. When the value of $r$ is small, $\ket{\Omega}$ has substantial weight on the states $\ket{0}$ and $\ket{2}$ (as opposed to higher occupation states). Thus, in this regime, the ansatz is powerful/expressive enough to yield small values of energy together with high values of fidelity. However, as $r$ becomes larger, the minimum-energy and maximum-fidelity ansatzes diverge. It is in this regime that the optimal values of the coefficients are not well constrained, and which optimization to choose may need to be informed by the subsequent use of the ansatz. In practice, computing fidelity quickly becomes intractable due to the large Hilbert-space sizes associated with systems of practical interest. In such cases, moment matching allows one to access such alternative choices for optimal parameters based on the knowledge of moments. 

\section{Euclidean Path-integral Monte Carlo}\label{app:pimc}
\noindent
This appendix details the Euclidean path-integral Monte-Carlo computations used in Sec.~\ref{sec:multimode}. Consider the dimensionless lattice Hamiltonian describing the lattice $\hat \phi^4$
theory in (1+1)D:
\begin{align}
    \hat{H} &= \sum_{j=0}^{N-1} \left[\frac{\pih_j^2}{2}+\frac{(\phih_{j+1}-\phih_j)^2}{2}+\frac{1}{2}m^2\phih_j^2+\frac{\lambda}{4}\phih_j^4\right].
\end{align}
We introduce the compact notation $\vec{\phi}\equiv (\phi_0,\phi_1,\ldots,\phi_{N-1})$ and $ \vec{\pi}\equiv (\pi_0,\pi_1,\ldots,\pi_{N-1})$ for the field operators on a spatial slice, and split the Hamiltonian into kinetic and potential terms $\hat H=K(\hat{\vec{\pi}})+V(\hat{\vec{\phi}})$. The partition function $Z_{T} \coloneq \mathrm{Tr}(e^{-T \hat H})$ can be approximated with the second-order Trotter decomposition $Z_{T,\theta}$ with (dimensionless) inverse temperature $T$,
\begin{align}
    Z_{T,\theta} &\coloneq \int \mathcal{D}\vec\phi_0 \bra{\vec{\phi}_0} \left(e^{-\theta V/2}e^{-\theta K}e^{-\theta V/2}\right)^M\ket{\Vec{\phi}_0} \nonumber \\
    &= \int \mathcal{D}\vec\phi_0\mathcal{D}\vec\phi_1\cdots \mathcal{D}\vec\phi_{M-1} \ e^{-\frac{\theta}{2} V(\vec{\phi}_0)}\bra{\vec{\phi}_0}e^{-\theta K}\ket{\vec{\phi}_1}e^{-\theta V(\vec{\phi}_1)}\ldots \bra{\vec{\phi}_{M-1}}e^{-\theta K}\ket{\vec{\phi}_0}e^{-\frac{\theta}{2} V(\vec{\phi}_0)} \nonumber \\ 
    &=  (2\pi\theta)^{-NM/2}\int \mathcal{D}\vec\phi_0\mathcal{D}\vec\phi_1\cdots \mathcal{D}\vec\phi_{M-1} \ e^{-S_{T,\theta}[\vec{\phi}_0,\ldots,\vec{\phi}_{M-1}]}.
\end{align}
Here, $\theta \coloneq T/M$ is the imaginary-time lattice spacing expressed in units of the spatial lattice spacing $a$, $\vec{\phi}_{t} \coloneq (\phi_{0,t},\phi_{1,t},\ldots,\phi_{N-1,t})$ denotes a spatial field vector at the imaginary time $\tau\coloneq t\theta$ with $ t\in\{0,1,\ldots,M-1\}$, and $\mathcal{D}\vec\phi_t \coloneq \Pi_{j=0}^{N-1}(d\phi_{j,t})$. Finally, the action $S_{T,\theta}$ is given by
\begin{align}
  S_{T,\theta}[\vec{\phi}_0,\ldots,\vec{\phi}_{M-1}] \equiv S_{T,\theta}(\{\phi_{j,t}\}) &= \sum_{t=0}^{M-1}\sum_{j=0}^{N-1} \frac{\theta}{2}\left[m^2\phi_{j,t}^2+ \frac{\lambda}{2}\phi_{j,t}^4 +  \frac{(\phi_{j,t+1}-\phi_{j,t})^2}{\theta^2} + (\phi_{j+1,t}-\phi_{j,t})^2\right]. 
\end{align}
To compute expectation values, one obtains sample configurations $\{\phi_{j,t}^{(\alpha)}\}_{T,\theta}$ with $\alpha \in \{1,2,\ldots,N_{\rm samples}\}$, with respect to the weight $e^{-S_{T,\theta}(\{\phi_{j,t}\})}$. The sample autocorrelation is estimated using the $(j,t)=(0,0)$ component of the samples as follows:
\begin{align}
    \mathrm{AC}(\Delta \alpha)\coloneq\frac{N_{\rm samples}}{N_{\rm samples}-\Delta \alpha}\frac{\sum_{\alpha=1}^{N_{\rm samples}-\Delta\alpha}\left(\phi_{0,0}^{(\alpha)}\phi^{(\alpha+\Delta \alpha)}_{0,0}\right)}{\sum_{\alpha=1}^{N_{\rm samples}}\left(\phi_{0,0}^{(\alpha)}\right)^2}
\end{align}
Using this, the integrated autocorrelation time is estimated as follows:
\begin{align}
    \mathrm{IAC}\coloneq \frac{1}{2}+\sum_{\Delta\alpha=0}^{100} \mathrm{AC}(\Delta \alpha).
\end{align}
The sum above is not taken all the way to $N_{\rm samples}$ because the autocorrelation estimates tend to be dominated by noise at large values of 
$\Delta \alpha$; we verify that the autocorrelation decays sufficiently close to zero for $\Delta\alpha < 100$.
The statistical independence of the samples $\{\phi_{j,t}^{(\alpha)}\}_{T,\theta}$ is ensured by arranging their integrated autocorrelation time to be close to 1. This original ensemble of configurations is then resampled to obtain $N_{\rm bootstrap}$ sets of samples $\{\phi_{j,t}^{I_b(\alpha)}\}_{T,\theta}$ with $b \in \{1,2,\ldots,N_{\rm bootstrap}\}$. The values of $I_b(\alpha)$ are chosen uniformly at random from the set $\{1,\ldots,N_{\rm samples}\}$. 
For each computation, the value of $N_{\rm bootstrap}$ is chosen to ensure the stabilization of the bootstrap 
mean and standard error within a reasonable threshold. We also verify the normality of the bootstrap means by checking their skewness and kurtosis.\\

The field operators on a spatial slice are denoted as $\hat{\vec{\phi}}\coloneq(\phih_0,\ldots,\phih_{N-1})$. Thus, the operator at Euclidean time $\tau$ is obtained as $\hat{\vec{\phi}}(\tau=t\theta) \equiv \hat{\vec{\phi}}_t = e^{\tau \hat H}\hat{\vec{\phi}}e^{-\tau\hat H}$. The ground-state expectation value of an observable $\mathcal{O}\big(\hat{\vec{\phi}}_{t_1},\ldots,\hat{\vec{\phi}}_{t_k}\big)$ can be expressed as
\begin{align}
    \bra{\Omega}\mathcal{O}\big(\hat{\vec{\phi}}_{t_1},\ldots,\hat{\vec{\phi}}_{t_k}\big)\ket{\Omega}&= \lim_{T\rightarrow \infty}\lim_{\substack{\theta \rightarrow 0 \\ M\theta=T}}\frac{\int\mathcal{D}\vec\phi_0\mathcal{D}\vec\phi_1\cdots \mathcal{D}\vec\phi_{M-1} \ \mathcal{O}\big(\vec{\phi}_{t_1},\ldots,\vec{\phi}_{t_k}\big) \ e^{-S_{T,\theta}[\vec{\phi}_0,\ldots,\vec{\phi}_{M-1}]}}{\int\mathcal{D}\vec\phi_0\mathcal{D}\vec\phi_1\cdots \mathcal{D}\vec\phi_{M-1} \ e^{-S_{T,\theta}[\vec{\phi}_0,\ldots,\vec{\phi}_{M-1}]}} \nonumber \\
    &\approx \lim_{T\rightarrow \infty}\lim_{\substack{\theta \rightarrow 0 \\ M\theta=T}} \ \frac{1}{N_{\rm bootstrap}}\sum_{b=1}^{N_{\rm bootstrap}}\left[\frac{1}{N_{\rm samples}}\sum_{\alpha=1}^{N_{\rm samples}}\mathcal{O}\big(\vec{\phi}_{t_1}^{I_b(\alpha)},\ldots,\vec{\phi}_{t_k}^{I_b(\alpha)}\big)\right],
\end{align}
where the statistical error in the Monte-Carlo estimate of the above integral can be measured by the standard deviation of the bootstrap means contained inside the square brackets. To take the $\theta \to 0$ limit, we perform the Monte-Carlo computation at various values of temporal lattice spacing $\theta$ and for a fixed value of inverse temperature $T$. For the temporal discretization $\theta$ to be meaningful, one needs to ensure $\theta<\Delta E^{-1}$, where $\Delta E$ is the energy gap measured in units of spatial lattice spacing, and then suitably take the $\theta\rightarrow 0$ limit at fixed $T$. Furthermore, to extract ground-state properties, sampling needs to take place in the limit $T \rightarrow \infty$. Instead of taking this limit explicitly, it suffices to work with a single value of $T \gg \Delta E^{-1}$. Since the energy gap is not known a priori, one can pick an arbitrary value for this fixed $T$ and for the temporal lattice spacings $\theta$, and check if the conditions $\theta\Delta E<1$ and $T\Delta E \gg 1$ are satisfied at the end of the computation. In this study, we use $T=10$ and $\theta \in \{0.4,0.2,0.1\}$. We further set $N=10$, and consider $(m^2,\lambda)\in\{(0.6,1.5),(0.4,1.0),(0.2,0.5),(0.1,0.25)\}$. \\

\begin{figure}
    \includegraphics[width=5in]{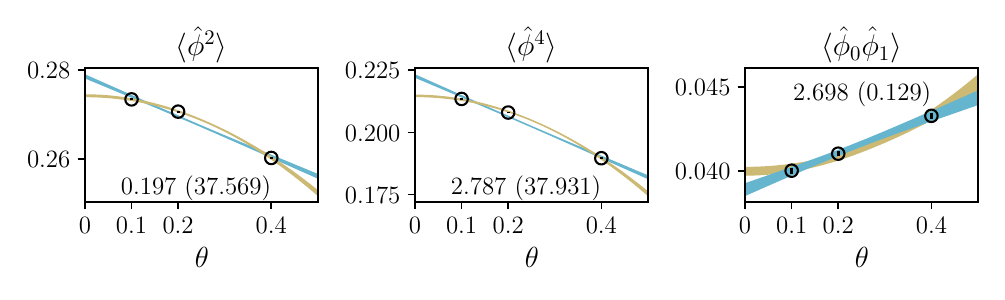}
    \caption{Extrapolations of the one- and two-point functions of $\phi$ that feature in the Hamiltonian to zero temporal lattice spacing $\theta$ for $(m^2,\lambda)=(0.6,1.5)$. The gold and blue bands are $68\%$ confidence bands for quadratic $(a+b\theta^2)$ and linear $(a+b\theta)$ fitting functions, respectively. The value of the reduced $\chi$-squared for the quadratic (linear) fit is listed without (with) the parenthesis. We take the final continuum value to be the one obtained with the fit with a lower $\chi$-squared value.}
    \label{fig:pimc_ham_pt0}
\end{figure}
\begin{figure}
    \includegraphics[width=5in]{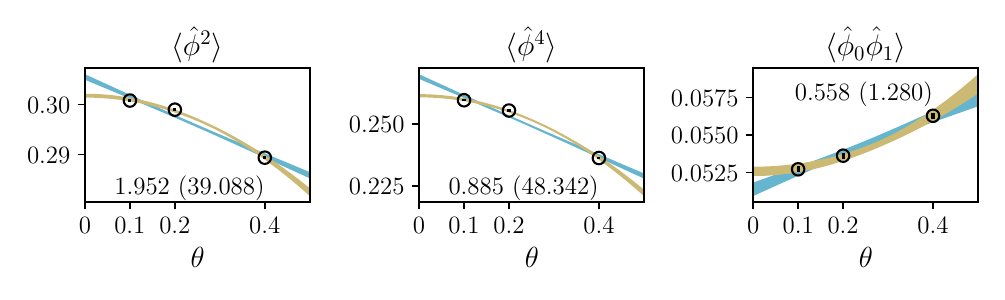}
    \caption{Extrapolations of the one- and two-point functions of $\phi$ that feature in the Hamiltonian to zero temporal lattice spacing $\theta$ for $(m^2,\lambda)=(0.4,1.0)$. The gold and blue bands are $68\%$ confidence bands for quadratic $(a+b\theta^2)$ and linear $(a+b\theta)$ fitting functions, respectively. The value of the reduced $\chi$-squared for the quadratic (linear) fit is listed without (with) the parenthesis. We take the final continuum value to be the one obtained with the fit with a lower $\chi$-squared value.}
    \label{fig:pimc_ham_pt1}
\end{figure}

\begin{figure}
    \includegraphics[width=5in]{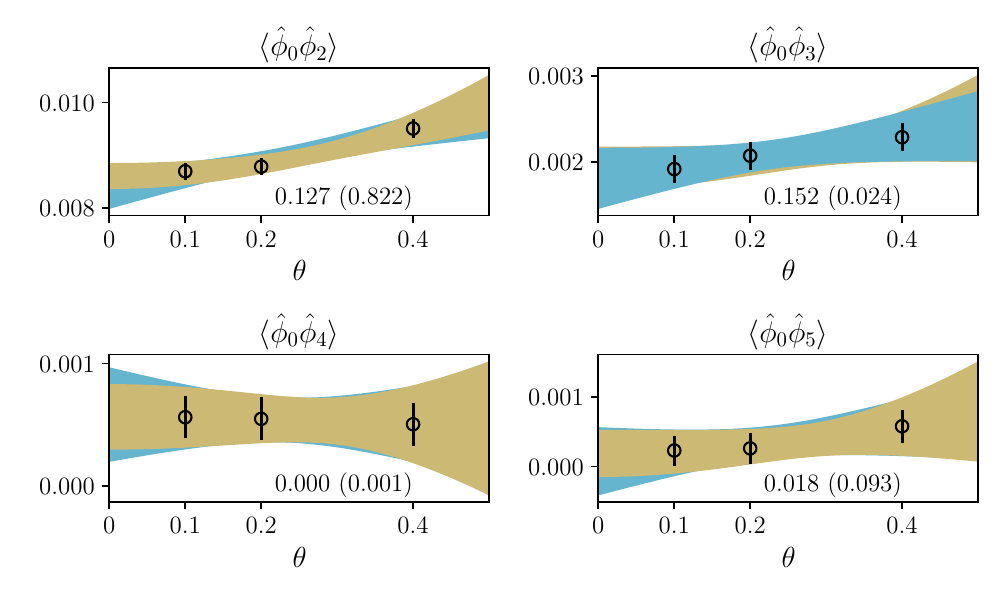}
    \caption{Extrapolations of the two-point correlation functions to zero temporal lattice spacing $\theta$ for $(m^2,\lambda)=(0.6,1.5)$. The gold and blue bands are $68\%$ confidence bands for quadratic $(a+b\theta^2)$ and linear $(a+b\theta)$ fitting functions, respectively. The value of the reduced $\chi$-squared for the quadratic (linear) fit is listed without (with) the parenthesis. We take the final continuum value to be the one obtained with the fit with a lower $\chi$-squared value.}
    \label{fig:pimc_twopt_pt0}
\end{figure}

\begin{figure}
    \includegraphics[width=5in]{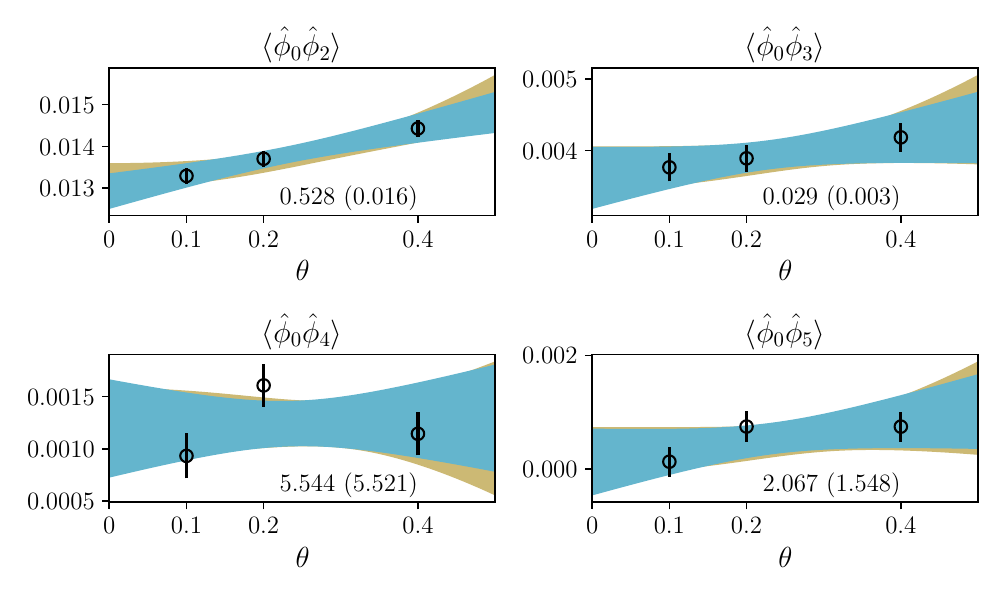}
    \caption{Extrapolations of the two-point correlation functions to zero temporal lattice spacing $\theta$ for $(m^2,\lambda)=(0.4,1.0)$. The gold and blue bands are $68\%$ confidence bands for quadratic $(a+b\theta^2)$ and linear $(a+b\theta)$ fitting functions, respectively. The value of the reduced $\chi$-squared for the quadratic (linear) fit is listed without (with) the parenthesis. We take the final continuum value to be the one obtained with the fit with a lower $\chi$-squared value.}
    \label{fig:pimc_twopt_pt1}
\end{figure}

\begin{figure}
    \includegraphics[width=5in]{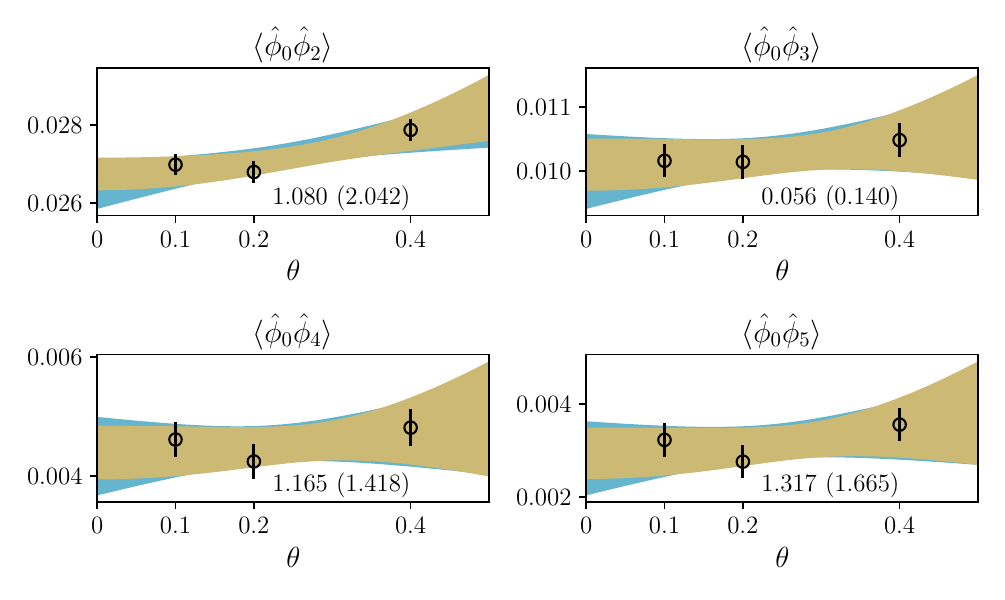}
    \caption{Extrapolations of the two-point correlation functions to zero temporal lattice spacing $\theta$ for $(m^2,\lambda)=(0.2,0.5)$. The gold and blue bands are $68\%$ confidence bands for quadratic $(a+b\theta^2)$ and linear $(a+b\theta)$ fitting functions, respectively. The value of the reduced $\chi$-squared for the quadratic (linear) fit is listed without (with) the parenthesis. We take the final continuum value to be the one obtained with the fit with a lower $\chi$-squared value.}
    \label{fig:pimc_twopt_pt2}
\end{figure}

\begin{figure}
    \includegraphics[width=5in]{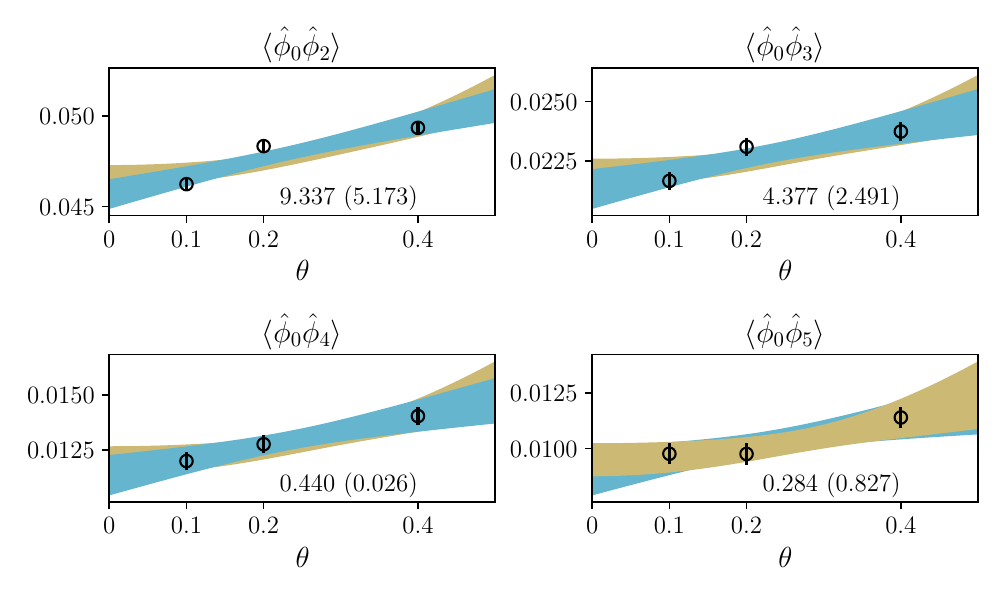}
    \caption{Extrapolations of the two-point correlation functions to zero temporal lattice spacing $\theta$ for $(m^2,\lambda)=(0.1,0.25)$. The gold and blue bands are $68\%$ confidence bands for quadratic $(a+b\theta^2)$ and linear $(a+b\theta)$ fitting functions, respectively. The value of the reduced $\chi$-squared for the quadratic (linear) fit is listed without (with) the parenthesis. We take the final continuum value to be the one obtained with the fit with a lower $\chi$-squared value.}
    \label{fig:pimc_twopt_pt3}
\end{figure}

\begin{figure}
    \includegraphics[width=5in]{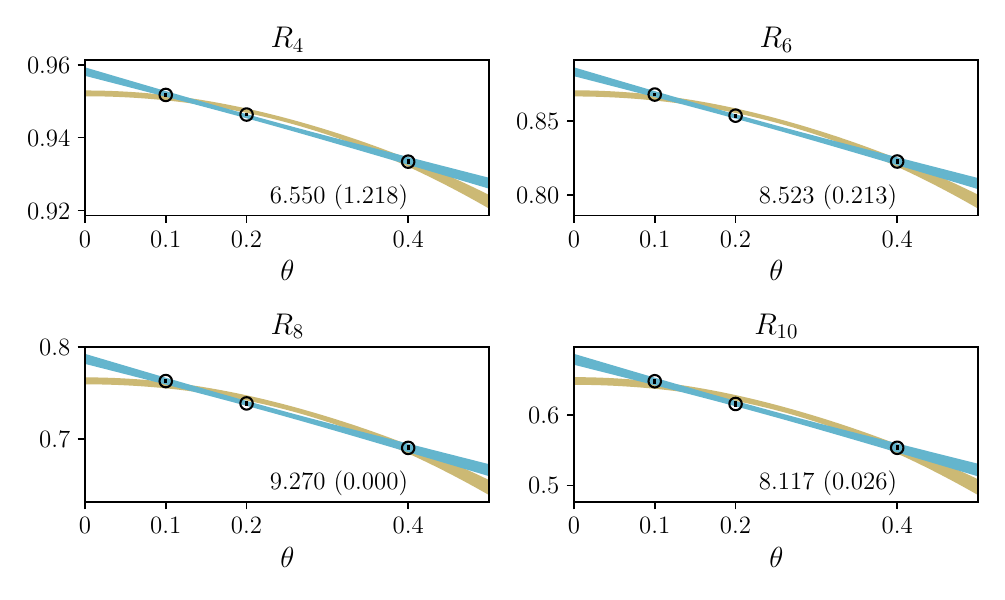}
    \caption{Extrapolations of the moment ratios to zero temporal lattice spacing $\theta$ for $(m^2,\lambda)=(0.6,1.5)$. The gold and blue bands are $68\%$ confidence bands for quadratic $(a+b\theta^2)$ and linear $(a+b\theta)$ fitting functions, respectively. The value of the reduced $\chi$-squared for the quadratic (linear) fit is listed without (with) the parenthesis. We take the final continuum value to be the one obtained with the fit with a lower $\chi$-squared value.}
    \label{fig:pimc_ratio_pt0}
\end{figure}

We will now describe the results for the computation of various ground-state moments, followed by non-linear functions of these moments, i.e., the moment ratio and energy gap. Figures~\ref{fig:pimc_ham_pt0} to \ref{fig:pimc_ham_pt1} show the $\theta\rightarrow 0$ extrapolations for the one- and two-point $\phi$-moments that feature in the Hamiltonian, i.e, $\langle \phih^2 \rangle$, $\langle \phih^4 \rangle$, and $\langle \phih_0\phih_1 \rangle$. Using these extrapolated values, the momentum variance $\langle \hat{\pi}_j^2 \rangle $ is estimated using the virial theorem, which together with translation invariance implies that
\begin{align}
    \langle \pih_j^2\rangle
    = \Bigg\langle \phih_j \frac{dV(\hat{\vec{\phi}})}{d\phih_j}\Bigg\rangle = (2+m^2)\langle \phih_j^2\rangle    +\lambda\langle \phih_j^4\rangle.
\end{align}
All of these values are then combined together to get the ($\theta \to 0$) extrapolated value of energy. Furthermore, Figs.~\ref{fig:pimc_twopt_pt0} to \ref{fig:pimc_twopt_pt3} show the extrapolation plots for the two-point correlation functions $\langle \phih_0\phih_j \rangle$ with $j=2,3,4,5$, while Figs.~\ref{fig:pimc_ratio_pt0} to \ref{fig:pimc_ratio_pt3} show the extrapolation for the moment ratios $R_{n}$ with $n={4,6,8,10}$.
\begin{figure}
    \includegraphics[width=5in]{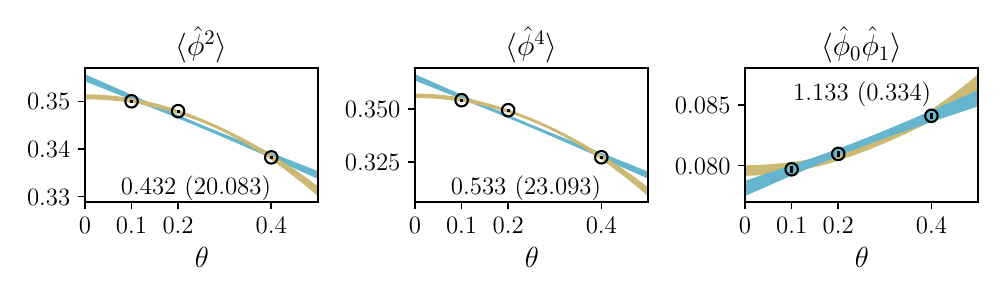}
    \caption{Extrapolations of the one- and two-point functions of $\phi$ that feature in the Hamiltonian to zero temporal lattice spacing $\theta$ for $(m^2,\lambda)=(0.2,0.5)$. The gold and blue bands are $68\%$ confidence bands for quadratic $(a+b\theta^2)$ and linear $(a+b\theta)$ fitting functions, respectively. The value of the reduced $\chi$-squared for the quadratic (linear) fit is listed without (with) the parenthesis. We take the final continuum value to be the one obtained with the fit with a lower $\chi$-squared value.}
    \label{fig:pimc_ham_pt2}
\end{figure}
\begin{figure}[t!]
    \includegraphics[width=5in]{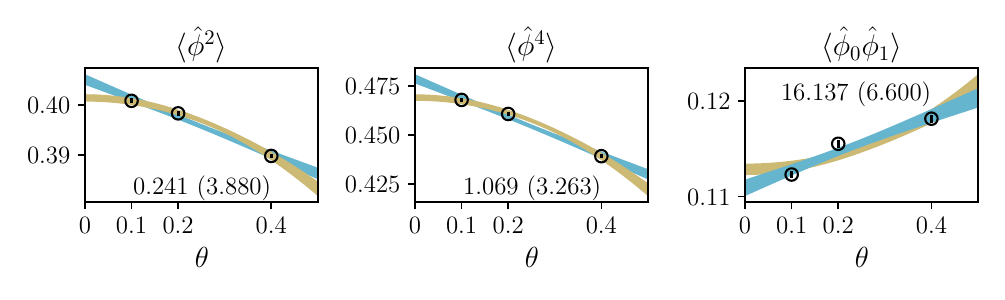}
    \caption{Extrapolations of the one- and two-point functions of $\phi$ that feature in the Hamiltonian to zero temporal lattice spacing $\theta$ for $(m^2,\lambda)=(0.1,0.25)$. The gold and blue bands are $68\%$ confidence bands for quadratic $(a+b\theta^2)$ and linear $(a+b\theta)$ fitting functions, respectively. The value of the reduced $\chi$-squared for the quadratic (linear) fit is listed without (with) the parenthesis. We take the final continuum value to be the one obtained with the fit with a lower $\chi$-squared value.}
    \label{fig:pimc_ham_pt3}
\end{figure}
\begin{figure}
    \includegraphics[width=5in]{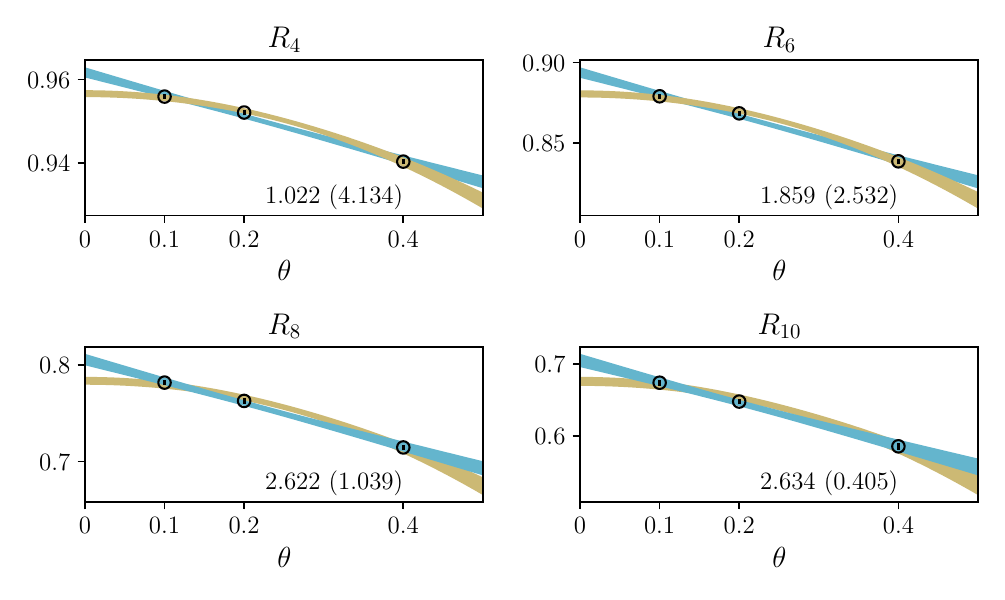}
    \caption{Extrapolations of the moment ratios to zero temporal lattice spacing $\theta$ for $(m^2,\lambda)=(0.4,1.0)$. The gold and blue bands are $68\%$ confidence bands for quadratic $(a+b\theta^2)$ and linear $(a+b\theta)$ fitting functions, respectively. The value of the reduced $\chi$-squared for the quadratic (linear) fit is listed without (with) the parenthesis. We take the final continuum value to be the one obtained with the fit with a lower $\chi$-squared value.}
    \label{fig:pimc_ratio_pt2}
\end{figure}
\begin{figure}
    \includegraphics[width=5in]{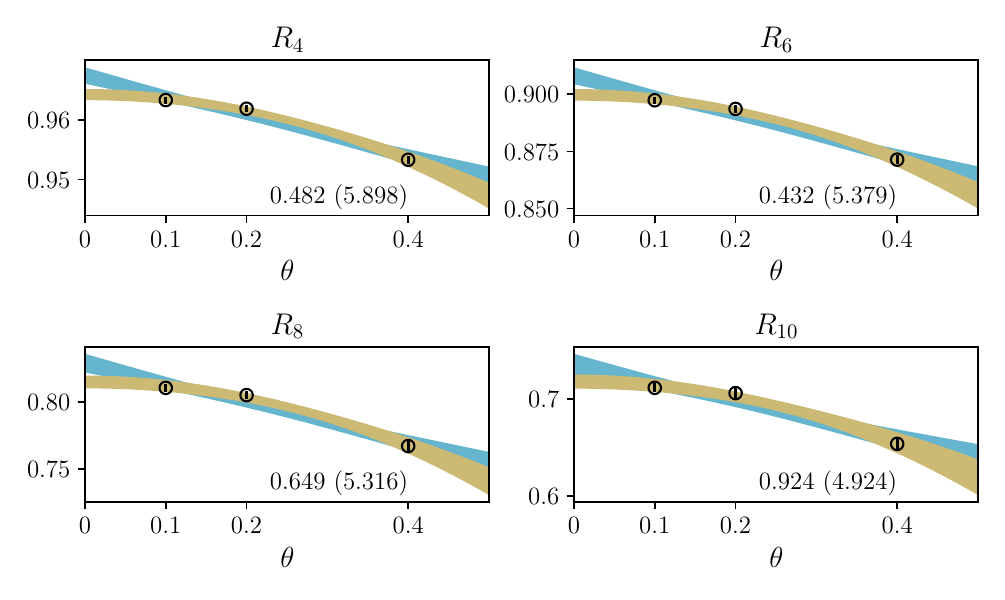}
    \caption{Extrapolations of the moment ratios to zero temporal lattice spacing $\theta$ for $(m^2,\lambda)=(0.2,0.5)$. The gold and blue bands are $68\%$ confidence bands for quadratic $(a+b\theta^2)$ and linear $(a+b\theta)$ fitting functions, respectively. The value of the reduced $\chi$-squared for the quadratic (linear) fit is listed without (with) the parenthesis. We take the final continuum value to be the one obtained with the fit with a lower $\chi$-squared value.}
    \label{fig:pimc_ratio_pt3}
\end{figure}
\begin{figure}
    \includegraphics[width=5in]{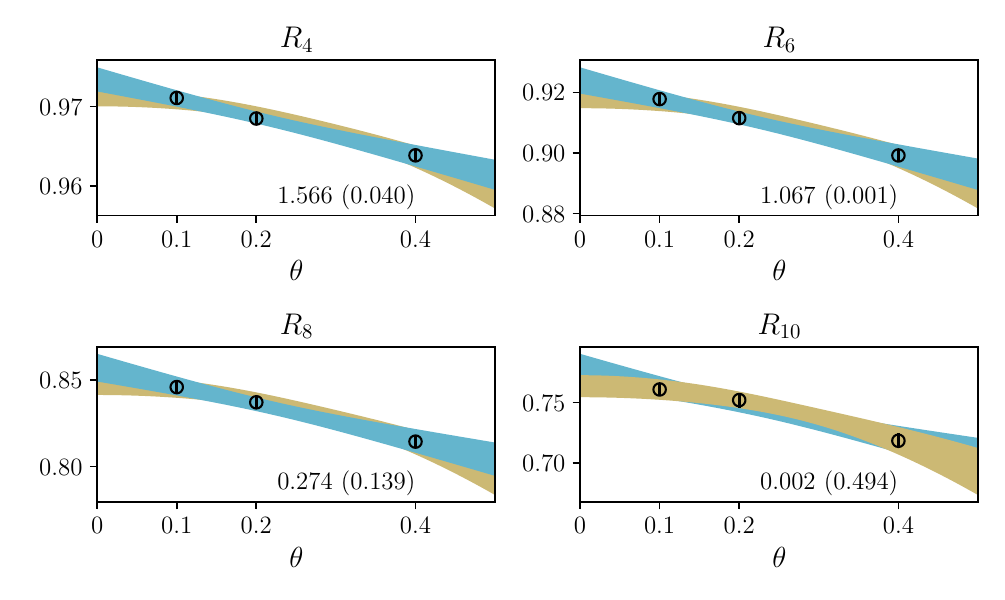}
    \caption{Extrapolations of the moment ratios to zero temporal lattice spacing $\theta$ for $(m^2,\lambda)=(0.1,0.25)$. The gold and blue bands are $68\%$ confidence bands for quadratic $(a+b\theta^2)$ and linear $(a+b\theta)$ fitting functions, respectively. The value of the reduced $\chi$-squared for the quadratic (linear) fit is listed without (with) the parenthesis. We take the final continuum value to be the one obtained with the fit with a lower $\chi$-squared value.}
    \label{fig:pimc_ratio_pt3}
\end{figure}

\begin{figure}
    \includegraphics[width=5in]{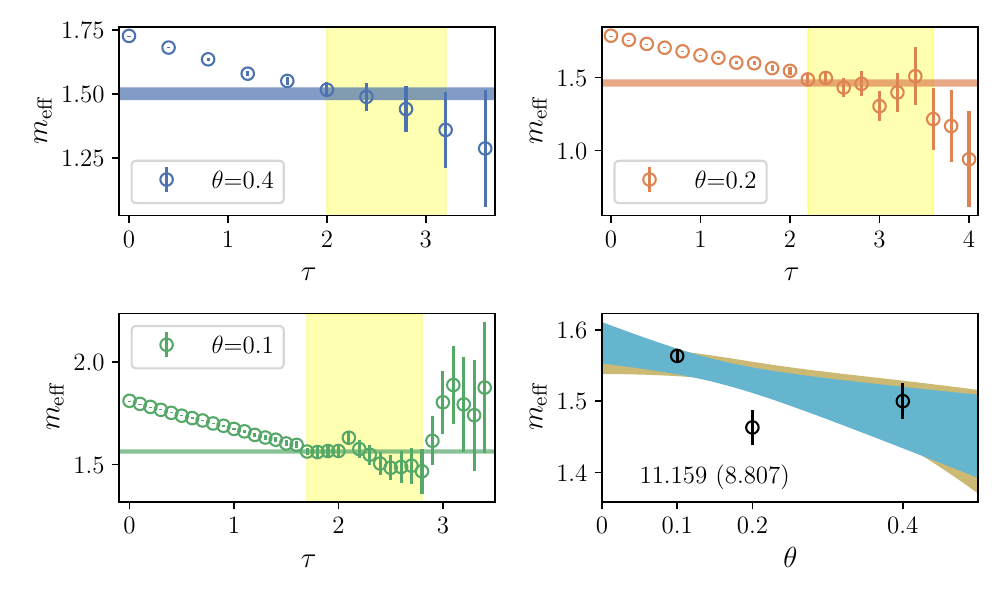}
    \caption{Effective-mass plots corresponding to $(m^2,\lambda)=(0.6,1.5)$. The bottom right plot shows the $m_{\rm eff}$ value from each plot averaged over the yellow window as a function of $\theta$. The gold and blue bands are $68\%$ confidence bands for quadratic $(a+b\theta^2)$ and linear $(a+b\theta)$ fitting functions, respectively. The value of the reduced $\chi$-squared for the quadratic (linear) fit is listed without (with) the parenthesis. We take the final continuum value to be the one obtained with the fit with a lower $\chi$-squared.}
    \label{fig:pimc_temp_pt0}
\end{figure}

\begin{figure}
    \includegraphics[width=5in]{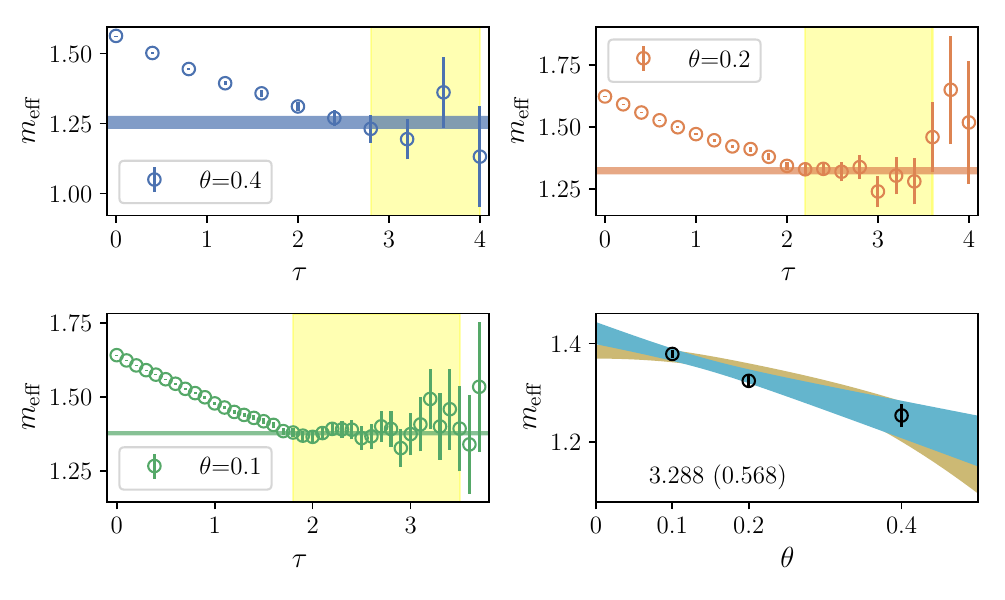}
    \caption{Effective-mass plots corresponding to $(m^2,\lambda)=(0.4,1.0)$. The bottom right plot shows the $m_{\rm eff}$ value from each plot averaged over the yellow window as a function of $\theta$. The gold and blue bands are $68\%$ confidence bands for quadratic $(a+b\theta^2)$ and linear $(a+b\theta)$ fitting functions, respectively. The value of the reduced $\chi$-squared for the quadratic (linear) fit is listed without (with) the parenthesis. We take the final continuum value to be the one obtained with the fit which has a lower $\chi$-squared.}
    \label{fig:pimc_temp_pt1}
\end{figure}

\begin{figure}
    \includegraphics[width=5in]{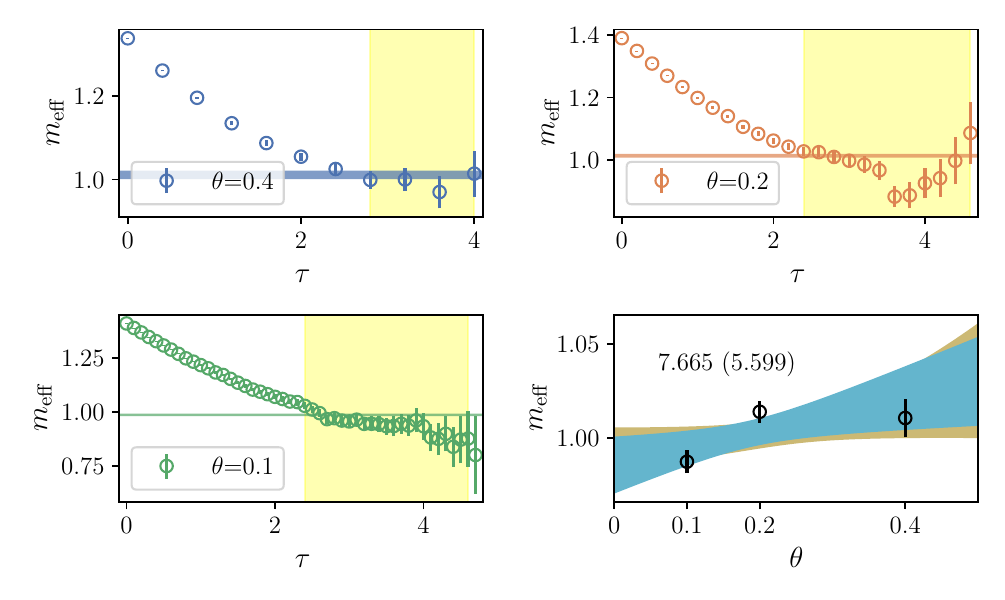}
    \caption{Effective-mass plots corresponding to $(m^2,\lambda)=(0.2,0.5)$. The bottom right plot shows the $m_{\rm eff}$ value from each plot averaged over the yellow window as a function of $\theta$. The gold and blue bands are $68\%$ confidence bands for quadratic $(a+b\theta^2)$ and linear $(a+b\theta)$ fitting functions, respectively. The value of the reduced $\chi$-squared for the quadratic (linear) fit is listed without (with) the parenthesis. We take the final continuum value to be the one obtained with the fit which has a lower $\chi$-squared.}
    \label{fig:pimc_temp_pt2}
\end{figure}

\begin{figure}
    \includegraphics[width=5in]{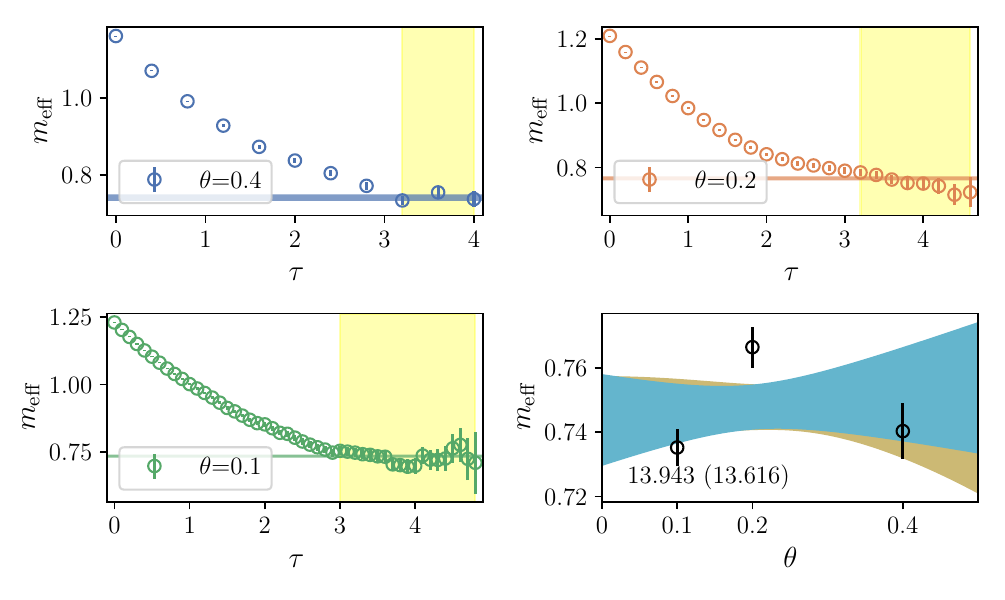}
    \caption{Effective-mass plots corresponding to $(m^2,\lambda)=(0.1,0.25)$. The bottom right plot shows the $m_{\rm eff}$ value from each plot averaged over the yellow window as a function of $\theta$. The gold and blue bands are $68\%$ confidence bands for quadratic $(a+b\theta^2)$ and linear $(a+b\theta)$ fitting functions, respectively. The value of the reduced $\chi$-squared for the quadratic (linear) fit is listed without (with) the parenthesis. We take the final continuum value to be the one obtained with the fit which has a lower $\chi$-squared.}
    \label{fig:pimc_temp_pt3}
\end{figure}

The energy gap can be estimated using the autocorrelation function. Suppose that $\hat H$ has a non-degenerate ground state $\ket{\Omega}$, and denote its eigenstates by $\ket{E_n}$, with $\ket{E_0}=\ket{\Omega}$, and the gap $\Delta E=E_1-E_0$. Recalling that $\hat{\phi}_j(\tau=t\theta)\coloneq \phih_{j,t}=e^{\tau \hat H}\hat{\phi}_j(0)e^{-\tau \hat H}$, one can write
\begin{align}
    \bra{\Omega}\hat{\phi}_j(\tau)\hat{\phi}_j(0)\ket{\Omega} &= \sum_{n=0}^{\infty} \ e^{-\tau(E_n-E_0)}|\langle\Omega | \hat{\phi}_j(0) | E_n \rangle  |^2 =e^{-\tau(E_1-E_0)}|\langle\Omega | \hat{\phi}_j(0) | E_1 \rangle |^2 + \cdots,
\end{align}
for $\tau < \frac{T}{2}$. The first term in the sum, $|\langle \Omega | \hat{\phi}_
j(0) | \Omega \rangle |^2$, has been set to zero due to the ($\mathbb{Z}_2$) parity-invariance of the state. Due to the periodic nature of this lattice, the dominant contribution to this correlator in the long Euclidean time limit is, in fact, given by
\begin{align}
    \bra{\Omega}\hat{\phi}_j(\tau)\hat{\phi}_j(0)\ket{\Omega}&=\frac{1}{2}\left[\bra{\Omega}\hat{\phi}_j(\tau)\hat{\phi}_j(0)\ket{\Omega}+\bra{\Omega}\hat{\phi}_j(T-\tau)\hat{\phi}_j(0)\ket{\Omega}\right] \nonumber \\
    &= |\langle \Omega | \hat{\phi}_j(0) | E_1 \rangle|^2 \, \frac{e^{-\tau\Delta E}+e^{-(T-\tau)\Delta E}}{2}+\cdots \nonumber \\
    &= |\langle \Omega | \hat{\phi}_j(0) | E_1 \rangle|^2 \, e^{-\frac{T\Delta E}{2}}\mathrm{cosh}\left[\Delta E\left(\frac{T}{2}-\tau\right)\right] +\cdots .
\end{align}
To better identify a range of $\tau$ where the above form is the dominant contribution, one can define the effective mass $m_{\rm eff}(\tau)$ at each point $\tau=t\theta$ on the discrete and periodic Euclidean lattice through the relation
\begin{align}
    \frac{\bra{\Omega}\phih_{j,t}\phih_{j,0}\ket{\Omega}}{\bra{\Omega}\phih_{j,t+1}\phih_{j,0}\ket{\Omega}} = \frac{\mathrm{cosh}\left[m_{\rm eff}(t\theta)\left(\frac{T}{2}-t\theta\right)\right]}{\mathrm{cosh}\left[m_{\rm eff}(t\theta)\left(\frac{T}{2}-(t+1)\theta\right)\right]}.
\end{align}
The resulting plots for $m_{\rm eff}$ are shown in Figs.~\ref{fig:pimc_temp_pt0} to \ref{fig:pimc_temp_pt3}. The shaded window in yellow represents the time period used to average over for the final gap estimate, $\Delta E$. The gap estimates for different $(m^2,\lambda)$ values vary from $\Delta E\approx 0.7$ to $\Delta E\approx 1.6$. They are all consistent with the requisite conditions stated before, i.e., $\frac{1}{T}=0.1 < \Delta E < \frac{1}{\theta_{\rm max}}=2.5$, where $\theta_\text{max}$ is the largest lattice spacing used in our computations. \\

To propagate the above errors through moment optimization, we perform the optimization procedure multiple times: each time, we source a random value for the ground-state moments by sampling from a normal distribution specified by the continuum extrapolated values and their associated errors. The plots in the main text show the mean and 68\% confidence bands obtained from these samples. In particular, all plots in this work are produced using 100 samples.

\section{Core-state polynomial with quadrature argument}\label{app:core_quad}

In Sec.~\ref{sec:multi-ansatze} of the main text, we asserted that the core state $\ket{C}\coloneq C(\hat{\bm{a}}^{\dagger})\ket{\bs{0}}$ can equivalently be expressed as $\ket{C}\coloneq C^{\phi}(\hat{\bm{\phi}})\ket{\bs{0}}$. There is a one-to-one mapping between the terms and independent coefficients of these two polynomials. While this mapping is easy to establish, for completeness, we provide details of the mapping in this appendix.

Consider an $N$-mode core state $\ket{C}$ with
\begin{align}
    C^{\phi}(\hat{\bm{\phi}})&
    \coloneq \sum_{\substack{\bs{n} \\ 
    \mathrm{sum}(\bs{n})\leq R}} \ d_{\bs{n}}\hat{\bm{\phi}}^{\bs{n}}\ket{\bs{0}} 
    = \sum_{\substack{\bs{n} \\ \mathrm{sum}(\bs{n})\leq R}} \ d_{n_0,\ldots,n_{N-1}} \ \phih_0^{n_0}\ldots\phih_{N-1}^{n_{N-1}}\ket{\bs{0}}. 
\end{align}
Now since $\hat a_j \ket{\bm{0}}=0$ and $\hat a_j\hat{a}^\dagger_j=\hat{a}^\dagger_j\hat a_j+1$,
\begin{align}
    \phih_j^{n_j}\ket{\bs{0}}
    &= \frac{1}{2^{n_j}}\left[(\hat a_j^{\dagger})^{n_j}+A_{n_j,2}(\hat a_j^{\dagger})^{n_j-2}+A_{n_j,4}(\hat a_j^{\dagger})^{n_j-4}+\cdots + A_{n_j,n_j-2\lfloor\frac{n_j}{2}\rfloor}(\hat a_j^{\dagger})^{n_j-2\lfloor\frac{n_j}{2}\rfloor}\right]\ket{\bs{0}}.
\end{align}
Thus, the core state can be expressed as
\begin{align}
    \ket{C}
    &= \sum_{\substack{\bs{n} \\ \mathrm{sum}(\bs{n})\leq R}} \ \frac{d_{\bs{n}}}{2^{\mathrm{sum}(\bs{n})}}\bigotimes_{j=0}^{N-1}\left[(\hat a_j^{\dagger})^{n_j}+A_{n_j,2}(\hat a_j^{\dagger})^{n_j-2}+A_{n_j,4}(\hat a_j^{\dagger})^{n_j-4}+\cdots + A_{n_j,n_j-2\lfloor\frac{n_j}{2}\rfloor}(\hat a_j^{\dagger})^{n_j-2\lfloor\frac{n_j}{2}\rfloor}\right]\ket{\bs{0}}\nonumber \\
    &= \sum_{\substack{\bs{n} \\ \mathrm{sum}(\bs{n})\leq R}} \ c'_{\bs{n}}(\hat{\bm{a}}^{\dagger})^{\bs{n}}\ket{\bs{0}} \equiv C(\hat {\bm{a}}^{\dagger})\ket{\bs{0}}.
\end{align}
Define $\zeta=0$ or $1$. Comparing the coefficients of $(\hat{\bm{a}}^{\dagger})^{\bs{n}}$ in the first and second lines, one gets that for $\bs{n}$ with $\mathrm{sum}(\bs{n})=R-\zeta$, 
\begin{align}
    c'_{\bs{n}}=\frac{d_{\bs{n}}}{2^{R-\zeta}}. 
\end{align}
Having solved for the coefficients $c'_{\bs{n}}$ accompanying the terms with the highest weights $\mathrm{sum}(\bs{n})=R-\zeta$,
one can now work out the ``descending coefficients." For instance, for $\bs{n}$ with $\mathrm{sum}(\bs{n})=R-\zeta-2$,
\begin{align}
    c'_{\bs{n}}=\frac{d_{\bs{n}}}{2^{R-\zeta-2}}+\sum_{j=0}^{N-1} \ \frac{d_{\bs{n}+2\bs{e}_j}}{2^{R-\zeta}}A_{n_j+2,2},
\end{align}
where $\bs{e}_j$ is the unit vector in the direction $j$. For $\mathrm{sum}(\bs{n})=R-\zeta-4$,
\begin{align}
    c'_{\bs{n}}=\frac{d_{\bs{n}}}{2^{R-\zeta-4}}+\sum_{j=0}^{N-1} \ \frac{d_{\bs{n}+2\bs{e}_j}}{2^{R-\zeta-2}}A_{n_j+2,2}+\sum_{j=0}^{N-1} \ \frac{d_{\bs{n}+4\bs{e}_j}}{2^{R-\zeta}}A_{n_j+4,4}+\sum_{j\neq j'}^{N-1} \ \frac{d_{\bs{n}+2\bs{e}_j+2\bs{e}_{j'}}}{2^{R-\zeta}}A_{n_j+2,2}A_{n_{j'}+2,2}.
\end{align}
This procedure can thus be used to determine the coefficients of the standard core-state polynomial $C(\hat{\bm{a}}^{\dagger})$ starting from the quadrature polynomial generator $C(\hat{\bm{\phi}})$. It is straightforward to invert the above equations. In other words, given any core-state polynomial $C(\hat{\bm{ a}}^{\dagger})$, one can obtain a unique polynomial $C^{\phi}(\hat{\bm{\phi}})$ which generates the same core state.

\section{Sparse quantum-state preparation}\label{app:sparse_prep}
\noindent
This appendix provides a brief overview of, and an illustrative example for, the sparse quantum-state preparation algorithm introduced in Ref.~\cite{gleinig2021efficient}. We proposed to use this method to prepare core states in Sec.~\ref{sec:core-state-prep}.

Suppose the goal is to prepare an $n$-qubit state $\ket{\psi}=\sum_{x \in S} \ c_{x}\ket{x}$, where $x$ are length-$n$ bit strings taking values in some set $S \subset \{0,1\}^{n}$. Reference~\cite{gleinig2021efficient} provides a method for deducing the unitary transformation that converts $\ket{\psi}$ to the state $\ket{\psi'}=\sum_{x \in S'} \ c'_{x}\ket{x}$ with $|S'|<|S|$. Thus, by successively applying this algorithm, the state $\ket{\psi}$ can be mapped to a single computational basis state $\ket{x}$. This state can, in turn, be mapped to the state $\ket{0\ldots 0}$ by the application of NOT (single-qubit Pauli-X) gates. Inverting the above series of transformations allows one to prepare the state $\ket{\psi}$ starting from the state $\ket{0 \ldots 0}$. This algorithm is especially efficient when the superposition is sparse, i.e., $|S|\ll 2^n$. As explained in the main text, this condition holds for the qubitzed $(R,Q)$ core state. \\

As an example, consider applying the algorithm to convert the state $\ket{\psi}=c_0\ket{0000}+c_1\ket{0101}+c_2\ket{0110}+c_3\ket{1011}+c_4\ket{1100}+c_5\ket{1101}$ into a superposition of a smaller number of computational basis states. Repeated application of this subroutine then maps $\ket{\psi}$ to $\ket{0000}$. (For the notation and pseudocode, refer to Ref.~\cite{gleinig2021efficient}, Algorithm 1.) 
\begin{enumerate}
    \item \textbf{Initialization:} Initialize
    $$T= S=[0000,0101,0110,1011,1100,1101], \ \mathit{dif\_qubits}=[], \ \mathit{dif\_vals}=[].$$ 
    \item \textbf{The first WHILE loop:} Scan over all qubits $b \in \{1,\ldots,n\}$ and look for the first qubit $b$ such that the sizes of the sets $T_0\coloneq \{x\in T| x[b]=0\}$ and $T_1\coloneq \{x\in T|x[b]=1\}$ are as unequal as possible (with neither set being empty). This turns out to be $b=1$; in this case, $T_0=[0000,1011]$ and $T_1=[0101,0110,1100,1101]$, i.e., $|T_0|=2$ and $|T_1|=4$. Identify the bit value $v \in \{0,1\}$ such that $T_v$ is the smaller of these two sets (if the two sets have the same size, pick $T_1$). Now, append the qubit index $b$ (i.e., $b=1$ in this case) to $\mathit{dif\_qubits}$, append the bit value $v$ (i.e, $v=0$ in this case) to $\mathit{dif\_vals}$, and update $T$ to the set of bit strings in $T$ which take the value $v$ at the identified qubit index $b$ (i.e., $T\rightarrow T_0$ in this case). Thus, at this stage, one has
    $$T=[0000,1011], \ \mathit{dif\_qubits}=[1], \ \mathit{dif\_vals}=[0].$$ 
    Since $|T|>1$, once again scan over all qubits $b$ to look for the first qubit that satisfies the above condition (now with the updated set $T$ which has a reduced number of strings). This turns out to be $b=0$; thus append this to $\mathit{dif\_qubits}$. Since $|T_0|=|T_1|$ in this case, update $T$ to $T=T_1$, and append 1 to $\mathit{dif\_vals}$. This yields
    $$T=[1011], \ \mathit{dif\_qubits}=[1,0], \ \mathit{dif\_vals}=[0,1].$$
    Since $|T|=1$, the first WHILE loop terminates.
    \item \textbf{Preparation for the second WHILE loop:}
    Remove the last value appended to $\mathit{dif\_qubits}$, i.e., $b=0$; store it as $\mathit{dif}=0$. Also remove the last value of $\mathit{dif\_vals}$. Finally, store the single element of $T$ as $x_1$. This leaves $$x_1=1011, \ \mathit{dif}=0, \ \mathit{dif\_qubits}=[1], \ \mathit{dif\_vals}=[0].$$
    Now, define a new set $T'$ which consists of all those strings $x\in S$ which take values in $\mathit{dif\_vals}$ on qubits in $\mathit{dif\_qubits}$, i.e., $T'=[0000,1011]$. Remove $x_1$ from $T'$ to have the final assignments
    $$T'=[0000], \ \mathit{dif\_qubits}=[1], \ \mathit{dif\_vals}=[0].$$
    \item \textbf{The second WHILE loop:} The second WHILE loop runs in exactly the same way as the first WHILE loop, except that the initial conditions now involve the set $T'$, and the updated values of $\mathit{dif\_qubits}$ and $\mathit{dif\_vals}$ defined above. Since $|T'|$ is already 1, the loop does not run in this case. 
    \item \textbf{Preparation for circuit building:} Assign the single element in $T'$ to the variable $x_2$. Thus, the relevant variable values at this stage are
    $$x_1=1011, \ x_2=0000, \ \mathit{dif}=0, \ \mathit{dif\_qubits}=[1], \ \mathit{dif\_vals}=[0].$$
    To summarize, the above steps identify two strings $x_1$ and $x_2$ which take equal values on the qubits in $\mathit{dif\_qubits}$, and no other strings in $S$ take the same values as $x_1$ and $x_2$ on $\mathit{dif\_qubits}$. This property allows the strings $x_1$ and $x_2$ to be combined into a single computational basis state using the operations described in the next few steps, without introducing additional strings in the remaining set of bit strings in $S$.
    \item \textbf{Circuit building:} 
    Initialize $\hat{\mathcal{U}}=\hat 1$.
    \begin{enumerate}
        \item If $x_1[\mathit{dif}]\neq 1$, add a NOT gate to the line $\mathit{dif}$. In this case, $x_1[\mathit{dif}]$ is already 1.    
        \begin{align}
            \hat{\mathcal{U}}=
            \begin{quantikz}
                \lstick{$\mathit{dif}\rightarrow 0$} & \qw \\
                \lstick{1} & \qw \\
                \lstick{2} & \qw \\
                \lstick{3} & \qw \\
            \end{quantikz}
        \end{align}
        \item For all qubits $b$ other than the $\mathit{dif}$ qubit, if $x_1[b]\neq x_2[b]$, apply a CNOT gate on qubit $b$ controlled on $\mathit{dif}$. 
        \begin{align}
            \hat{\mathcal{U}}=
            \begin{quantikz}
                \lstick{$\mathit{dif}\rightarrow 0$} & \ctrl{2} & \ctrl{3} & \qw \\
                \lstick{1} & \qw & \qw & \qw\\
                \lstick{2} & \targ{} & \qw & \qw\\
                \lstick{3} & \qw & \targ{} & \qw\\
            \end{quantikz}
        \end{align}
        $x_1[\mathit{dif}]=1$ and $x_2[\mathit{dif}]=0$ by design. Thus, the application of these CNOT gates leaves $x_2$ untouched while $x_1$ is transformed so that it agrees with $x_2$ on all qubits except the $\mathit{dif}$ qubit. More concretely,
        \begin{align}
            \hat{\mathcal{U}}\ket{\psi}&=\hat{\mathcal{U}}(c_0\mathbf{\ket{0000}}+c_1\ket{0101}+c_2\ket{0110}+c_3\mathbf{\ket{1011}}+c_4\ket{1100}+c_5\ket{1101}) \nonumber \\
            &= c_0\mathbf{\ket{0000}}+c_1\ket{0101}+c_2\ket{0110}+c_3\mathbf{\ket{1000}}+c_4\ket{1111}+c_5\ket{1110}.
        \end{align}
        Here and below, the strings $x_1$ and $x_2$, and their transformed versions, are highlighted.
        \item As mentioned earlier, $x_1$ and $x_2$ strings are the only strings that take the values $\mathit{dif\_vals}$ on the qubits $\mathit{dif\_qubits}$. This step ensures that these values are all set to 1 so that the merging operation can be activated in the next step. In other words, for all qubits $b$ in $\mathit{dif\_qubits}$, if $x_1[b]=x_2[b]\neq 1$, a NOT gate is applied on line $b$.
        \begin{align}
            \hat{\mathcal{U}}=
            \begin{quantikz}
                \lstick{$\mathit{dif}\rightarrow 0$} & \ctrl{2} & \ctrl{3} & \qw & \qw\\
                \lstick{1} & \qw & \qw & \targ{} & \qw\\
                \lstick{2} & \targ{} & \qw & \qw & \qw\\
                \lstick{3} & \qw & \targ{} & \qw & \qw\\
            \end{quantikz}
        \end{align}
        This gives 
        \begin{align}
            \hat{\mathcal{U}}\ket{\psi }&=c_0\mathbf{\ket{0100}}+c_1\ket{0001}+c_2\ket{0010}+c_3\mathbf{\ket{1100}}+c_4\ket{1011}+c_5\ket{1010} \nonumber \\
            &= (c_0\mathbf{\ket{0}}+c_3\mathbf{\ket{1}})\otimes\mathbf{\ket{100}}+c_1\ket{0001}+c_2\ket{0010}+c_4\ket{1011}+c_5\ket{1010}.
        \end{align}
        \item The highlighted strings can now be merged into one string by performing a suitable rotation $\hat{\mathcal{R}}$ on the qubit $\mathit{dif}$ controlled on all qubits in $\mathit{dif\_qubits}$. The rotation is defined so that
        $$\hat{\mathcal{R}}(c_0\ket{0}+c_3\ket{1})=\mathcal{N}_{c_0,c_3}\ket{0},$$
        where $\mathcal{N}_{c_0,c_3}$ is a complex amplitude.
        \begin{align}
            \hat{\mathcal{U}}=
            \begin{quantikz}
                \lstick{$\mathit{dif}\rightarrow 0$} & \ctrl{2} & \ctrl{3} & \qw & \gate{\hat{\mathcal{R}}}  & \qw\\
                \lstick{1} & \qw & \qw & \targ{} & \ctrl{-1}  & \qw\\
                \lstick{2} & \targ{} & \qw & \qw & \qw  & \qw\\
                \lstick{3} & \qw & \targ{} & \qw & \qw  & \qw\\
            \end{quantikz}
        \end{align} 
        This is the final form for $\hat{\mathcal{U}}$. It gives the desired state $\ket{\psi'}$,
        \begin{align}
            \hat{\mathcal{U}}\ket{\psi}&= 
            \mathcal{N}_{c_0,c_3}
            \mathbf{\ket{0100}}+c_1\ket{0001}+c_2\ket{0010}+c_4\ket{1011}+c_5\ket{1010} \equiv \ket{\psi'},
        \end{align}
        which satisfies $|S'|=5<|S|=6$.
    \end{enumerate}
\end{enumerate}

In each iteration of the WHILE loops, the sizes of the sets $T$ and $T'$ are at least halved. Thus, the number of WHILE loops is $O(\lceil \mathrm{log}(|S|)\rceil)$. Each WHILE loop takes $O(|S|n)$ time as it involves looking at at most $|S|$ strings of length $n$. Thus, the total classical runtime of the above algorithm is $O(|S| n \, \mathrm{log}(|S|))$.\\

Since each WHILE loop adds an element to $\mathit{dif\_qubits}$, $|\mathit{dif\_qubits}|$ is $O(\mathrm{log}(|S|))$. Step 6(a) involves at most one single-qubit gate while step 6(b) involves 
at most $n-1$ CNOT gates. In step 6(c), one adds $O(\mathrm{log}(|S|))$ one-qubit gates. In step 6(d), one must implement a rotation gate controlled on the $O(\mathrm{log}|S|)$ $\mathit{dif\_qubits}$, which can be implemented using $O(\mathrm{log}(|S|))$ CNOT gates. Thus, the total number of CNOT gates is $O(n)$, and the total number of single-qubit gates is $O(\mathrm{log}(|S|))$.\\

Now, one instance of the above algorithm reduces the number of states in the superposition by one. Thus, this algorithm must be repeated $|S|$ times to deduce the unitary which maps the state $\ket{\psi}$ to a single computational basis state. Thus, the total classical runtime for determining the full quantum circuit turns out to be $O(|S|^2n \, \mathrm{log}(|S|))$. The total number of CNOT gates is $O(|S|n)$, while the total number of one-qubit gates is $O(|S| \, \mathrm{log}(|S|)+n)$ (the addition of $n$ arises because converting a single basis state to $\ket{0\ldots 0}$ requires $O(n)$ single-qubit gates). In summary, if $|S| \ll 2^n$, this algorithm is efficient in converting the state $\ket{\psi}$ to $\ket{0\ldots 0}$ and vice versa.

\section{Digitization of the squeezing operator}

As discussed in Sec.~\ref{sec:qsqueeze} of the main text, in order to make the single-mode squeezing operator, $\hat S(r)$, amenable to a discrete-variable implementation, it must first be mapped to a squeezing operator acting just on the truncated Hilbert space, i.e, $\hat S^{\Lambda}(r)$. Thereafter, $\hat S^{\Lambda}(r)$ is approximated by a first-order product expansion, $\hat S^{\Lambda,K}(r)$: 
\begin{align}
    \hat S(r) = e^{\frac{r}{2}\left(\hat a^{\dagger 2} - \hat a^2\right)}
    \mapsto & \hat S^{\Lambda}(r) \coloneq e^{\frac{r}{2}\left[(\hat a^{\Lambda \dagger})^2 - (\hat a^{\Lambda})^2\right]} 
    \nonumber \\
    \mapsto & \hat S^{\Lambda,K}(r) \coloneq \left(e^{-i\frac{r}{K}\hat s_0^{\Lambda}} e^{-i\frac{r}{K}\hat s_2^{\Lambda}}\right)^K \left(e^{-i\frac{r}{K}\hat s_1^{\Lambda}} e^{-i\frac{r}{K}\hat s_3^{\Lambda}}\right)^K.
\end{align}
To compare $\hat S^{\Lambda}(r)$ and $\hat S^{\Lambda,K}(r)$ (which act only on the truncated Hilbert space) with the squeezing operator $\hat S(r)$ (which acts on the full single-mode Hilbert space), we will embed the former in the untruncated single-mode Hilbert space by suitably padding the operator matrix with zero matrix elements to ensure the truncated operators do not act on Fock states beyond the cutoff $\Lambda$. In this appendix, we describe the errors arising from this truncation and Trotterization for both the single-mode and multimode cases. 

For a single mode, the squeezing operator is applied to the rank-$R$ core state to obtain the rank-$R$ ansatz $\ket{\psi}_R=\hat S(r)\ket{C}_R$. We demand a bound on the distance:
\begin{align}
    \norm{\ket{\psi}_R-\ket{\psi^{\Lambda,K}}_R}\leq \norm{\ket{\psi}_R-\ket{\psi^{\Lambda}}_R}+\norm{\ket{\psi^{\Lambda}}_R-\ket{\psi^{\Lambda,K}}_R}\leq \epsilon^{(1)}\coloneq \epsilon^{(1)}_{\rm trunc} + \epsilon^{(1)}_{\rm trott},
\end{align}
where $\norm{\gket{\psi}}\coloneq \sqrt{\gbraket{\psi}{\psi}}$ is the inner-product norm acting on the full single-mode Hilbert space, $\ket{\psi^{\Lambda}}_R\coloneq\hat S^{\Lambda}(r)\ket{C}_R$, and $\ket{\psi^{\Lambda,K}}_R\coloneq\hat S^{q,\Lambda,K}(r)\ket{C}_R$. 
 
In the multimode case, the tensor product of single-mode squeezing operators is applied to the multimode rank-$R$, span-$Q$ core state $\ket{C}_{R,Q}$ to obtain the $(R,Q)$ ansatz $\ket{\psi}_{R,Q}=\otimes_{j=0}^{N-1}\hat S_j(r)\ket{C}_{R,Q}$. In this case, we demand a bound on the distance:
\begin{align}
    \norm{\ket{\psi}_{R,Q}-\ket{\psi^{\Lambda,K}}_{R,Q}}\leq \norm{\ket{\psi}_{R,Q}-\ket{\psi^{\Lambda}}_{R,Q}}+\norm{\ket{\psi^{\Lambda}_{R,Q}}-\ket{\psi^{\Lambda,K}}_{R,Q}}\leq \epsilon^{(N)} \coloneq \epsilon^{(N)}_{\rm trunc} + \epsilon^{(N)}_{\rm trott},
\end{align}
where we use the same notation $\norm{\ket{\psi}}$ to denote the inner-product norm over the full multimode Hilbert space. Moreover, $\ket{\psi^{\Lambda}}_{R,Q}\coloneq\otimes_{j=0}^{N-1}\hat S_j^{\Lambda}(r)\ket{C}_{R,Q}$, and $\ket{\psi^{\Lambda,K}}_{R,Q}\coloneq\bigotimes_{j=0}^{N-1}\hat S_j^{q,\Lambda,K}(r)\ket{C}_{R,Q}$.

\subsection{Truncation error}\label{app:truncation}

The goal is to compare the action of the full squeezing operator, $\hat{S}(r)$, and the approximate squeezing operator, $\hat S^{\Lambda}(r)$ on the single- and multi-mode core states. The single-mode rank-$R$ core state is given by $\ket{C}_R=\bar{\sum}_{n}c_{n}\ket{n}$. Here, the sum $\bar{\sum}$ is restricted to a sum of $\frac{R}{2}+1$ terms, such that $\ket{C}$ is a symmetric rank-$R$ core state as defined in Eq.~\eqref{eq:single_mode_decomp}. Introducing the notation $\ket{n,r,\Lambda}\coloneq \hat S^{\Lambda}(r)\ket{n}$, the distance between the two states for the single-mode case is bounded by
\begin{align}
    \norm{\ket{\psi}_{R}-\ket{\psi^{\Lambda}}_{R}} = \norm{\left(\hat S(r)-\hat S^{\Lambda}(r)\right)\ket{C}_R}&\leq \bar{\sum}_n \ |c_n| \norm{\ket{n,r}-\ket{n,r,\Lambda}} \leq \sqrt{\frac{R}{2}+1}g(r,R,\Lambda),
\end{align}
where $g(r,R,\Lambda)\coloneq \mathrm{argmax}_{n\leq R} \ \norm{\ket{n,r}-\ket{n,r,\Lambda}}$, and we have used the Cauchy-Schwartz inequality to bound $\bar{\sum}_n |c_n|=\bar{\sum}_n |c_n|\times 1\leq \sqrt{(\bar{\sum}_n |c_n|^2)(\bar{\sum}_n 1)}$. Similarly, the $N$-mode $(R,Q)$ core state can be expressed as $\ket{C}_{R,Q}=\bar{\sum}_{\bs{n}}c_{\bs{n}}\ket{\bs{n}}$, where the sum $\bar{\sum}$ is restricted to the $N|c_{R,Q}|$ terms as defined in Eq.~\eqref{eq:c-RQ-def}. It follows that,
\begin{align}
    \norm{\ket{\psi}_{R,Q}-\ket{\psi^{\Lambda}}_{R,Q}} &= \norm{\left(\otimes_{j=0}^{N-1}\hat S_j(r)-\otimes_{j=0}^{N-1} \hat S_j^{\Lambda}(r)\right)\ket{C}_{R,Q}} 
    \nonumber\\
    &\leq \bar{\sum}_{\bs{n}} |c_{\bs{n}}|\norm{\left(\otimes_{j=0}^{N-1}\hat S_j(r)-\otimes_{j=0}^{N-1} \hat S_j^{\Lambda}(r)\right)\ket{\bs{n}}} \nonumber \\
    &= \bar{\sum}_{\bs{n}} |c_{\bs{n}}|\norm{\otimes_{j=0}^{N-1}\left(\ket{n_j,r}\right)-\otimes_{j=0}^{N-1}\left(\ket{n_j,r,\Lambda}\right)} \nonumber \\
    &\leq \bar{\sum}_{\bs{n}} |c_{\bs{n}}| \sum_{j=0}^{N-1} \norm{\ket{n_j,r}-\ket{n_j,r,\Lambda}} \leq N\sqrt{N|c_{R,Q}|}g(r,R,\Lambda),
\end{align}
where in the last line, we have used the fact that for normalized states $\gket{\psi_0},\ldots,\gket{\psi_{m-1}},\gket{\phi_0},\ldots,\gket{\phi_{m-1}}$,
\begin{align}
    \norm{\otimes_{k=0}^{m-1}\gket{\psi_k}
    \ - \ \otimes_{k=0}^{m-1}\gket{\phi_k}
    }&=\norm{\sum_{k=0}^{m-1} \ (\gket{\psi_0}    \ldots \gket{\psi_{k-1}}) \ (\gket{\psi_k} -\gket{\phi_k}) \ \left(\gket{\phi_{k+1}}\ldots\gket{\phi_{m-1}}    \right)} \nonumber \\
    &\leq \sum_{k=0}^{m-1} \norm{\gket{\psi_0}    \ldots \gket{\psi_{k-1}}}\norm{\gket{\psi_k} -\gket{\phi_k} }\norm{\gket{\phi_{k+1}}\ldots\gket{\phi_{m-1}}}  \nonumber \\
    &= \sum_{k=0}^{m-1} \norm{\gket{\psi_k} -\gket{\phi_k}}.
\end{align}

Both the single-mode and multimode distances depend upon $g(r,R,\Lambda)$, which can be bounded above as follows:
\begin{align}
    g(r,R,\Lambda) &\coloneq \mathrm{argmax}_{n_1\leq R} \ \norm{\ket{n_1,r}-\ket{n_1,r,\Lambda}} \nonumber \\
    &= \mathrm{argmax}_{n_1\leq R}  \ \left[ \sum_{n_2=0}^{\infty}\left|\langle n_2 |\hat S(r)-\hat S^{\Lambda}(r)|n_1\rangle\right|^2\right]^{1/2} \nonumber \\
    &\leq  \underbrace{\mathrm{argmax}_{n_1\leq R}  \ \left[ \sum_{n_2=0}^{\Lambda}\left|\langle n_2 |\hat S(r)-\hat S^{\Lambda}(r)|n_1\rangle\right|^2\right]^{1/2}}_{g^{\rm disc}(r,R,\Lambda)} + \underbrace{\mathrm{argmax}_{n_1\leq R}  \ \left[ \sum_{n_2=\Lambda+1}^{\infty}\left|\langle n_2 |\hat S(r)|n_1\rangle\right|^2\right]^{1/2}}_{g^{\rm leak}(r,R,\Lambda)}.
    \end{align}
In the last line, we have used the fact that $\gbra{n_2}\hat S^{\Lambda}(r)\gket{n_1}=0$ for $n_2>\Lambda$.  We will refer to the term $g^{\rm disc}(r,R,\Lambda)$ as the matrix-element \emph{discrepancy} and the second term $g^{\rm leak}(r,R,\Lambda)$ as \emph{leakage} [as it describes how much of the weight of initial state $\gket{n_1}$, with $n_1\leq R$, is leaked outside the truncated Hilbert space by $\hat S(r)$]. 

\begin{figure}
    \centering \includegraphics[width=0.9\columnwidth]{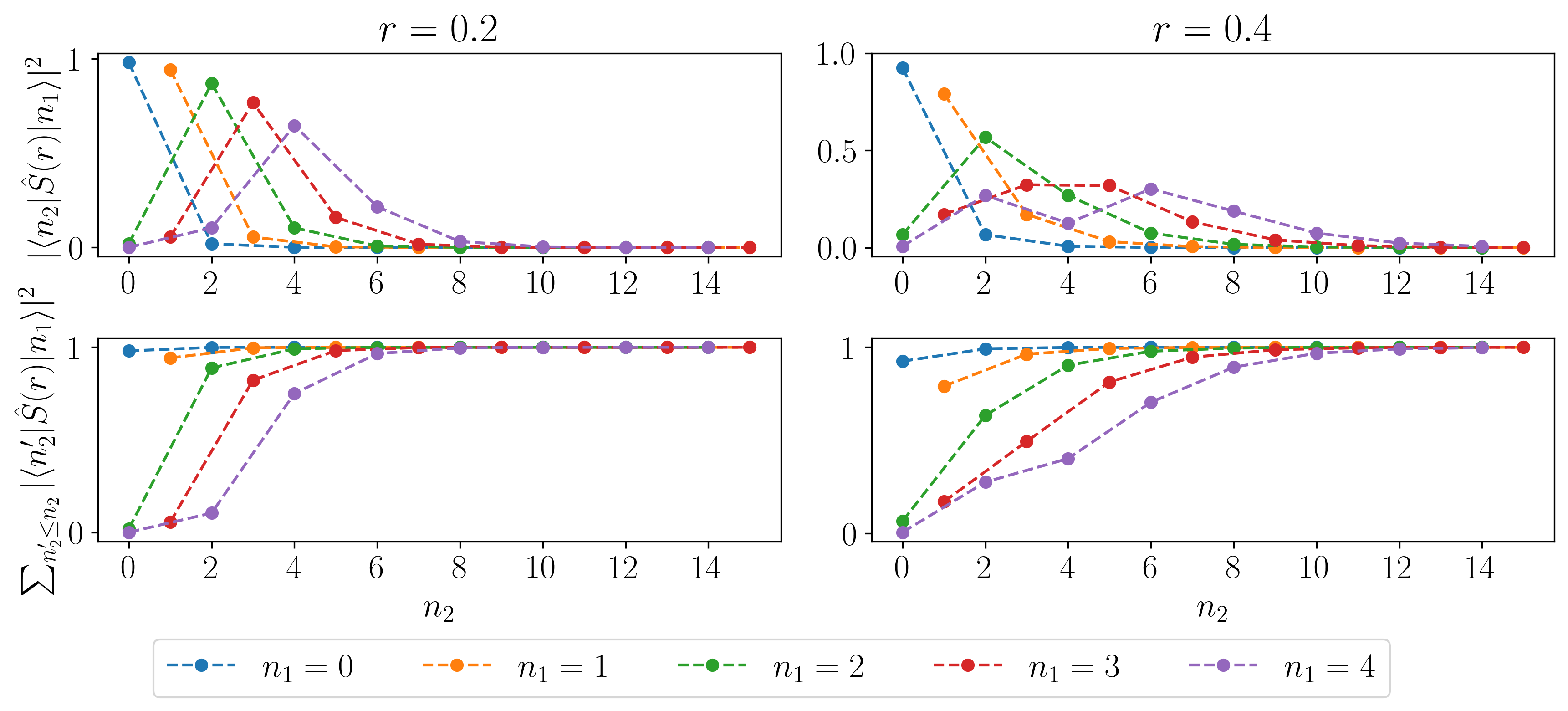}
    \caption{Boson-number distribution for squeezed Fock states. Top panels show $
    |\langle n_2 | \hat S(r) | n_1 \rangle|^2$ as a function of $n_2$ for $n_1$ values up to 4 (which is the maximum rank of the core states considered in this work). Bottom panels show the cumulative distributions, $\sum_{n'_2\leq n_2}|\langle n_2' | \hat S(r) | n_1 \rangle|^2$, as a function of $n_2$. These plots demonstrate that the squeezing operator acted on a state with initial occupations $n_1 \leq R$ has larger overlap with low-lying states $n_2$. In the top plots, for even (odd) values for $n_1$, only the non-vanishing matrix elements corresponding to even (odd) values of $n_2$ are plotted. In the bottom plots, the points corresponding to odd (even) $n_2$ values for even (odd) $n_1$ values are dropped as they equal their previous adjacent point. The dotted lines connecting the points are to guide the eye and do not represent a fit.
    }
    \label{fig:squeeze_spread}
\end{figure}

Using the analytical form for $\left|\langle n_2 |\hat S(r)|n_1\rangle\right|^2$ (see Ref.~\cite{de1990properties}) and the unitarity of the squeezing operator, the value of $g^{\rm leak}(r,R,\Lambda)$ can be calculated exactly.
Figure~\ref{fig:squeeze_spread} shows the boson-number distribution $|\langle n_2|\hat S(r)|n_1\rangle|^2$, together with the cumulative distribution $\sum_{n_2' \leq n_2}|\langle n_2'|\hat S(r)|n_1\rangle|^2$ for $n_1\leq R=4$, and for two characteristic values, $r=0.2,0.4$, encountered in this work. The plots illustrate that the distribution is peaked around small values of $n_2$, and the cumulative distribution quickly converges to one as $n_2$ is increased for these small $r$ values. 

We now upper bound $g^{\rm disc}(r,R,\Lambda)$. Any matrix elements between even and odd Fock states vanish for both the full and truncated squeezing operators. Thus, consider the matrix elements with both $n_1$ and $n_2$ being odd or even. Upon Taylor expanding the exponential in $\hat S(r)$, the matrix element becomes
\begin{align}
    \bra{n_2}e^{\frac{r}{2}\sum_{n=0}^{\infty} \ell_n\left(\ket{n+2}\bra{n}-\ket{n}\bra{n+2}\right)}\ket{n_1} &= \sum_{p=0}^{\infty} \ \frac{r^p}{p!2^p}\bra{n_2}\left[\sum_{n=0}^{\infty}\ell_n\left(\ket{n+2}\bra{n}-\ket{n}\bra{n+2}\right)\right]^p\ket{n_1} \nonumber \\
    &= \sum_{\Delta=0}^{\infty} \ \frac{r^{p_0+2\Delta}}{(p_0+2\Delta)!2^{p_0+2\Delta}}\underbrace{\bra{n_2}\left[\sum_{n=0}^{\infty}\ell_n\left(\ket{n+2}\bra{n}-\ket{n}\bra{n+2}\right)\right]^{p_0+2\Delta}\ket{  n_1}}_{f(\Delta)} \nonumber \\
    & \equiv \sum_{\Delta=0}^{\infty} t(\Delta).
    \label{eq:sum-path}
\end{align}
In the second equality, the sum over integer-valued $p \geq 0$ is exchanged with a sum over $\Delta$, where $p=p_0+2\Delta \geq 0$, and we have defined $p_0 \coloneq \frac{|n_2-n_1|}{2}$. This latter sum, by construction, involves only non-vanishing matrix elements from the former sum. 
The term $f(\Delta)$, in particular, receives contributions from ``paths" consisting of $p_0+2\Delta$ steps (of two units in Fock space) connecting $n_1$ and $n_2$. These steps can either be in the forward (corresponding to the $\ket{n+2}\bra{n}$ term) or backward (corresponding to the $\ket{n}\bra{n+2}$ term) direction. Each such path is composed of $p_0+\Delta$ steps in the direction from $n_1$ to $n_2$, and $\Delta$ steps in the opposite direction. Thus, the total number of such paths is $\binom{p_0+2\Delta}{\Delta}$, i.e., the number of ways in which one could intersperse the $\Delta$ steps in the opposite direction among the total number of steps. When $n_2 \geq n_1$, out of these paths, the highest numerical weight is carried by the one which goes all the way from $n_1$ to $n_2+2\Delta$, and then circles back to $n_2$. The contribution from this highest-weight path is 
\begin{align}
     w(n_2 \geq n_1)&\coloneq (-1)^{\Delta}
     \underbrace{\ell_{n_1}\ldots \ell_{n_2-2} \ell_{n_2}\ldots \ell_{n_2+2\Delta-2}}_{n_1\rightarrow n_2+2\Delta} \ \underbrace{\ell_{n_2+2\Delta-2} \ldots \ell_{n_2}}_{n_2+2\Delta \rightarrow n_2}=(-1)^{\Delta}\frac{(n_2+2\Delta)!}{\sqrt{n_1!n_2!}}.
\end{align}
When $n_1>n_2$, the highest-weight path goes from $n_1$ to $n_1+2\Delta$ and then circles back all the way to $n_2$. The contribution from this path is given by,
\begin{align}
     w(n_2 < n_1)&\coloneq (-1)^{\Delta}
     \underbrace{\ell_{n_1} \ldots\ell_{n_1+2\Delta-2} }_{n_1 \rightarrow n_1+2\Delta} \ \underbrace{\ell_{n_1+2\Delta-2}\ldots \ell_{n_1+2} \ell_{n_1}\ldots \ell_{n_2}}_{n_1+2\Delta \rightarrow n_2} \ =(-1)^{\Delta}\frac{(n_1+2\Delta)!}{\sqrt{n_1!n_2!}}.
\end{align}
Thus,
\begin{align}
     |f(\Delta)|\leq \binom{p_0+2\Delta}{\Delta}\frac{(\mathrm{max}(n_1,n_2)+2\Delta)!}{\sqrt{n_1!n_2!}},
\end{align}
and the contribution at $\Delta$ can be bounded as
\begin{align}
    |t(\Delta)| \leq s(\Delta)\coloneq\frac{r^{p_0}}{2^{p_0}\sqrt{n_1!n_2!}} \left(\frac{r}{2}\right)^{2\Delta}\frac{(\mathrm{max}(n_1,n_2)+2\Delta)!}{\Delta!\left(p_0+\Delta\right)!}.
\end{align}

The corresponding matrix element for $\hat S^{\Lambda}(r)$ is obtained by dropping all contributions in the sum in Eq.~\eqref{eq:sum-path} from paths visiting states $\ket{n}$ with $\ n>\Lambda$. Thus, we obtain an analogous expansion,
\begin{align}
    \langle  n_2|\hat S^{\Lambda}(r)|n_1\rangle = \sum_{\Delta=0}^{\infty}t^{\Lambda}(\Delta),
\end{align}
where $|t^{\Lambda}(\Delta)|\leq s(\Delta)$. Furthermore, $t(\Delta)=t^{\Lambda}(\Delta)$ for $\Delta \leq \lfloor\frac{\Lambda-\mathrm{max}(n_1,n_2)+1}{2}\rfloor$ because no paths of these lengths can reach states beyond the cutoff. Thus, the difference between the relevant matrix elements of $\hat S(r)$ and $\hat S^\Lambda(r)$ can be bounded as
\begin{align}
    \left|\langle n_2 |\hat S(r)-\hat S^{\Lambda}(r)|n_1\rangle\right|=\left|\sum_{\Delta=\lfloor\frac{\Lambda-\mathrm{max}(n_1,n_2)+1}{2}\rfloor+1}^{\infty}\left(t(\Delta)-t^{\Lambda}(\Delta)\right)\right|\leq \quad 2\sum_{\Delta=\lfloor\frac{\Lambda-\mathrm{max}(n_1,n_2)+1}{2}\rfloor+1}^{\infty} s({\Delta}),
\end{align}
for $n_1\leq R$ and $n_2 \leq \Lambda$. Consider the ratio
\begin{align}
    b(\Delta)\coloneq\frac{s(\Delta+1)}{s(\Delta)}=\left(\frac{r}{2}\right)^2\frac{\left(\mathrm{max}(n_1,n_2)+2\Delta+1\right)\left(\mathrm{max}(n_1,n_2)+2\Delta+2\right)}{(\Delta+1)(p_0+\Delta+1)}.
\end{align}
Observe that $b(\Delta)\rightarrow r^2<1$ as $\Delta \rightarrow \infty$. Furthermore, denoting $q_0\coloneq\mathrm{max}(n_1,n_2)$, we see that
\begin{align}
    b(\Delta+1)-b(\Delta)&=\frac{a_2(q_0,p_0)\Delta^2+a_1(q_0,p_0)\Delta+a_0(q_0,p_0)}{(\Delta+1)(p_0+\Delta+1)(\Delta+2)(p_0+\Delta+2)},
\end{align}
with
\begin{align}
    a_2(q_0,p_0)&\coloneq 2-4q_0 + 4p_0, \nonumber \\
    a_1(q_0,p_0)&\coloneq 6-10q_0-2q_0^2 + 12p_0, \nonumber \\
    a_0(q_0,p_0)&\coloneq 4-5q_0 -3q_0^2+8p_0+q_0p_0-q_0^2p_0.
\end{align}
Thus, when $n_1=n_2=0$ (or equivalently when $q_0=0$ and $p_0=0$), the sequence of ratios $b(\Delta)$ increases monotonically with $\Delta$. Since $b(\Delta)$ is upper bounded by $r^2$, in this case we can bound the above sequence by a geometric series as
\begin{align}
    \left|\langle 0 |\hat S(r)-\hat S^{\Lambda}(r)|0\rangle\right| \leq \tilde{G}_0(r,R,\Lambda) \coloneq\frac{2s\left(\lfloor\frac{\Lambda-\mathrm{max}(n_1,n_2)+1}{2}\rfloor+1\right)}{1-r^2}.
\end{align}
In all other cases, $b(\Delta)$ decreases monotonically with $\Delta$ and is lower bounded by $r^2$. Thus, one can find the smallest value of $\Delta\geq \lfloor\frac{\Lambda-\mathrm{max}(n_1,n_2)+1}{2}\rfloor+1$ at which the ratio satisfies $\frac{s(\Delta+1)}{s(\Delta)}\leq q$ for some $r^2<q\leq 1$. Suppose this value is denoted by $\Delta_q$. Then, the above series can be bounded as,
\begin{align}
    \sum_{\Delta=\lfloor\frac{\Lambda-\mathrm{max}(n_1,n_2)+1}{2}\rfloor+1}^{\infty} s({\Delta})=\sum_{\Delta=\lfloor\frac{\Lambda-n_2+1}{2}\rfloor+1}^{\Delta_q-1} s({\Delta}) + \sum_{\Delta=\Delta_q}^{\infty}s(\Delta)\leq \left(\sum_{\Delta=\lfloor\frac{\Lambda-\mathrm{max}(n_1,n_2)+1}{2}\rfloor+1}^{\Delta_q-1} s({\Delta})\right)+\frac{s(\Delta_q)}{1-q}.
\end{align}
Because $b(\Delta)$ can never drop below $r^2$ in this case, we set the threshold to be just $10\%$ above this lower bound, i.e., $q=1.1r^2$. This choice keeps the geometric-tail prefactor $\frac{1}{1-q}$ small and the resulting bound tight. With this choice of $q$, we have:
\begin{align}
    \left|\langle n_2 |\hat S(r)-\hat S^{\Lambda}(r)|n_1\rangle\right| \leq \tilde{G}(r,R,\Lambda,n_1,n_2) \coloneq 2\left(\sum_{\Delta=\lfloor\frac{\Lambda-\mathrm{max}(n_1,n_2)+1}{2}\rfloor+1}^{\Delta_q-1} s({\Delta})\right)+\frac{2s(\Delta_q)}{1-q}; \ n_1+n_2>0.
\end{align}
Further, we define $\tilde{G}(r,R,\Lambda,0,0)=\tilde{G}_0(r,R,\Lambda)$. Therefore,
\begin{align}
    g^{\rm disc}_{r,R,\Lambda}\leq G(r,R,\Lambda)\coloneq \mathrm{argmax}_{n_1\leq R}  \ \left[ \sum_{n_2=0}^{\Lambda}\tilde{G}(r,R,\Lambda,n_1,n_2)^2\right]^{1/2}.
\end{align}
Thus, the final expressions for the truncation error are given by
\begin{align}
    &\norm{\left(\hat S(r)-\hat S^{\Lambda}(r)\right)\ket{C}_R}\leq \epsilon^{(1)}_{\rm trunc}(r,R,\Lambda)\coloneq\sqrt{\frac{R}{2}+1}\Big[G(r,R,\Lambda)+g^{\rm leak}(r,R,\Lambda)\Big], \nonumber \\
    &\norm{\left(\otimes_{j=0}^{N-1}\hat S_j(r)-\hat \otimes_{j=0}^{N-1}S_j^{\Lambda}(r)\right)\ket{C}_{R,Q}} \leq \epsilon^{(N)}_{\rm trunc}(r,R,\Lambda,Q)\coloneq N\sqrt{N|c_{R,Q}|}\Big[G(r,R,\Lambda)+g^{\rm leak}(r,R,\Lambda)\Big].
\end{align}

\subsection{Trotter error}\label{app:trotter}

While we presented a state-dependent, i.e., $(R,Q)$-dependent, bound for the truncation error, it is common and often more straightforward to establish a state-independent bound for the Trotter error.  Thus, we will bound,
\begin{align}
    \norm{\ket{\psi^{\Lambda}}_R-\ket{\psi^{\Lambda,K}}_R} \leq \norm{\hat S^{\Lambda,K}(r)-\hat S^{\Lambda}(r)},
\end{align}
where $\norm{\hat A}$ denotes the spectral norm of the operator which is induced by the inner-product norm $\norm{\ket{\psi}}$ on the full single-mode or multimode bosonic Hilbert space. The starting point is Eq.~\eqref{eq:T-trotter-decom} of the main text, which is the first-order Trotter decomposition of $\hat S^{q,\Lambda}(r)$,
\begin{align}
    \hat S^{\Lambda,K}(r) = \left(e^{-i\frac{r}{K}\hat s_0^{\Lambda}} e^{-i\frac{r}{K}\hat s_2^{\Lambda}}\right)^K \left(e^{-i\frac{r}{K}\hat s_1^{\Lambda}} e^{-i\frac{r}{K}\hat s_3^{\Lambda}}\right)^K,
\end{align}
with the $\hat s_{0,\cdots,3}^{\Lambda}$ operators defined in Eq.~\eqref{eq:s-q-lambda-m-def}. Noting that for unitary operators $\hat A_0,\ldots,\hat A_{m-1},\hat B_0,\ldots,\hat B_{m-1}$,
\begin{align}
    \norm{\hat A_0\ldots \hat A_{m-1}-\hat B_0\ldots\hat B_{m-1}}&=\norm{\sum_{k=0}^{m-1} \ (\hat A_0\ldots \hat A_{k-1})(\hat A_k-\hat B_k)(\hat B_{k+1}\ldots\hat B_{m-1})} \nonumber \\
    &\leq \sum_{k=0}^{m-1} \norm{\hat A_1\ldots \hat A_{k-1}}\norm{\hat A_k-\hat B_k}\norm{\hat B_{k+1}\ldots\hat B_{m-1}}  
    \nonumber \\
    &= \sum_{k=0}^{m-1} \norm{\hat A_k-\hat B_k},
\end{align}
one obtains,
\begin{align}
    \norm{\hat S^{\Lambda,K}(r)-\hat S^{\Lambda}(r)}\leq K\left(\norm{e^{-i\frac{r}{K}\hat s_0^{\Lambda}} e^{-i\frac{r}{K}\hat s_2^{\Lambda}}-e^{-i\frac{r}{K}(\hat s_0^{\Lambda}+\hat s_2^{\Lambda})}}+\norm{e^{-i\frac{r}{K}\hat s_1^{\Lambda}} e^{-i\frac{r}{K}\hat s_3^{\Lambda}}-e^{-i\frac{r}{K}(\hat s_1^{\Lambda}+\hat s_3^{\Lambda})}}\right).
\end{align}
The Trotter error bound, following Ref.~\cite{childs2021theory}, is calculated as:
\begin{align}
    \norm{e^{-i\frac{r}{K}\hat s_m^{\Lambda}} e^{-i\frac{r}{K}\hat s_{m+2}^{\Lambda}}-e^{-i\frac{r}{K}(\hat s_m^{\Lambda}+\hat s_{m+2}^{\Lambda})}} \leq \frac{r^2}{2K^2}\norm{[\hat s_m^{\Lambda},\hat s_{m+2}^{\Lambda}]},
\end{align}
with $m=0,1$. This gives the following expression for the Trotter error bound for the single-mode squeezing operator:
\begin{align}
    \norm{\hat S^{\Lambda,K}(r)-\hat S^{\Lambda}(r)}\leq \ \epsilon^{(1)}_{\rm trott}(r,\Lambda,K) \coloneq \frac{r^2}{2K}\left(\norm{[\hat s_0^{\Lambda},\hat s_{2}^{\Lambda}]}+\norm{[\hat s_1^{\Lambda},\hat s_{3}^{\Lambda}]}\right),
    \label{eq:epsilon-def}
\end{align}
and ultimately for the multimode product squeezing operator,
\begin{align}
    \norm{\bigotimes_{j=0}^{N-1} \hat S_j^{\Lambda,K}(r)-\bigotimes_{j=0}^{N-1} \hat S_j^{\Lambda}(r)}\leq \ \epsilon^{(N)}_{\rm trott}(r,\Lambda,K) \coloneq N\epsilon^{(1)}_{\rm trott}(r,\Lambda,K).
\end{align}
While we will eventually compute the above commutator norms exactly to arrive at the bound, it is also useful to evaluate a looser but analytic bound. We first observe that the commutator evaluates to:
\begin{align}
    [\hat s^\Lambda_m,\hat s^\Lambda_{m+2}]&= \sum_{n=0}^{\lfloor \frac{\Lambda-2-m}{4}\rfloor}\sum_{n'=0}^{\lfloor \frac{\Lambda-4-m}{4}\rfloor}[\hat T_{4n+m},\hat T_{4n'+m+2}] \nonumber \\
    &= -\frac{1}{4}\sum_{n=0}^{\lfloor \frac{\Lambda-2-m}{4}\rfloor}\sum_{n'=0}^{\lfloor \frac{\Lambda-4-m}{4}\rfloor}\ell_{4n+m}\ell_{4n'+m+2}\Big( \ket{4n+m+2}\bra{4n'+m+2}\delta_{n,n'+1}+
    \nonumber\\
    & \hspace{6.2 cm} \ket{4n+m}\bra{4n'+m+4}\delta_{n,n'} \Big) - \mathrm{h.c.} \nonumber \\
    &= -\frac{1}{4}\sum_{n=0}^{\lfloor \frac{\Lambda-2-m}{4}\rfloor}\Big(\ell_{4n+m}\ell_{4n+m-2}\ket{4n+m+2}\bra{4n+m-2}+ \ell_{4n+m}\ell_{4n+m+2}\ket{4n+m}\bra{4n+m+4}\Big) - \mathrm{h.c.}
\end{align}
\noindent Thus, the norm of this commutator can be bounded as
\begin{align}
    \norm{[\hat s^\Lambda_m,\hat s^\Lambda_{m+2}]} \leq \beta({\Lambda})\coloneq\frac{1}{2}\sum_{n=0}^{\lfloor \frac{\Lambda-2-m}{4}\rfloor} \ell_{4n+m}(\ell_{4n+m-2}+\ell_{4n+m+2}),
\end{align}
\noindent which leads to
\begin{align}
    \norm{\bigotimes_{j=0}^{N-1} \hat S_j^{q,\Lambda,K}(r)-\bigotimes_{j=0}^{N-1} \hat S_j^{q,\Lambda}(r)}\leq \ \epsilon_{\rm trott}^{(N)}(r,\Lambda,K) \leq \tilde\epsilon_{\rm trott}^{(N)}(r,\Lambda,K) \coloneq \frac{Nr^2\beta(\Lambda)}{2K}.
    \label{eq:epsilon-def}
\end{align}
Thus, $\epsilon_{\rm trott}^{(N)}(r,\Lambda,K)$ is $O\left(\frac{Nr^2\Lambda^3}{K}\right)$.

\subsection{Combined errors}\label{app:combined}

The distance between the true state $\ket{\psi}_{R,Q}$ and the state obtained by the application of the truncated, Trotterized squeezing operator(s) satisfies
\begin{align}
    \norm{\ket{\psi}_{R,Q}-\ket{\psi^{\Lambda,K}}_{R,Q}}^2=2\left[1-\mathrm{Re}({}_{R,Q}\langle \psi|\psi^{\Lambda,K}\rangle_{R,Q})\right]\leq \left(\epsilon^{(N)}\right)^2.
\end{align}
Since the core state, as well as the squeezing operator(s), are characterized by real coefficients, the overlap $\langle \psi|\psi^{\Lambda,K}\rangle$ is real. Further, we work in regimes where the net error $\epsilon^{(N)}<1$. This means that the fidelity $F=|\langle \psi|\psi^{\Lambda,K}\rangle|^2$ can be bounded as
\begin{align}
    F \geq \left[1-\frac{\left(\epsilon^{(N)}\right)^2}{2}\right]^2.
\end{align}
Similar results hold for the single-mode case with $\ket{\psi}_{R,Q}$ and $\ket{\psi^{\Lambda,K}}_{R,Q}$ replaced by $\ket{\psi}_R$ and $\ket{\psi^{\Lambda,K}}_{R}$, respectively, and $\epsilon^{(N)}$ replaced by $\epsilon^{(1)}$.

Assuming $\epsilon_{\rm trunc}^{(N)}=\epsilon_{\rm trott}^{(N)}$, we report in Table~\ref{tab:cutoff_layers} of the main text, the smallest truncation $\Lambda$ and the number of Trotter layers $K$ that guarantee a fidelity $F\geq F_0$ based on the above computed bounds, for the system parameters considered in this work. The above estimates for the minimum value of cutoff and the number of Trotter layers result from fairly conservative estimates of the errors, and thus, there is substantial scope for tightening these bounds. In particular, the Trotter error bound is state-independent, and a potentially tighter bound could be obtained by specializing the derivation to the highly constrained core state.

\section{On the implementation of the squeezing operator
\label{app:implementation}}

In Sec.~\ref{sec:qsqueeze}, two implementation algorithms were introduced for implementing individual Trotter layers of the Trotterized squeezing operator in Eq.~\eqref{eq:T-trotter-decom}: the direct and SVD implementations. For a binary map of the bosonic register, direct implementation can introduce additional Trotter error (not accounted for by the analysis of the previous section), and is hence less favorable. An alternative method which works for both the unary and binary representations is based on the singular-value-decomposition (SVD) technique of Ref.~\cite{davoudi2022general}. We describe in this appendix how this algorithm can be applied to implement the squeezing operator. Additionally, the full squeezing operator can be implemented without Trotterization on a hybrid analog-digital device like that used in Ref~\cite{andersen2025thermalization}. We further describe this scheme in this appendix.

\subsection{Implementation of the squeezing operator using a singular-value decomposition
\label{app:svd}}
Each operator $\hat s_m^{q,\Lambda}$ with $m \in \{0,\cdots,3\}$ in Eq.~\eqref{eq:s-q-lambda-m-def} can be explicitly expressed as
\begin{align}
    \hat s_m^{q,\Lambda} &= \hat s_m^{q,\Lambda,+}+\hat s_m^{q,\Lambda,-},
\end{align}
where
\begin{align}
    \hat s_m^{q,\Lambda,+}& \coloneq \frac{i}{2}\sum_{n=0}^{\lfloor \frac{\Lambda-2-m}{4} \rfloor} \ell_{4n+m} \ket{q^{\Lambda}(4n+m+2)}\bra{4n+m}, \ s_m^{q,\Lambda,-} \coloneq \left(s_m^{q,\Lambda,+}\right)^{\dagger}.
\end{align}
Note importantly that $\left(s_m^{q,\Lambda,+}\right)^2=\left(s_m^{q,\Lambda,-}\right)^2=0$. The above decomposition of $\hat s_m^{q,\Lambda}$ makes it amenable to an SVD implementation. Following Ref.~\cite{davoudi2022general}, we introduce an ancilla qubit, and note that the operator $\hat s^{q,\Lambda}_m\otimes |0 \rangle\langle 0|_{\rm anc}$ can be decomposed as
\begin{align}
    \hat s^{q,\Lambda}_m
    |0 \rangle\langle 0|_{\rm anc} &= \ \hat{\mathfrak{U}}^{\dagger} 
    \hat Z_{\rm anc} 
    \hat D_m 
    \hat{\mathfrak{U}},
\end{align}
where $\hat Z_{\rm anc}$ is the Pauli-$Z$ operator acting on the ancilla, $\hat D_m$ is a singular-value matrix which will be defined below, and the diagonalizing unitary is given by
\begin{align}
    \hat{\mathfrak{U}}&=\hat{\tt H}_{\rm anc}
    \left(|0 \rangle\langle 0|_{\rm anc}
    \hat V^{\dagger}+ |1  \rangle\langle 1|_{\rm anc}
    \hat W^{\dagger}\right)
    \hat P.
\end{align}
\noindent Here, $\hat{\tt H
}_{\rm anc}$ is the Hadamard gate acting on the ancilla qubit. The operators $\hat V$ and $\hat W$ feature in the SVD of $\hat s_a^{q,\Lambda,+}$:
\begin{align}
    \hat s_m^{q,\Lambda,+}
    &= \left(\hat I^+\right)^2 
    \left(\frac{i}{2}\sum_{n=0}^{\lfloor\frac{\Lambda-2-m}{4}\rfloor} \ell_{4n+m} \ \ket{q^{\Lambda}(4n+m)}
    \bra{q^{\Lambda}(4n+m)}
    \right) 
    \hat 1 
    \equiv \hat V\hat D_m\hat W^{\dagger},
\end{align}
\noindent with the singular-value matrix $\hat D_m$. Above, we have introduced the incrementor operator, $ \hat I^{+}\ket{q^{\Lambda}(n)}=(1-\delta_{n,\Lambda})\ket{q^{\Lambda}(n+ 1)}$ (with the decrementor operator being $\hat I^{-}\ket{q^{\Lambda}(n)}=(1-\delta_{n,0})\ket{q^{\Lambda}(n-1)}$). The operator $\hat P$ is defined by the action $\hat P\ket{0}_{\rm anc}\ket{\psi}=\ket{b}_{\rm anc}\ket{\psi}$, where the bit $b$ is 0 when the state $\ket{\psi} \in \mathrm{ker}\left(\hat s^{q,\Lambda,+}\right)$, and 1 otherwise. Thus, one finally arrives at
\begin{align}
    e^{-i\frac{r}{K} \ \ket{0}\bra{0}_{\rm anc} 
    \hat s^{q,\Lambda}_m } = \hat{\mathfrak{U}}^{\dagger} \ e^{-i\frac{r}{K} \ \hat Z_{\rm anc} 
    \hat D_m} \ \hat{\mathfrak{U}},
    \label{eq:s-SVD-form}
\end{align}
which implements the desired unitary operator acting on the bosonic qubit register upon initializing the ancilla qubit in state $\ket{0}_{\rm anc}$. 

The cost of implementing the diagonalizing transformation $\mathfrak{U}$ is dominated by the cost of the incrementor (controlled on the ancilla state). Using near-term strategies (which amount to introducing no or minimal additional ancilla qubits), this operation can be achieved with $O(n_q(\Lambda)^2)$ CNOT gates and $O(n_q(\Lambda)^2)$ single-qubit rotations~\cite{davoudi2023general}. A near-term strategy for implementing the diagonal unitary $\hat D_m$ is to decompose (using classical preprocessing) the exponent of the diagonal operator to a sum of products of Pauli $Z$ operators acting on the qubits of the bosonic register (i.e., to find the Walsh-series expansion of the exponent~\cite{golubov2012walsh}). In general, there are $2^{n_q(\Lambda)}$ such terms in the decomposition with up to $n_q(\Lambda)$ Pauli-$Z$ operators in each term. Each of these terms can be exponentiated separately with no Trotter error, each costing $O(n_q(\Lambda))$ CNOT gates and single-qubit rotation gates. The total CNOT and single-qubit gate costs of implementing Eq.~\eqref{eq:s-SVD-form} are, therefore, $O(n_q(\Lambda)2^{n_q(\Lambda)})$ and $O(2^{n_q(\Lambda)})$, respectively. This exponential cost can be traded for a polynomial cost using far-term algorithms that employ a quantum phase-kickback routine and extra ancillary registers. Quantumly evaluating the coefficients $\ell_n$ can, in general, be costly~\cite{davoudi2023general}. Nonetheless, one can classically evaluate and store these values when acting on a given $(R,Q)$ core state, and subsequently hardcode them in the quantum circuit.

Table~\ref{tab:squeezing} summarizes the cost associated with various (near-term) strategies for implementing each Trotter layer of the squeezing operator, i.e., $e^{-i\frac{r}{K}\hat s^{q,\Lambda}_m }$.\\
\begin{table}[t!]
    \centering
    \begin{tabular}{c|c|c|c}
    	 Count & Direct (unary) & SVD (unary) & SVD (binary) \\
    	\hline
    	\begin{tabular}{c}
        Qubits in bosonic register, i.e., $n_q(\Lambda)$  \end{tabular} & $\Lambda +1$ & $\Lambda +1$ & $\lceil \mathrm{log}_2(\Lambda+1)\rceil$  \\
    	\hline
        Ancilla qubits & 0 & 1 & 1\\
        \hline
        \begin{tabular}{c} CNOT gates per Trotter layer  \end{tabular} & $
        O(\Lambda)$ & $O(\Lambda \, 2^{\Lambda})$ & $O(\Lambda \, \mathrm{log}\Lambda)$ \\
        \hline
        \begin{tabular}{c} Single-qubit gates per Trotter layer  \end{tabular} & $
        O(\Lambda)$ & $O(2^{\Lambda})$ & $O(\Lambda)$ \\
        \hline
    \end{tabular}
    \caption{Summary of various qubit implementations of the single-mode squeezing operator. The above table details the complexity of a single Trotter layer. The Trotter error associated with all three methods is the same, and is discussed in the text and the previous appendix. When implementing a tensor product of such single-mode squeezing operators using the SVD method, one could reuse the ancilla.
    }
    \label{tab:squeezing}
\end{table}

\subsection{Implementation of the squeezing operator using a hybrid analog-digital simulator}\label{app:fsim}
Recall that according to Eq.~\eqref{eq:qsqueeze_map} of the main text, the single-mode squeezing operator in the unary map can be expressed as:
\begin{align}
    \hat S^{\text{u},\Lambda}(r) 
    &= e^{\frac{r}{2}\sum_{n=0}^{\Lambda-2} \sqrt{(n+1)(n+2)} \ket{\text{u}^\Lambda(n+2)}\bra{\text{u}^\Lambda(n)} - \ket{\text{u}^\Lambda(n)}\bra{\text{u}^\Lambda(n+2)}}.
\end{align}
Given the presentation of the unary-encoded bosonic state in Eq.~\eqref{eq:unary-state}, $\hat S^{\text{u},\Lambda}$ can be regraded as evolution by a ``squeezing Hamiltonian''
\begin{align}
   \hat H_{\text{squeezing}} 
   &= \sum_{n=0}^{\Lambda-2} \sqrt{(n+1)(n+2)}(\hat Y_{n+2}\hat X_{n}-\hat X_{n+2}\hat Y_{n})
\end{align}
for time $r/4$. 

In architectures such as that used in Ref.~\cite{andersen2025thermalization}, tunable couplings between transmon qubits result in a nearest-neighbor XY Hamiltonian,
\begin{align}
    \hat H_{\text{device}} = \sum_j \omega_j \hat Z_j +  \sum_{\langle j,k\rangle} \frac{g_{j,k}} 2 (\hat X_j\hat X_k + \hat Y_j\hat Y_k) + \cdots,
\end{align}
where the ellipsis denotes residual terms consisting of parametrically suppressed farther-neighbor $XX + YY$ terms.
One can match the connectivity of the effective squeezing Hamiltonian to the connectivity of the device by mapping the unary encoding onto a ladder with legs 1 and 2.
Next, one should match the Pauli content of the analog-device Hamiltonian by applying a site-dependent basis transformation.

The matching requires a new, modified unary representation, which we denote as $\tilde{\text{u}}$.
In this modified unary representation, the mapping of states reads
\begin{align}
\begin{split}
    & \ket{2j} \mapsto \ket{\tilde{\text{u}}^\Lambda(2j)} = (-i)^j \hat{L}^+_{1,j} 
    \ket{0}^{\otimes\, n_\text{u}(\Lambda)},\\
    &\ket{2j+1} \mapsto \ket{\tilde{\text{u}}^\Lambda(2j+1)} = (-i)^j \hat{L}^+_{2,j} 
    \ket{0}^{\otimes\, n_\text{u}(\Lambda)},
\end{split}
\end{align}
where $\hat{L}^+_{1,j}$ and $\hat{L}^+_{2,j}$ are raising operators on leg 1 rung $j$ and leg 2 rung $j$ of the ladder, respectively. 
This modified representation is related to the previous unary representation $\text{u}$ by an onsite basis change. One can check that in this basis
\begin{align}
    \hat X_{1,j} \hat X_{1,j+1} + \hat Y_{1,j} \hat Y_{1,j+1} = 2i\ketbra{\tilde u^\Lambda(2j+2)}{\tilde u^\Lambda(2j)} + \text{h.c.},
\end{align}
(and analogously for rung 2), as desired. Therefore, by setting $\omega_j=0$ and tuning the couplings $g_{j,k}$ in experiment such that only nearest-neighbor transmons on the same leg couple, the squeezing Hamiltonian can be engineered. Consequently, evolving this system continuously for time $t$ (with $g_{j,j+1}=\frac{r\ell_{2j}}{2t}$ on leg 1 and $g_{j,j+1}=\frac{r\ell_{2j+1}}{2t}$ on leg 2), implements the squeezing operator exactly, up to the errors associated with the residual terms.

In this work, the Gaussian part of the stellar ansatz does not involve displacement operators due to symmetry constraints. However, as discussed in the next appendix, the implementation of the core state itself may involve displacements. Under the unary map, the displacement operator takes the form,
\begin{align}
    \hat D^{u,\Lambda}(\alpha) = e^{\sum_{n = 0}^{\Lambda-1} \sqrt{n+1}
    \Big(\alpha \hat L^{+,u,\Lambda}_{n+1} \hat L^{-,u,\Lambda}_{n} - \alpha^* \hat L^{+,u,\Lambda}_{n} \hat L^{-,u,\Lambda}_{n+1}
    \Big)}.
\end{align}
To implement this operator via a unary mapping to the ladder geometry above, thus, requires not only couplings between legs at the same rung, but also diagonal couplings $2,j \to 1,j+1$.
Those diagonal couplings between qubits would require either diagonal couplings between transmons,
or (potentially) careful Floquet engineering of the residual second-neighbor terms in the device Hamiltonian $\hat H_{\text{device}}$.

\section{On the continuous-variable approach to implementing the $(R,Q)$ ansatz
\label{app:CV}}
\noindent
In this appendix, we discuss what ingredients are needed to implement our finite-rank $R$, span-$Q$ ansatz on a continuous-variable quantum platform. The advantage of continuous-variable platforms is that the scalar field can be faithfully encoded into the simulator's degrees of freedom without the need for any Hilbert-space truncation. Thus, one can avoid the truncation errors encountered in discrete-variable platforms, and the need to perform any extrapolation to the infinite truncation cutoff. Quantum simulations of quantum field theories with continuous-variable platforms have gained significant attention~\cite{marshall2015quantum,davoudi2021toward,jha2023toward,ale2024quantum,briceno2023toward,crane2024hybrid}. Bosonic degrees of freedom occur in photonic quantum simulators, phononic modes in trapped ion systems, and electromagnetic fields in superconducting microwave cavities and circuit QED. Here, we assume a generic bosonic platform that can perform:
\begin{enumerate}
    \item \textit{Single-mode Gaussian operations}: These include the squeezing $\hat S(\xi)=e^{\frac{1}{2}\left[\xi(\hat a^{\dagger})^2-\xi^*\hat a^2\right]}$ and displacement $\hat D(\alpha)=e^{\alpha\hat a^{\dagger}-\alpha^*\hat a}$ operators with $\xi,\alpha\in \mathbb{C}$.
    \item \textit{Multimode Gaussian operations}: These include passive rotations $R(\Phi)=e^{i\sum_{j,j'=0}^{N-1}\hat a_j^{\dagger}\Phi_{j,j'}\hat a_{j'}}$.
    \item \textit{Single-boson additions} $\hat a_j^{\dagger}$.
\end{enumerate}
Among the above Gaussian operations, single-mode displacements and passive rotations are often easier to perform than single-mode squeezing. For example, in a photonic device, the latter is limited to smaller squeezing-parameter magnitudes $|\xi| = r$. Roughly, this is because as $r$ increases, the mean photon number in the squeezed mode grows as $\langle \hat a^{\dagger}\hat a\rangle=\mathrm{sinh}^2(r)$. States with such a large mean photon number are significantly more susceptible to decoherence from photon loss, and interactions with the environment. Gaussian operations form a continuous-variable analog of the Clifford group, and they can be efficiently simulated classically~\cite{bartlett2002efficient, weedbrook2012gaussian}. (In the field of quantum optics, squeezed states are often referred to as non-classical because they exhibit reduced-noise quadratures. Yet, they possess Gaussian Wigner functions and \emph{can} be simulated efficiently classically.) Thus, non-Gaussian operations are needed to achieve universal continuous-variable quantum computing. The non-Gaussian boson-addition primitive is considerably more demanding than any Gaussian gate because it typically relies on probabilistic schemes~\cite{marco2010manipulating} or requires strong, precisely controlled interactions with ancillary systems (like qubits or other bosons), which are difficult to implement with high fidelity and efficiency~\cite{hofheinz2008generation}. In contrast, as demonstrated in Sec.~\ref{sec:DV}, preparing our $(R,Q)$ ansatz core state on a discrete-variable quantum computer is rather straightforward.

We will now explore the feasibility of implementing the $(R,Q)$ ansatz with the above operations. Following Ref.~\cite{chabaud2020stellar}, the single-mode finite-rank ansatz can be decomposed into the above elementary operators as 
\begin{align}
    \ket{\psi}_R &= \hat S(r) \left(c_0\ket{0} + c_2\ket{2} + \cdots + c_R\ket{R}\right) \nonumber \\
    &= \hat S(r) \left(c'_0+c'_2(\hat a^{\dagger})^2 + \cdots + c'_R(\hat a^{\dagger})^R \right) \ket{0} \nonumber \\
    &= \hat S(r) \, c'_R \prod_{n=0,2,\ldots,R}(\hat a^{\dagger}-\beta_n)\ket{0} \nonumber \\
    &=  \hat S(r) \, c'_R \prod_{n=0,2,\ldots,R}\hat D(\beta_n^*)\hat a^{\dagger}\hat D^{\dagger}(\beta_n^*)\ket{0},
\end{align}
where $\beta_n\in \mathbb{C}$ are the roots of the core-state polynomial $C(\hat a^{\dagger})=c_0'+c_2'(\hat a^{\dagger})^2 +\ldots +c'_R(\hat a^{\dagger})^R$. Therefore, the single-mode core state can be prepared by performing a series of displaced single-boson additions 
on
the vacuum state. The value of these displacements is given by the roots of the core-state polynomial. Thereafter, the full symmetric rank-$R$ ansatz can be prepared by applying the real squeezing operator to this core state. \\

Consider now the multimode case. Just like the single-mode case, a multimode core state is generated by a polynomial in the creation operators acting on the multimode vacuum, i.e., $\ket{C}={C}(\hat{\bm{a}}^{\dagger})\ket{\bs{0}}$. Since $C(\hat{\bm{a}}^{\dagger})=C(a_0^{\dagger},\cdots,\hat a_{N-1}^{\dagger})$ is a multivariate polynomial, in general, it does not possess a discrete set of roots (unlike the single-mode core-state polynomial described above). Consequently, in general, an arbitrary multimode core state cannot be prepared by performing displaced single-boson additions to the vacuum. However, as shown in Ref.~\cite{chabaud2021classical}, the above single-mode state-preparation recipe can be generalized to prepare a smaller class of multimode finite-rank states. These states are prepared by interleaving single-boson additions with multimode Gaussian operations:
\begin{align}
    \ket{\psi}_R^{\rm IPAG}& \coloneq \hat U_G^{(R)}\hat a_0^{\dagger}\hat U_G^{(R-1)}\hat a_0^{\dagger}\cdots \hat U_G^{(1)}\hat a_0^{\dagger}\hat U_G^{(0)}\ket{\bs{0}} \nonumber \\
    &= \left( \hat U_G^{(R)}\cdots \hat U_G^{(0)}\right) \left(\hat U_G^{(0)\dagger}\cdots\hat U_G^{(R-1)\dagger}\hat a_0^{\dagger}\hat U_G^{(R-1)}\cdots\hat U_G^{(0)}\right)\cdots\left(\hat U_G^{(0)\dagger}\hat U_G^{(1)\dagger}\hat a_0^{\dagger}\hat U_G^{(1)}\hat U_G^{(0)}\right)\left(\hat U_G^{(0)\dagger}\hat a_0^{\dagger}\hat U_G^{(0)}\right)\ket{\bs{0}} \nonumber \\
    &= \hat U_G\left[\alpha_0^{(R-1)}+\sum_{j=0}^{N-1}\left(u_{0,j}^{(R-1)}\hat a_j^{\dagger}+v^{(R-1)}_{0,j}\hat a_j\right)\right]\ldots\left[\alpha_0^{(0)}+\sum_{j=0}^{N-1}\left(u_{0,j}^{(0)} \hat a_j^{\dagger}+v^{(0)}_{0,j}\hat a_j\right)\right]\ket{\bs{0}} \nonumber \\
    &= \hat U_G\ket{C}^{\rm IPAG}_R,
\end{align}
where $\hat U_G^{(n)}$ with $n \in \{0,1,\ldots,R\}$ are multimode Gaussian unitary operations, and the Gaussian $\hat U_G$ is defined as $\hat U_G \coloneq \hat U_G^{(R)}\hat U_G^{(R-1)}\cdots \hat U_G^{(0)}$. In the third line, we have defined the action of the Gaussian operations $\left(\hat U_G^{(n)} \ldots \hat U_G^{(0)}\right)^{\dagger}$, with $n \in \{0,\cdots,R\}$, by the Boguliubov transformation
\begin{align}
    \hat A_j^{(n)\dagger}\coloneq\left(\hat U_G^{(n)} \cdots \hat U_G^{(0)}\right)^{\dagger}\hat a_j^{\dagger}\left(\hat U_G^{(n)}\cdots\hat U_G^{(0)}\right)=\alpha_j^{(n)}+\sum_{j'=0}^{N-1}\left(u_{j,j'}^{(n)}\hat a_{j'}^{\dagger}+v_{j,j'}^{(n)}\hat a_j\right).
    \label{eq:multimode_gaussian}
\end{align}
Here, $\alpha_j^{(n)} \in \mathbb{C}$ are complex displacement parameters while the matrices $U^{(n)}=[u^{(n)}_{j,j'}]$ and $V^{(n)}=[v^{(n)}_{j,j'}]$ with $u,v \in \mathbb{C}^{N \times N}$ satisfy $U^{(n)}U^{(n)\dagger}-V^{(n)}V^{(n)\dagger}=\mathbb{I}$ and $U^{(n)}V^{(n)T}=V^{(n)}U^{(n)T}$ to ensure the bosonic commutation relations $[A_j^{(n)},A_{j'}^{(n)\dagger}]=\delta_{j,j'}$ and $[A_j^{(n)},A_{j'}^{(n)}]=[A_j^{(n)\dagger},A_{j'}^{(n)\dagger}]=0$. States which exhibit the above decomposition are referred to as rank-$R$ interleaved photon-added Gaussian (IPAG) states (photon-added since these were originally introduced in the context of quantum optics~\cite{chabaud2021classical}). Similar to the single-mode case, the IPAG ansatz can be prepared using a series of single-boson additions sandwiched between (multimode) Gaussian operations. The interleaved multimode Gaussian operations can be implemented by means of the Bloch-Messiah decomposition~\cite{serafini2023quantum}, which involves single-mode squeezings, single-mode displacements, and passive rotations. \\

As mentioned above, rank-$R$ IPAG states constitute a strictly smaller subspace of rank-$R$ states. This was shown in Ref.~\cite{chabaud2021classical} by explicitly constructing a rank-2 state that cannot be written in the IPAG form. Thus, we must determine which $(R,Q)$ states (if any) belong to the rank-$R$ IPAG subspace. This amounts to checking if there exists a choice of Gaussian unitary transformations $\hat U_G^{(0)},\cdots,\hat U_G^{(R)}$ in the IPAG state which makes it equal to the $(R,Q)$ state of interest. While making general statements about the ability of an arbitrary $(R,Q)$ state to be represented by an IPAG state is challenging, we provide indications for the simpler case of $R=2$ with a simpler IPAG state with vanishing displacements, i.e., $\alpha_0^{(n)}=0$. With these choices, the rank-2 IPAG core state has the form
\begin{align}
    \ket{C}_2^{\rm IPAG}&=\left[\sum_{j=0}^{N-1}\left(u^{(1)}_{0,j}\hat a_j^{\dagger}+v^{(1)}_{0,j}\hat a_j\right)\right]\left[\sum_{j'=0}^{N-1}\left(u^{(0)}_{0,j'}\hat a_{j'}^{\dagger}+v^{(0)}_{0,j'}\hat a_{j'}\right)\right]\ket{\bs{0}} \nonumber \\
    &= \left[\sum_{j=0}^{N-1}v_{0,j}^{(1)}u_{0,j}^{(0)}+\sum_{j=0}^{N-1}u_{0,j}^{(1)}u_{0,j}^{(0)}\hat a_j^{\dagger 2}+\sum_{j=0}^{N-1}\sum_{j'=0}^{j-1}\left\{u_{0,j}^{(1)}u_{0,j'}^{(0)}+u_{0,j'}^{(1)}u_{0,j}^{(0)}\right\}\hat a_j^{\dagger}a_{j'}^{\dagger}\right]\ket{\bs{0}} \nonumber \\
    &= \left[\sum_{j=0}^{N-1}v_{0,j}^{(1)}u_{0,j}^{(0)}+\sum_{j=0}^{N-1}u_{0,j}^{(1)}u_{0,j}^{(0)}\hat a_j^{\dagger 2}+\sum_{q=1}^{\frac{N}{2}}\sum_{j=0}^{N-1}\left\{u_{0,j}^{(1)}u_{0,j+q}^{(0)}+u_{0,j+q}^{(1)}u_{0,j}^{(0)}\right\}\hat a_{j}^{\dagger}a_{j+q}^{\dagger}\right]\ket{\bs{0}},
\end{align}
while the $(R,Q)=(2,Q)$ (symmetric) core state takes the form
\begin{align}
    \ket{C}_{2,Q}=\left[A+\sum_{q=0}^Qb_q\sum_{j=0}^{N-1}\hat a_{j}^{\dagger}\hat a_{j+q}^{\dagger}\right]\ket{\bs{0}}.
\end{align}
To make the above states equal to each other, first, we must have $u_{0,j}^{(1)}u_{0,j}^{(
0)}=b_0$ with $j \in \{0,\cdots,N-1\}$. Assuming $b_0\neq 0$, it follows that $u_{0,j}^{(1)},u_{0,j}^{(0)}\neq 0$. Now, consider the equivalence of the coefficients of the terms $\hat a_j^{\dagger}\hat a_{j+1}^{\dagger}$ in the $(2,Q)$ state, 
\begin{align}
    &u_{0,j}^{(1)}u_{0,j+1}^{(0)}+u_{0,j+1}^{(1)}u_{0,j}^{(0)}=\frac{b_0}{u_{0,j}^{(0)}}u_{0,j+1}^{(0)}+\frac{b_0}{u_{0,j+1}^{(0)}}u_{0,j}^{(0)}=b_1,
\end{align}
which simplifies to
\begin{align}
    & \left(u_{0,j+1}^{(0)}\right)^2 + \left(u_{0,j}^{(0)}\right)^2 = b_{1,0}u_{0,j}^{(0)}u_{0,j+1}^{(0)}, 
\end{align}
with $j \in \{0,\cdots,N-1\}$ and $b_{1,0} \coloneq \frac{b_1}{b_0}$. Thus, the problem is reduced to 
the problem of $N$ beads, $u^{(0)}_{0,0},\cdots,u_{0,N-1}^{(0)}$, arranged on a closed necklace obeying an identical quadratic nearest-neighbor gluing rule $f_1(u^{(0)}_{0,j},u^{(0)}_{0,j+1})=0$ with $f_1(x,y) \coloneq x^2+y^2-b_{1,0} xy$. Thus, one may write $u_{0,j+1}^{(0)}=\lambda_{\sigma_{j}}u_{0,j}^{(0)}$, where $\lambda_{\pm}=\frac{b_{1,0}\pm\sqrt{b_{1,0}^{2}-4}}{2}$ and $\sigma_{j}\in\{\pm1\}$. The closed ring condition is given by $\prod_{j=0}^{N-1}\lambda_{\sigma_{j}}=1$. Since $\lambda_+\lambda_-=1$, this condition is satisfied when exactly $\frac{N}{2}$ of the $\sigma_j$ values are chosen to be $+1$. Once such a suitable choice of signs is picked,  $u_{0,j}^{(0)}=u_{0,0}^{(0)}\lambda_{\sigma_0}\cdots\lambda_{\sigma_j}$ is a solution with arbitrary $u_{0,0}^{(0)}\in\mathbb{C}$.  

The next step is to demand the equivalence of the $q=2$ term, i.e., to consider the coefficients of the terms $\hat a_j^{\dagger}\hat a_{j+2}^{\dagger}$. This amounts to imposing the next-to-nearest-neighbor gluing constraint given by
\begin{align}
    & \left(u_{0,j+2}^{(0)}\right)^2 + \left(u_{0,j}^{(0)}\right)^2 = b_{2,0}u_{0,j}^{(0)}u_{0,j+2}^{(0)}, 
\end{align}
where $b_{2,0} \coloneq \frac{b_2}{b_0}$. Substituting the form of the solution obtained earlier, one gets
\begin{align}
    u_{0,0}^{(0)}\left(\lambda_{\sigma_{j+2}}^2-b_{2,0}\lambda_{\sigma_{j+2}}+1\right)=0.
\end{align}
Thus, a non-trivial solution only exists when the coefficients $b_2$, as well as $b_1$ and $b_0$ (which determine the $\lambda_{\sigma}$ factors), of the $(R,Q)$ ansatz obey the above constraint. In other words, we cannot find a solution for a generic $(R=2,Q)$ ansatz.

This result should not be too surprising since the IPAG is a very restricted class of states. It may be more illuminating to consider a symmetric form of the IPAG state, or to directly use the IPAG state as the ground-state ansatz. We did not consider this latter possibility in the present work given the greater classical cost of optimizing the IPAG states as opposed to the $(R,Q)$ states. This direction, nonetheless, is worth exploring since the IPAG states greatly simplify circuit translation 
and implementation on bosonic quantum platforms.

\bibliography{references.bib}

\end{document}